\documentclass[fleqn,usenatbib]{mnras}

\usepackage{newtxtext,newtxmath}
\usepackage{float}
\usepackage[T1]{fontenc}
\usepackage{upgreek}

\DeclareRobustCommand{\VAN}[3]{#2}
\let\VANthebibliography\thebibliography
\def\thebibliography{\DeclareRobustCommand{\VAN}[3]{##3}\VANthebibliography}

\usepackage{graphicx}	
\usepackage{amsmath}	
\usepackage{pifont}
\usepackage{xcolor}
\usepackage[most]{tcolorbox}
\usepackage{placeins}

\newcommand{\cmark}{\ding{51}}%
\newcommand{\xmark}{\ding{55}}%
\newcommand{\dtg}{$\mathcal{DTG}$}
\newcommand{\dtz}{$\mathcal{DTZ}$}
\newcommand{\micrometer}{\textmu m}
\defcitealias{Chabrier03}{C03}
\defcitealias{Wiersma09}{W09}
\defcitealias{Correa25}{C25}
\defcitealias{Tremonti04}{T04}
\defcitealias{Ploeckinger25}{P25}

\title[Dust Evolution in COLIBRE]{Modelling the evolution and influence of dust in cosmological simulations that include the cold phase of the interstellar medium}

\author[Trayford et al.]{James W. Trayford$^{1}$\thanks{E-mail:  \href{mailto:james.trayford@port.ac.uk}{\tt james.trayford@port.ac.uk} (JWT)}, Joop Schaye$^{2}$, Camila Correa$^{2}$, Sylvia Ploeckinger$^3$, Alexander J. Richings$^{4,5}$, \newauthor{Evgenii Chaikin$^2$, Matthieu Schaller$^{6,2}$, Alejandro Ben\'{i}tez-Llambay$^{7}$, Carlos Frenk$^8$, Filip Hu\v{s}ko$^{2}$}
\\
$^{1}$Institute of Cosmology and Gravitation, University of Portsmouth, Dennis Sciama Building, Burnaby Road, Portsmouth PO1 3FX, UK\\
$^{2}$Leiden Observatory, Leiden University, PO Box 9513, NL-2300 RA Leiden, The Netherlands\\
$^{3}$Department of Astrophysics, University of Vienna, T\"{u}rkenschanzstrasse 17, 1180 Vienna, Austria\\
$^{4}$Centre for Data Science, Artificial Intelligence and Modelling, University of Hull, Cottingham Road, Hull, HU6 7RX, UK\\
$^{5}$E. A. Milne Centre for Astrophysics, Department of Physics and Mathematics, University of Hull, Cottingham Road, Hull, HU6 7RX, UK\\
$^{6}$Lorentz Institute for Theoretical Physics, Leiden University, PO Box 9506, NL-2300 RA Leiden, The Netherlands\\
$^{7}$University of Milano-Bicocca, Piazza della Scienza, 3, 20126 Milano MI, Italy\\
$^{8}$Institute for Computational Cosmology, Department of Physics, Durham University, South Road, Durham, DH1 3LE, United Kingdom}

\date{Accepted XXX. Received YYY; in original form ZZZ}

\pubyear{2025}

\begin{document}
\label{firstpage}
\pagerange{\pageref{firstpage}--\pageref{lastpage}}
\maketitle

\begin{abstract}
  While marginal in mass terms, dust grains play an outsized role in both the physics and observation of the interstellar medium (ISM). However, explicit modelling of this ISM constituent remains uncommon in large cosmological simulations. In this work, we present a model for the life-cycle of dust in the ISM that couples to the forthcoming COLIBRE galaxy formation model, which explicitly simulates the cold ISM. We follow 6 distinct grain types: 3 chemical species, including carbon and two silicate grains, with 2 size bins each. Our dust model accounts for seeding of grains from stellar ejecta, self-consistent element-by-element metal yields and growth by accretion, grain size transfer (shattering and coagulation) and destruction of dust by thermal sputtering in the ISM. We detail the calibration of this model, particularly the use of a clumping factor, to account for unresolved gas clouds in which dust readily evolves. We present a fiducial run in a 25$^3$~cMpc$^3$ cosmological volume that displays good agreement with observations of the cosmic evolution of dust density, as well as the $z=0$ galaxy dust mass function and dust scaling relations. We highlight known tensions between observational datasets of the dust-to-gas ratio as a function of metallicity depending on which metallicity calibrator is used; our model favours higher-normalisation metallicity calibrators, which agree with the observations within 0.1~dex for stellar masses $>10^9 \; {\rm M_\odot}$. We compare the grain size distribution to observations of local galaxies, and find that our simulation suggests a higher concentration of small grains, associated with more diffuse ISM and the warm-neutral medium (WNM), which both play a key role in boosting H$_2$ content. Putting these results and modelling approaches in context, we set the stage for upcoming insights into the dusty ISM of galaxies using the COLIBRE simulations.
\end{abstract}


\newcommand{\mgsilchem}{{Mg$_{2}$\,Si\,O$_{4}$}}
\newcommand{\fesilchem}{{Fe$_{2}$\,Si\,O$_{4}$}}
\newcommand{\silchem}{{(Fe$_{\, \rm X}$\,Mg$_{\, \rm 1-X}$)$_{\, 2}$\,Si\,O$_{4}$}}

\section{Introduction}

While the interaction of stars and multi-phase gas have been a primary focus in the field of galactic astrophysics, cosmic dust, a vital piece of the puzzle of how galaxies form, has received less attention. Despite comprising a marginal fraction of the baryonic mass within galaxies, these small, solid-state particulates in the interstellar medium (ISM) play a prominent role in our understanding of the physics of galaxies and their evolution. The influence of dust belies its marginal mass contribution and is felt in a variety of ISM physics; influencing cooling and heating of gas, fostering conditions for star formation, and various dynamical effects. Dust depletes metals from the gas phase \citep[e.g.][]{Jenkins09, DeCia16}, catalyses molecular hydrogen (${\rm H}_2$) formation \citep{Cazaux02}, can function as a `critical coolant' in astrophysical gas (including the intra-cluster medium, the ISM of elliptical galaxies, protostellar clouds e.g. \citealt{Matthews03, Montier04, Dopcke11}), and mediates radiative transfer through absorption and re-emission of stellar radiation \citep[e.g.][]{Whitaker17, Barisic20, Galliano21}.

In addition to its physical importance, dust shapes much of what we can learn from galaxy observations, due to the reprocessing and modulating of radiation. Dust directly shapes spectral energy distributions (SEDs) through extinction curves and dust-to-gas ratios, which are essential nuisance parameters in deducing key physical properties, such as stellar masses and star formation rates, that dictate galaxy evolution across cosmic time. This is exacerbated by the fact that converting from the \textit{attenuated} emergent SED to the  \textit{intrinsic} SED (that may be decomposed into simple stellar populations) is also a complex function of the star-dust geometry in galaxies \citep[e.g.][]{Narayanan18, Trayford20}. Approximately half of the starlight produced in galaxies throughout cosmic time having been processed by dust and re-radiated in the far infrared (FIR) \citep{Finke10}, and even higher proportions in high-redshift star-forming galaxies, where the processing of UV/optical radiation into infrared emission dominates the integrated light of star-forming galaxies \citep{Blain02}. ALMA has revealed significant dust content in distant galaxies ($z > 6$, e.g. \citealt{Palla24}), underscoring its early role in obscuring and shielding star forming gas. Dust grain populations and abundances may vary widely across high-redshift systems and have differential effects on observables, further highlighting the need for detailed models of dust evolution \citep[e.g.][]{Hou19,Granato21}.

The dust grain size distribution, composition, and spatial configuration in galaxies is critical for understanding how radiation propagates through galaxies, but remain poorly constrained outside of the Milky Way and its satellites due to the complex interplay between dust production/destruction mechanisms and the ISM environment. The majority of dust is thought to comprise carbonaceous (C) grains and silicate-based grains (dominated by O, Si, Mg, Fe). Dust forms in stellar ejecta (e.g., supernovae, AGB stars), grows via accretion of gas-phase metals, and is destroyed by sputtering in hot plasmas, while shattering and coagulation modulate the grain size distribution in turbulent regions \citep[e.g.][]{Hirashita12, Asano13}. These processes drive a transition in dominant dust growth mechanisms with metallicity: at low metallicity, stellar ejecta dominate, while accretion becomes critical at higher metallicities ($Z \gtrsim 0.1 \; Z_\odot$), explaining observed breaks in dust-to-gas vs. metallicity relations \citep[e.g.][]{RemyRuyer14, Hou19}.

Despite its importance, modelling dust evolution poses significant challenges due to additional computational complexity and profound uncertainties in physical processes like grain growth and destruction, often happening deep within dense, obscured media. Early modelling efforts focused on one-zone chemical evolution models \citep[e.g.][]{Dwek98}, but recent advances have allowed incorporation of dust into (cosmological) hydrodynamical
simulations, including dust models using either a minimal grain size distribution (via e.g. a simple \textit{two-size} approximation, \citealt[e.g.][]{Hirashita15, Hou19, Granato21}), or more extensive size distribution, modelling many grain sizes \citep[e.g.][]{Mckinnon18}. The two-size approximation, explored in \textit{smoothed-particle hydrodynamics} (SPH) and \textit{adaptive mesh refinement} (AMR) codes (across idealised galaxy, zoom-in and even cosmological, hydrodynamical simulations), tracks separate populations of small ($r_{\rm grain} \lesssim 0.03 \; {\rm \mu m}$) and large
grains ($r_{\rm grain} \gtrsim 0.1 \; {\rm \mu m}$) to capture size-dependent effects critical for modelling extinction curves and cooling rates, while forgoing excessive computational cost and the memory footprint associated with tracking a full size distribution per resolution element \citep{Hirashita15, Aoyama17, Hou19}. However, spatial inhomogeneity of dust properties, evident in Milky Way extinction curve variations and observations of nearby galaxies \citep[e.g.][]{Gordon03}, and the need for multi-species treatments (carbon vs. silicate grains, e.g. \citealt{Shivaei20}) remain important challenges \citep{Granato21, Dubois24}. 

The dust properties that emerge across galaxies and galaxy populations are a product of galaxy evolution, with dust itself driving ongoing galaxy-scale processes through its physical influence in the ISM. To capture this connection, models should consider dust evolution in a realistic galaxy formation context, and include the interdependence of dust and the ISM wherever relevant. Idealised simulations of Milky Way like discs can provide a detailed and high-cadence test bed for dust evolution, where it is feasible to explore variations and compare to a wealth of observational data. Including modelling for the dust life-cycle in this context provides insight into the build up of dust in our galaxy, and the processes needed to yield the extinction observed in the Milky Way \citep[e.g.][]{McKinnon15, Aoyama17, Hou19, Granato21}. When also paired with a multi-phase model for the ISM, we begin to capture the range of environments relevant for dust evolution self-consistently; from the cold, dense ISM in which dust grows, to the hot, diffuse shocks in which it is destroyed \citep[e.g.][]{Choban22, Dubois24}. Simpler dust models may also provide insights in this multi-phase ISM context. For example \citet{Richings22} use an empirical, instantaneous model for dust depletion to show that it is necessary to reproduce FIR emission lines. Overall, however, idealised simulations lack the cosmological context, and thus the co-evolution of dust with metal enrichment and the structural assembly of galaxies over cosmic time.

Cosmological zoom-in simulations provide a means to self-consistently capture dust-gas co-evolution over cosmic timescales. These can follow how dust properties emerge with the enrichment of the ISM, avoiding the artificial initial conditions of isolated disc models \citep{Gjergo18, Dubois24, Choban24, Choban25}. Zoom-in simulations can afford more computationally and memory intensive representations of the ISM, allowing more sophisticated modelling of physical processes and finer spatial resolution. However, these simulations lack the statistical population sampling of larger volume simulations. Large cosmological volume simulations including dust allow galaxy dust scaling relations and distributions to be tested, providing an exacting test of dust models \citep[e.g.][]{Dayal10, Bekki13, Mckinnon18, Li19}. Previously, volume simulations lacked the physics to explicitly trace the cold ISM, imposing an equation of state on the ISM gas, so cannot explicitly model the cool and cold ($\ll 10^4~{\rm K}$) ISM that is critically important for dust evolution processes, pushing more of the relevant physics to sub-grid modelling. Given these limitations for  population scales for simulations, semi-analytic models have provided insight into the potential balance of dust evolutionary processes, and have provided a means to explore the poorly understood parameter space of dust modelling on a population scale \citep[e.g.][]{Bennassuti14, Hirashita15, Popping17, Vijayan19, Yates24}.

Despite progress with simulations and insights from semi-analytical models, discrepancies persist between models and observations of high-redshift dust-rich galaxies, suggesting models may be insufficient at early cosmic epochs \citep[e.g.][]{Parente25}. Future efforts must integrate detailed grain physics with multi-phase ISM modelling to explain the observed diversity in extinction curves, e.g., the Milky Way's prominent 2175\AA{} bump versus smoother features in Magellanic Clouds \citep[e.g.][]{Noll09}, and reconcile dust evolution across metallicity and redshift regimes \citep[e.g.][]{Dubois24}.  

In this work, we present a dust evolution model for incorporation into the upcoming COLIBRE suite of next-generation cosmological hydrodynamical simulations \citep{Schaye25, Chaikin25}, which include cold gas physics in the multiphase ISM. We track dust via a relatively lightweight grain representation, comprising two grain sizes and three material species (one carbonaceous and two silicate subspecies).  In \S\ref{sec:colibre}, we first briefly describe the elements of the COLIBRE galaxy formation model outside the dust model presented here, focussing on those most relevant or connected to our dust implementation. In \S\ref{sec:dust} we then describe the components of our dust model itself (in terms of dust composition and life-cycle, with gas-dust and dust-dust interaction processes). We then explore the connection between dust and ISM physics, and salient model parameters mediating this in \S\ref{sec:couple}. We present results of this model in our fiducial test simulation run in \S\ref{sec:results}, focussing on dust properties and scaling relations of the low-redshift ($z=0.1$) population. We discuss our overall findings and present a summary of the paper and main conclusions in \S\ref{sec:discussion}.

We aim to describe our dust model in a standalone context, but we note that due to the interconnected nature of the dust presented in this work with other novel modules constituting the COLIBRE galaxy formation model, and their co-development, this is not completely possible. We describe elements of other works or refer to them where necessary, and use some forward referencing to motivate choices made in the dust model which were made in the context of developing the COLIBRE model. 

\section{The COLIBRE model \& Simulations}
\label{sec:colibre}

The dust evolution model we present below (\S\ref{sec:dust}) is integrated into the COLIBRE model of galaxy formation physics \citep{Schaye25}, run using the gravity and hydrodynamics solver SWIFT \citep{Schaller24}. In the multi-resolution parlance of COLIBRE, we use m6 to refer to a gas (DM) particle mass of $1.84\times10^6 \; {\rm M_\odot}$ ($2.42\times10^6 \; {\rm M_\odot}$). We note that COLIBRE employs 4$\times$ more dark matter particles than gas, achieving relatively high DM resolution. In this section, we first describe the general COLIBRE model, and then provide technical details of the simulation runs and parameter variations used in this work for context (\S\ref{sec:sims}).

\subsection{Energy feedback}
\label{sec:fb}

The COLIBRE model accounts for the influence of energetic processes in the ISM that transfer energy to the gas phase, termed \textit{feedback}. These processes are contributed by other baryonic phases; particularly stellar and black hole particles.

For a nascent stellar population, the very first feedback processes are the ionizing radiation, stellar winds, and radiation pressure exerted by the most massive stars. These processes are collectively termed early stellar feedback and detailed in \citet{BenitezLlambay25}. In short, the stellar age and metallicity dependent rates for stellar wind momentum injection and hydrogen ionization are from BPASS \citep{Eldridge17, Stanway18}, assuming a Chabrier (C03) IMF. The BPASS spectra are further processed through a pressure-dependent column density using CHIMES \citep{Richings14a, Richings14b} to calculate the absorbed radiation, and therefore radiation pressure, around the young star particle. The available momentum from stellar winds and radiation pressure is injected stochastically as kinetic energy with a target velocity of $50 \; {\rm km \; s^{-1}}$. The hydrogen ionization rate is used to calculate the time-dependent size of H\textsc{ii} regions (Str\"omgren spheres) around young star particles, for which a temperature floor of 10$^4$~K is imposed.

Once a massive star reaches the end of its life, it may erupt in a violent CCSN. While individual SNe are below the resolution of cosmological simulations, when acting in concert these CCSNe may drive large-scale galactic winds, forcing gas away from star forming regions and out of the galaxy itself. COLIBRE implements this `superbubble feedback', where each feedback event comprises many CCSN to mitigate numerical cooling losses. The CCSN energy budget is computed per timestep, where the number of stars expiring in an SSP is inferred from tabulated lifetimes  of stars \citep{Portinari98} with masses of $\geq 8 \; {\rm M_\odot}$ given our \citetalias{Chabrier03} IMF. This energy can be released back into surrounding gas following the thermal, stochastic approach of EAGLE \citep{DallaVecchia12}. This approach is refined following \citet{Chaikin23}, where thermal \textit{and} kinetic injection are combined. The kinetic component, constituting 10$\%$ of the energy, goes into diametrically opposed, low-velocity (target of $\Delta v=50$~${\rm km \; s^{-1}}$) kicks for particle pairs. The relatively low energy increment required for these kicks ensures good sampling of feedback events in the kinetic component, and has been shown to induce turbulence in the surrounding media, helping to reduce star formation \citep{Chaikin23}.

For the thermal part, the target heating temperature, $\Delta T_{\rm SN}(n_{\rm H})$, is an increasing function of local gas density with resolution-dependent floor and ceiling values ($6.5 < \log_{10} \Delta T_{\rm SN, min}/{\rm K}  < 7$ and $7.5 < \log_{10} \Delta T_{\rm SN, max}/{\rm K} < 8$) calibrated from test runs. In the intermediate range between these extrema, $\Delta T_{\rm SN}(n_{\rm H}) = 10^{6.5} \; {\rm K} (n_{\rm H}/ n_{\rm H,pivot})^{2/3}$. This approach mitigates overly destructive feedback and extensive bubbles in low-density gas. Finally, we adopt the statistically isotropic gas particle particle selection for energy injection of \citet{Chaikin22}, with the thermal feedback also using a variable fraction of energy per single CCSN in units of $10^{51}~\mathrm{erg}$. This is subject to a variable coupling efficiency, $f_{\rm E}$, to avoid over-cooling at high density \citep{Crain15}. In COLIBRE this is implemented as a sigmoid in SSP birth pressure, calibrated to reproduce the $z=0$ galaxy stellar mass function and galaxy sizes \citep{Chaikin25}.

For completeness, COLIBRE also implements SNIa feedback. This is purely thermal, following the CCSN thermal component implementation, with the exception that a fixed $f_{\rm E}=1$ is used. Rather than using stellar lifetimes, an exponential \textit{delay time distribution} (DTD) provides the number of SNIa per timestep, calibrated to reproduce the observed cosmic SNIa rate \citep{Nobels25}. We note that SNIa have marginal impact relative to CCSN.

Alongside stars, COLIBRE accounts for feedback by \textit{supermassive black holes} (SMBHs). SMBH particles are first seeded for friends-of-friends halos (identified on-the-fly) in a resolution-dependent way, using an SMBH particle of fixed mass ($3\times10^4 \; {\rm M_\odot}$ at m6), for halos above a halo FoF mass ($10^{10}\;{\rm M_\odot}$ at m6). These particles can then accrete gas, merge and feed energy back to the surrounding media. The COLIBRE model includes two prescriptions for AGN feedback: purely thermal energy injections and a hybrid model combining thermal and kinetic jet feedback. In this work, we employ the purely thermal feedback prescription. Following similar arguments for effective feedback in CCSN, feedback is quantised to heat surrounding gas by a target temperature increment, $\Delta T_{\rm AGN}$, to mitigate numerical cooling losses. SMBHs build up an energy reservoir until enough is accrued for a single feedback event. The energy for feedback comes from gas accretion; mass energy is converted to a luminosity, that couples back to the gas with fixed efficiency ($5\%$ at m6 resolution). In COLIBRE, $\Delta T_{\rm AGN}$ is a function of black hole mass, again enforcing floor and ceiling values ($ \log_{10} \Delta T_{\rm AGN, min} / {\rm K}  = 6.5$ and a resolution-dependent  $9.5 < \log_{10} \Delta T_{\rm SN, max} / {\rm K} < 10$) helping to maintain a reasonable sampling and regularity of BH feedback across the mass range of AGN. Feedback is meted out to the gas particle closest to the SMBH at feedback time.  

Following \citet{Ploeckinger20} and \citet{Ploeckinger25}, we also account for ambient energetic fields acting on gas particles, particularly the UV/X-ray background \citep[via][]{FaucherGiguere20}, an interstellar radiation field (ISRF) and an ionising cosmic ray background. The ISRF normalisation and cosmic ray rate, as well as the shielding column density, vary with gas pressure to reflect environments with different levels of nearby star formation and with different coherence lengths. 

\subsection{Star formation}
\label{sec:sf}

Our star formation (SF) prescription is detailed in \citet{Nobels24}. Star formation rates are applied assuming a Schmidt law, with an SF efficiency per free-fall time of $1\%$, which yields good agreement with the Kennicutt-Schmidt law observed locally \citep[e.g.][]{Ochsendorf17, Utomo18, Pokhrel21}. To be eligible for star formation, gas particles must satisfy a gravitational instability criterion, comparing turbulent and thermal velocity support to the gravitational field within their SPH kernel.

 Star formation rates are computed for eligible gas particles, which correspond to a probability of wholesale conversion into stars over a given time increment. Star formation then proceeds stochastically, sampling gas particles that are to be converted to star particles in the following timestep.

\subsection{Stellar mass loss and chemical enrichment}
\label{sec:enrichment}

As star particles form in our simulations, their stellar mass loss and chemical enrichment are modeled following the approach described in \citet{Correa25}, building on \citet{Wiersma09}. We track the evolution of 14 chemical elements: H, He, C, N, O, Ne, Mg, S, Si, Ca, Fe, Sr, Ba, and Eu. For simplicity, S and Ca are assumed to trace Si at fixed solar abundance ratios. Each star particle represents a simple, coeval stellar population (SSP). We follow the evolution of the relevant enrichment phases: core-collapse supernovae (CCSN) and asymptotic giant branch (AGB) evolution, based on the mass- and metallicity-dependent stellar lifetimes from \citet{Portinari98}. These lifetimes are used at each timestep to determine when an SSP injects mass and metals into the surrounding gas particles.

Stellar enrichment is split into two components: nucleosynthetic (new elements formed during the star's lifetime), and throughput (pre-existing elements present at birth that are carried out with the ejected material, and assumed to be well mixed within the star). AGB enrichment is attributed to stars with initial masses of 1-8~${\rm M_{\odot}}$. The AGB yields are drawn from a composite of studies, including \citet{Karakas10, Fishlock14, Doherty14, Karakas16}, and \citet{Cinquegrana22}. CCSN enrichment comes from stars born with masses of 8-40~${\rm M_{\odot}}$. These stars also undergo significant mass loss through stellar winds both during and after the main sequence. We adopt CCSN and pre-supernova wind yields from \citet{Nomoto13} and \citet{Kobayashi06}, respectively. \citet{Wiersma09} noted that relative abundance ratios predicted from nucleosynthesis yields may only be accurate within a factor of $\approx 2$, even assuming a fixed IMF. To account for this uncertainty, we apply `boost factors' to the CCSN yields, increasing the yields of C and Mg from massive stars by a factor of 1.5, calibrated to better match observed stellar abundance ratios, especially those relative to Fe from APOGEE \citep{Jonsson18}. Our simulations use the \citetalias{Chabrier03} IMF spanning the mass range 0.1-100~${\rm M}_{\odot}$. Stars born with masses greater than 40~${\rm M}_{\odot}$  are assumed to collapse directly into black holes, re-accreting all their material and locking it out of the ISM.

For SNIa, we adopt an exponential DTD. The minimum delay time is set at 40~Myr, corresponding to the time required for the first stars in binary systems to evolve into compact remnants capable of accreting from a companion. The DTD timescale and normalization are calibrated to match the observed cosmic SNIa rate
\citep{Nobels25}. SNIa yields are taken from \citet{Leung18}. Rare r-process elements are treated similarly, with a DTD approach, but their production is attributed to neutron star mergers, common envelope jet supernovae, and collapsars. These sources use different calibrations and functional forms, along with a stochastic implementation, to reflect the rarity and variability of such enrichment events.

Enrichment timesteps for star particles are designed to balance accurate sampling of stellar evolution within each SSP against computational cost. To capture the early, most active phase of enrichment, we use fine timesteps ($\leq 1$~Myr) for SSPs younger than 40 Myr. For older populations ($>100$~Myr), the time intervals after which mass transfer is performed increase proportionally to age.

Dust seeding is integrated into this enrichment framework, budgeting material from both CCSN and AGB yields to self-consistently form (and deplete, see \S\ref{sec:depletion}) dust, as described in \S\ref{sec:yields}, providing a unified picture of the build-up of metals through cosmic time. 

\subsection{COLIBRE heating and cooling without live dust}
\label{sec:cool}

A key aspect for this work is the treatment of dust in the cooling module. The fiducial treatment of cooling processes in COLIBRE is fully described in \citet{Ploeckinger25} (hereafter, \citetalias{Ploeckinger25}). In short, the species abundances of H, He, and free electrons are evolved for each gas particle\footnote{While this limited set was chosen to reduce runtime, we note our model is compatible with the with tracking all elements present in the full CHIMES chemical network in non-equilibrium.} in non-equilibrium throughout the simulation by the chemical network CHIMES \citep{Richings14a, Richings14b}, integrated in SWIFT. The species abundances of C, N, O, Ne, Mg, S, Si, Ca, and Fe are pretabulated with the stand-alone version of CHIMES, assuming ionisation equilibrium and steady-state chemistry. The total cooling rates are the sum of the cooling rates from the non-equilibrium and the equilibrium species. The equilibrium cooling rates are rescaled to account for the non-equilibrium densities of free electrons from H and He species. Of the physics modelled in this module, dust-related processes are naturally most relevant for this work, and are a key aspect in certain physical conditions.

Given the important role of dust in gas heating and cooling, the model of \citetalias{Ploeckinger25} itself requires a prescription for dust content to compute cooling and heating rates if a live dust model (as described in \S\ref{sec:dust}) is \textit{not} used. This is handled through an assumed instantaneous dust-to-gas ratio, $\mathcal{DTG}_{\rm hyb}$, based on the physical state of the gas\footnote{For the sake of brevity, we forgo a full derivation of these relationships here, but see \S2.1 (and in particular equation 16) of \citetalias{Ploeckinger25}.},
which we hereafter refer to as \citetalias{Ploeckinger25} dust. This is the dust prescription used by the cooling model of our \textit{uncoupled} dust runs (i.e. \texttt{Uncoupled} and \texttt{SeedOnly}), while in other runs we use the model described in \S\ref{sec:dust}. 

For \citetalias{Ploeckinger25} dust, it is assumed that grains are immediately destroyed at $T > 10^5\;{\rm K}$ and do not exist in relatively diffuse gas, represented by a minimum hydrogen column density, ($N_{\rm min} = 3.1\times 10^{15}\; {\rm cm}^{-2}$), such that $\mathcal{DTG}_{\rm hyb}(T>10^5\;{\rm K}) = 0$ and $\mathcal{DTG}_{\rm hyb}(N_{\rm H} \leq N_{\rm min}) = 0$. This corresponds to a range in volumetric hydrogen number densities at which shielding processes can occur of $n_{\rm H} > 10^{-8}\;{\rm cm^{-3}}$. At high column densities ($N_{\rm H} > 10^{20} \; {\rm cm}^{-2}$), a constant dust-to-metal ratio is assumed, such that\footnote{As in \citetalias{Ploeckinger25}, we use the $Z_\odot = 0.0134$ value of \citet{Asplund09} where appropriate.} $\mathcal{DTG}_{\rm hyb} \propto Z_{\rm gas}/{\rm Z_\odot}$. For intermediate column densities ($3.1\times 10^{15} < N_{\rm H}/{\rm cm}^{-2} < 10^{20}$), an additional taper is introduced to smooth the transition between the metallicity-dependent $\mathcal{DTG}_{\rm hyb}(N_{\rm H} = 10^{20} \; {\rm cm}^{-2}) \propto Z_{\rm gas}/{\rm Z_\odot}$ and $\mathcal{DTG}_{\rm hyb}(N_{\rm H} = 3.1\times 10^{15})  = 0$, assuming a power law in column density with index $\alpha = 1.4$. This is chosen based on the Kennicutt-Schmidt relation \citep{Kennicutt98}, and assuming a scaling between the star formation rate density and dust production. At high density and solar metallicity, the \dtg{} is normalised to a representative MW value of \textcolor{black}{$\mathcal{DTG}_{\rm MW} = 6.6 \times 10^{-3}$}.

One way dust influences the thermal and chemical state of the gas is through grain-gas collisions. This plays a particularly important role for molecule formation, which can happen readily on grain surfaces. For the H$_2$ formation rate on dust grains, $R_{\rm H_2}$, CHIMES adopts the approach of \citet{Cazaux02} (their equation 18)

\begin{equation}
    R_{\rm H_2} = \frac{1}{2} n_{\rm HI} \nu_{\rm HI} n_{\rm d} \sigma_{\rm d} \epsilon_{\rm H_2} S_{\rm H},
\end{equation}

\noindent where $\nu_{\rm HI}$ and $n_{\rm HI}$ are the thermal velocity and the physical number density of atomic hydrogen\footnote{We note that dust properties assumed by the cooling module do not make use of the density boost introduced in \S\ref{sec:clump}}, $\epsilon_{\rm H_2}$ is the recombination efficiency of molecular hydrogen, $S_{\rm H}$ is a dimensionless sticking coefficient and $n_{\rm d} \sigma_{\rm d}$ is the number density of dust grains times the grain cross section, i.e. the total cross section per unit volume. While $S_{\rm H}$ and $\sigma_{\rm d}$ are temperature-dependent, a constant dust temperature of 10~K is assumed\footnote{The H$_2$ formation rate was shown to be approximately constant for $6  \; {\rm K} \lessapprox T_{\rm dust} \lessapprox 50 \; {\rm K}$ in fig.~A1 of \citet{Richings14a}.}. It is then the $n_{\rm d} \sigma_{\rm d}$ term that depends on the local mass fraction and size distribution of grains, with the other terms depending on the material properties and local gas properties. Dust grains can also contribute to catalysis of ion recombination via available dust surfaces \citep[see e.g.][]{Weingartner01}, and transfer energy between the dust and gas \citep{Richings14b}. 

The influence of dust can also be felt indirectly by processing radiative energy via extinction. A particularly important example of this is photoelectric heating by dust grains; absorption of FUV photons by dust produces energetic free electrons that contribute to a heating of the gas component. In CHIMES, the volumetric heating rate, $\Gamma_{\rm PE}$, follows that of \citet{Wolfire03} (their equations 19 and 20\footnote{This is an update of what is described in \citet{Richings14a}, described by \citet{Richings18}.}), and is given by
\begin{equation}
  \Gamma_{\rm PE} = \left( 1.3\times10^{-24} \; {\rm erg \; cm^{-3} \; s^{-1}} \right) \epsilon G_0,
\end{equation}
\noindent where $G_0$ is the integrated incident UV radiation energy density, and $\epsilon$ represents the heating efficiency of the gas\footnote{Tabulated values, depending on the gas temperature $T$, the electron number density $n_{\rm e}$, and incident UV energy density, $G_0$.}.  Here, the pre-factor $1.3\times10^{-24} \; {\rm cm^{-3} s^{-1}}$ depends on the abundance as well as mass and material properties of grains (see \citetalias{Ploeckinger25}). Similarly, incidence of photodissociating and photoionising photons are reduced through shielding of gas \citep[][]{Richings14a, Richings14b}. The extinction, $A_V/N_{\rm}$, also scales with the $n_{\rm d} \sigma_{\rm d}$ term to first order.

With the grain cross-section per unit volume the relevant quantity for all of these processes, we can modulate them using the masses and sizes of grains traced through our live dust model. We describe their coupling in \S\ref{sec:couple}. A final dust-sensitive aspect to consider is the role of depletion. The same chemical elements that constitute dust also play an important cooling role in the ISM when existing in the gas-phase, particularly through metal-line cooling. These processes are handled via per-ion cooling rates tabulated for COLIBRE and based upon \citet{Oppenheimer13} (\citetalias{Ploeckinger25} equation 19\footnote{Updated from \citet{Ploeckinger20}}). These rates are based on gas-phase metals, so it is necessary to assume a per-element depletion alongside the $\mathcal{DTG}_{\rm hyb}$, in lieu of a live dust model. These assume depletion in the fixed relative proportions of the Milky Way as inferred by \citet{Jenkins09}. We discuss how we then adjust these rates for the case of live dust in \S\ref{sec:couple}, consistent with the depletions we calculate for our dust model (\S\ref{sec:depletion}). 

\subsection{Simulation Runs}
\label{sec:sims}

\begin{table*}
 \caption{Simulation variations used in this work. All runs represent a 25$^3$ Mpc$^3$ cosmological box, initialised with 376$^3$ gas particles of mass $m_{\rm g} = 1.8\times10^{6} \; {\rm M_\odot}$ (and 4 times as many dark matter particles, with mass $m_{\rm DM} = 2.4\times10^{6} \; {\rm M_\odot}$).  We adopt `m6' as shorthand for this resolution (from the COLIBRE parlance). Simulations are run using the COLIBRE galaxy formation model (Schaye et al. \textit{in prep.}), using the \texttt{swift} code \citep[][see section \ref{sec:sims}]{Schaller24}. The columns from left to right indicate: 1) the simulation name, 2) if the dust model is coupled to the cooling (see \S~\ref{sec:cool}), 3) if the model includes small grains rather than large, if we include processes of 4) accretion, 5) destruction 6) or size transfer, 7) the maximum $C$ value 8) if $C$ has a fixed value 9) the size representing the small grain bin 10) the chemical group assumed for silicates and 11) the diffusion constant (\S~\ref{sec:diff}).}
 \begin{tabular}{lllllllllll}
   \hline
   1) & 2) & 3) & 4) & 5) & 6) & 7) & 8) & 9) & 10) & 11)\\
  Name & Coupled?  & Size  & Accretion & Destruction & Size transfer & Max. $C$ & Const. $C$ & Small grain $r_{\rm grain}$ & Silicates & $C_{\rm d}$ \\
  &&&&&& (a) & & [\micron] & & \\
  \hline
    \texttt{Fiducial} & \cmark  & \cmark  & \cmark  & \cmark & \cmark & 100. & \xmark & 0.01 & Olivine & 0.01 \\ 
    \texttt{FidUncoupled} & \xmark  & \cmark  & \cmark  & \cmark & \cmark & 100. & \xmark & 0.01 & Olivine & 0.01\\
    \texttt{SeedOnly} & \xmark  & \cmark  & \xmark  & \xmark & \xmark & n/a & \xmark & 0.01 & Olivine & 0.01\\ 
    \texttt{NoDestruction} & \cmark  & \cmark  & \cmark  & \xmark & \cmark & 100. & \xmark & 0.01 & Olivine & 0.01 \\ 
    \texttt{OneSize} & \cmark  & \xmark  & \cmark  & \cmark & \xmark & 100. & \xmark & 0.01 & Olivine & 0.01 \\
    \texttt{MaxC10} & \cmark  & \cmark  & \cmark  & \cmark & \cmark & 10. & \xmark & 0.01 & Olivine & 0.01 \\
    \texttt{constC30} & \cmark  & \cmark  & \cmark  & \cmark & \cmark & 30. & \cmark & 0.01 & Olivine & 0.01 \\ 
    \texttt{NoC} & \cmark  & \cmark  & \cmark  & \cmark & \cmark & 1. & \cmark & 0.01 & Olivine & 0.01 \\
    \texttt{PyroSil} & \cmark  & \cmark  & \cmark  & \cmark & \cmark & 100. & \cmark & 0.01 & Pyroxene & 0.01 \\
    \texttt{SmallerGrains} & \cmark  & \cmark  & \cmark  & \cmark & \cmark & 100. & \xmark & 0.005 & Olivine & 0.01 \\
    \texttt{NoDiff} & \cmark  & \cmark  & \cmark  & \cmark & \cmark & 100. & \xmark & 0.01 & Olivine & 0 \\
    \texttt{LoDiff} & \cmark  & \cmark  & \cmark  & \cmark & \cmark & 100. & \xmark & 0.01 & Olivine & 0.001 \\
    \texttt{HiDiff} & \cmark  & \cmark  & \cmark  & \cmark & \cmark & 100. & \xmark & 0.01 & Olivine & 0.1 \\
  \hline
  \hline
 \end{tabular}
 (a) i.e. the clumping factor \hfill $\;$ \\
 \label{table:sims}
\end{table*}

Prior to describing our dust model in detail (\S\ref{sec:dust}, next), we describe the simulations used in this work and our choice of parameter variations. \textcolor{black}{Throughout this work we assume the $3\times2$~pt cosmology, combining all data and constraints presented with the DES Y3 results \citep{Abbott22}; $h=0.681$, $\Omega_{\rm m} = 0.306$, $\Omega_{\rm b} = 0.0486$, $\sigma_8 = 0.807$, $n_{\rm s}=0.967$.}

We use cosmological volume initial conditions for the simulations in this work. This choice allows us to evolve a population of galaxies in a cosmological context, and inspect galaxy scaling relations, which is important to illustrate the behaviour of the model over the cosmological remit of COLIBRE. 
We focus on a $25^3 \; {\rm cMpc^{3}}$ volume, with an initial $376^3$ gas particles of $m_{\rm gas} = 1.8\times10^6 \; {\rm M_{\odot}}$ and $4\times376^3$ dark matter particles with $m_{\rm DM} = 2.4\times10^{6} \; {\rm M_\odot}$, representing matched (4$\times$ higher) baryonic (dark matter) resolution relative to fiducial EAGLE \citep{Schaye15}.

We identify collapsed structures using the friends-of-friends (FoF) algorithm post processed using the publically available {\tt HBT-HERONS} subhalo finder described by \citet{Forouhar25} to assign particles to self-bound substructures. This builds on the {\tt HBT+} code for $N$-body simulations \citep{Han18}, with critical improvements for the hydrodynamical context, considering the histories and hierarchical formation of subhalos. In particular, the {\tt HBT-HERONS} code exhibits more consistent allocation of mass between satellite and central galaxies during close passes, as compared to various other phase-space halo finders \citep{Chandro25}. The subhalo locations and, optionally, the particles are then taken up by the {\tt SOAP} tool \citep{McGibbon25}, computing a wide range of aggregated properties using a range of 3D and projected apertures of varying physical size.

Here, we use gravitationally bound particles within a 50~pkpc, 3D aperture about the most-bound particle to define `galaxy' properties, unless stated otherwise. These apertures are defined about the \textcolor{black}{most-bound particle (i.e. closest to the minimum of the local potential well), as defined by the {\tt HBT-HERONS} code.}

The simulations run for analysis in this work and their defining parameters are listed in Table~\ref{table:sims}. The \texttt{Fiducial} run uses the full dust model described in \S\ref{sec:dust}, and also includes the effects of \textit{``coupling''} our dust model to gas cooling, heating and phase transition processes, detailed in \S\ref{sec:couple}. Our runs incorporate dust coupling by default (excepting \texttt{SeedOnly} where dust fractions are too unphysical), with an uncoupled counterpart to isolate the influence of the dust evolution model on the cooling, e.g. \texttt{FidUncoupled}. The \texttt{SeedOnly} and \texttt{NoDestruction} simulations isolate evolutionary processes. Size transfer effects are tested by \texttt{LargeGrainsOnly}. A number of runs are dedicated to exploring the influence of parameters; particularly a subgrid clumping factor, $C$, (described in \S\ref{sec:clump}; \texttt{NoC}, \texttt{ConstC30} and \texttt{MaxC10}) and the turbulent diffusion constant, $C_{\rm d}$ (see \S\ref{sec:diff}; \texttt{NoDiff}, \texttt{LoDiff} and \texttt{HiDiff}). We also present runs exploring pyroxene silicate chemistry, \texttt{PyroSil}, and the representative size of small grains, \texttt{SmallerGrains}. The use of coupled cooling physics by default means we see the impact of these dust model changes on the physics and resultant galaxy populations. We will explore the most pertinent variations alongside our primary low-redshift results in \S\ref{sec:results}, as well as presenting some additional variations in Appendix~\ref{sec:parvar}. 

\section{Modelling the formation \& evolution of Dust}
\label{sec:dust}

We now describe the constituent elements of our dust model, including the assumed properties of grains and the evolutionary processes of grain seeding/nucleation, grain growth, grain destruction and grain size transfer. Particular focus is placed on aspects of the modelling that differ from previous studies. Our dust model is designed for compatibility with SWIFT SPH simulations \citep{Schaller16, Schaller24}, particularly the upcoming COLIBRE simulations, as detailed above. All the processes detailed below (seeding, growth and destruction of grains), are implemented on a particle-by-particle basis, using the properties of a given gas resolution element. Throughout, we characterise the dust content in terms of the mass fractions, either as the dust-to-gas ratio (\dtg{}) or dust-to-metal ratio (\dtz{}). With the exception of direct destruction by supernovae (SNe) or astration, all our processes depend on gas density. For dense-gas processes (such as grain growth), we consider an \textit{effective} density indicated using $^\prime$ (i.e. $\rho^\prime$, $n^\prime_{\rm H}$), which is related to the physical density by a \textit{clumping factor}, C, as detailed in \S\ref{sec:clump}. The \textit{total} element abundances include both the abundances of metals that remain in the gas-phase, which we refer to as the \textit{gas-phase} component, and the abundance that is \textit{depleted} into dust grains. Throughout, we use the fiducial COLIBRE stellar initial mass function (IMF) of \citealt{Chabrier03}, with a range $\in [0.1,100] \; {\rm M}_{\odot}$ (hereafter \citetalias{Chabrier03}).

\subsection{Types of dust grains}
\label{sec:diversity}

Interstellar dust grains are diverse, in terms of both their structure and chemical composition. Practically, representing the full diversity of grains is infeasible within a simulation, so a number of simplifying assumptions are necessary to represent dust. Here we discuss and motivate the assumptions that we make.

The chemical composition of grains (see Table~\ref{table:grains}) is potentially highly complex.
For simplicity, we consider two primary species; \textit{silicate} and \textit{graphite} grains. This dichotomous representation is well established \citep[e.g.][]{Weingartner01}. Carbonaceous grains are generally thought to be either homonuclear (i.e. graphite or diamond) or hydrocarbons. Consequently, the vast majority of their mass comprises carbon. We therefore assume pure carbon grains. This simplifies the budgeting of dust- and gas-phase carbon. 

Conversely, silicates are generally heteronuclear, with a variety of different elements contributing significantly to their mass. Strong candidate minerals from stoichiometry are \textit{olivine}, \textit{pyroxene} and intermediate/composite grains \citep[e.g.][]{Fogerty16}. We explore both \textit{olivine} and \textit{pyroxene} mineral series, adopting \textit{olivine} as our fiducial silicate. We then treat Mg and Fe end-members as distinct subspecies tracked independently: \textit{forsterite} (Mg\textsubscript{2}SiO\textsubscript{4}) and \textit{fayalite} (Fe\textsubscript{2}SiO\textsubscript{4}) in the case of olivine\footnote{Enstatite (MgSiO\textsubscript{3}) and ferrosilite (FeSiO\textsubscript{3}) for pyroxenes.}.  While we limit the silicates to a particular mineral series, the choice to track Mg and Fe end-member silicates separately allows the ratio of these species to freely float given the local chemical abundances, such that they are not artificially limited by a fixed ratio of the two. In particular, the ratio of Mg to Fe species can then reflect variations in $\alpha$-enrichment, $[\rm{\alpha/Fe}]$.

\begin{table*}
 \caption{Composition and nucleation properties of the chemical species of grain adopted in our model. While graphite is homonuclear, the silicates are heteronuclear, comprising \textit{fayalite} (Fe\textsubscript{2}SiO\textsubscript{4}) and \textit{forsterite} (Mg\textsubscript{2}SiO\textsubscript{4}) sub-species. For seeding purposes, we assume an equal abundance of molecules in the \textit{fayalite} and \textit{forsterite} sub-species \textit{fayalite}  (i.e. an effective molecule FeMgSiO\textsubscript{4}, \S \ref{sec:yields}), yielding a value of $A_{\rm G} = 171.853$ used in equation~\ref{eq:nucleation}.}

 \label{tab:mathssymbols}
 \begin{tabular}{lllllllllll}
  \hline
  Species & Sub-species & Seed Composition  &  $A_{\rm G}$ & $\eta_{\rm CCSN}$ & $j$ & $A_j$ & $\tau_{\rm G}$ [Myr]\\ 
  \hline
  Carbonaceous & Graphite & C & 12.01 & 0.15 & C & 12.01 & 180 \\
  Silicate (Olivine) & Fayalite, Forsterite & FeMgSiO\textsubscript{4} & 171.85
  &  3.5$\times10^{-4}$ &  Si | Fe | Mg &  28.09 | 55.85 | 24.31 & 99.3\\

  \hline
 \end{tabular}
 \label{table:grains}
\end{table*}

A common simplifying assumption for grain structure is that of spherical grains. This allows a grain's structure to be characterised by a single value, its radius $r_{\rm grain}$. While this greatly simplifies the equations for dust evolution, the assumption of sphericity has certain limitations. Some degree of asphericity in cosmic dust is evident in harvested samples \citep[e.g. the STARDUST mission,][]{Brownlee14}, and by the polarisation of light in the ISM \citep[e.g.][]{Cho07}. In addition, grains may have a degree of porosity, increasing their surface area and collisional and optical properties \citep{Hirashita22}. While the degree of asphericity and porosity across interstellar environments is hard to constrain, the assumption of spherical, aporous grains is extreme in that it minimises their surface-to-volume ratio, with implications for their interaction with gas or other grains. However, sphericity is degenerate with other physical properties modulating the growth/destruction rates of dust, which are themselves typically empirically determined or calibrated. For our purposes, the assumption of an effective spherical grain is adopted for simplicity.

Given our spherical grain paradigm, grain sizes are characterised solely by their radii, $r_{\rm grain}$,  and are another fundamental property of the dust. As for the chemical composition, representing an arbitrary number of grain sizes is not feasible for our simulations. Instead, we adopt fixed grain size bins. We employ a two-size grain model assuming a 0.1~\textmu m and 0.01~\textmu m radius for large and small grains, respectively.

In combination, this  leads to 3 distinct chemical bins for dust grains; \texttt{Carbon}, \texttt{Mg-Silicate} and \texttt{Fe-Silicate}, each with two possible sizes \texttt{Large} and \texttt{Small}. We adopt this 6-component dust model in all of our simulations.

\textcolor{black}{We note that polycyclic aromatic hydrocarbon (PAH) grains are not explicitly traced in our modelling}. These are extremely small (< 1~nm), and as such, subjected to different forces \citep[e.g.][]{Madden06, Tielens08, Murga19}, more in the realm of molecular chemistry. The creation mechanism for PAHs is uncertain (whether through grain shattering, condensation in AGB winds or spontaneously through chemical reactions; \citealt{Li20}), and the process of aromatisation itself is driven by local-scale radiation \citep{Rau19}, requiring radiative transfer calculations. With these caveats in mind, PAH presence could be inferred using empirical fractions relative to the carbon depletion in the simulations.

\subsection{Seeding grains}
\label{sec:yields}
With our assumed grain properties characterised, we now first consider how grains are \textit{seeded} before considering physical processes that act upon dust grains and the influence of grains in the ISM. We detail the channels of grain seeding below. 

A requirement is that any dust yields should be compatible with the underlying nucleosynthetic yields from stars, such that the abundances of constituent elements of the dust are conserved overall. In the context of the COLIBRE galaxy formation model (see \S\ref{sec:sims}) we build on the composite yields of \citet{Correa25}, and budget dust-phase metals from overall metal yields and rates of mass return from stars of differing mass and chemical composition. In the \citet{Correa25} chemical enrichment scheme, yields are sensitive to the particular enrichment pattern of the enriching star particle, due to their implementation as a linear combination of throughput elements (present at the creation of a star particle) and a net change owing to creation or destruction by stellar evolutionary processes. For a given total metal mass fraction ($Z$), we assume a fixed solar abundance pattern when displaying these dust yields, based on the \citet{Asplund09} solar values used by \citet[][]{Correa25}, hereafter \citetalias{Correa25}.

\subsubsection{AGB grain production}
Although asymptotic giant branch (AGB) and Super-AGB (SAGB) stars are a relatively transient phase of intermediate mass stars $(2\; {\rm M_\odot} < M_{\rm init} < 8\; {\rm M_\odot})$, they are key to the enrichment of the ISM due to their prodigious mass loss via strong stellar winds. In addition, the circumstellar envelopes around AGB stars are sufficiently cool and dense
at distances of 3-10 stellar radii  to facilitate the formation of dust in substantial quantities \citep[e.g.][]{Ferrarotti06}.

Empirical studies of dust associated with (S)AGB have historically been limited to low-metallicity, extragalactic stars, due to the difficulty in obtaining accurate distances locally. \textit{Gaia} parallax measurements should eventually serve to improve this situation \citep{DiCriscienzo16}, probing (S)AGB up to solar metallicity. In the meantime, modelling of AGB evolution, ejecta chemistry and circumstellar formation of different dust grain species have been extended to higher metallicities through enhanced theoretical insights and observations.

We adapt the dust yields of \citet{DellAgli17}, which include AGB models up to solar metallicity. These are illustrated in their figure 5. Dust yields are computed using the ATON stellar models \citep{Canuto91} for AGB stars with zero-age main sequence (ZAMS) stellar masses over the range $1 \lessapprox M_{\rm init}/{\rm M_\odot} \lessapprox 8$ and fixed metallicities.  We use the yields for three absolute stellar metallicities; $Z_\ast \in [0.004,0.008,0.018]$. These are consistent with the \citet{Correa25} nucleosynthetic yields for AGB stars, with the exception of the highest metallicities, which use a slightly higher value of $Z=0.019$. In this case, dust yields are scaled up with metallicity from $Z=0.018$. For consistency with \citet{Correa25}, we match AGB yields to stars in the initial mass range $1 \leq M_{\rm init}/{\rm M_\odot} \leq 8$, assuming no dust is yielded below the \citet{DellAgli17} range (given yields fall precipitously towards the lowest ZAMS masses).

With the dust mass yields in place, \textit{silicate} and \textit{carbon} grains (see Table~\ref{table:grains}) are partitioned by the ZAMS stellar mass of AGB progenitors, assuming AGB with $ M_{\rm init}/{\rm M_\odot} \leq 3.5$ produce carbon grains, with more massive stars producing silicates. In the lower mass subset, this is attributed to the formation of carbon enriched outer atmospheres and the formation of carbon grains in their circumstellar envelopes. In higher mass stars, hot bottom burning destroys surface carbon, and favours silicate production in stronger winds \citep{Ventura14, Schneider14, DiCriscienzo16, DellAgli17}

\subsubsection{SN grain production}
In addition to synthesising large quantities of metals, the shocks from SNe are thought to facilitate significant dust grain formation. Local SN remnants provide valuable case studies for these processes \citep[e.g.][]{Dunne09, Shahbandeh24, Milisavljevic24}. However, the overall dust yields from SNe are particularly uncertain, as strong reverse shocks can also destroy grains (e.g. \citealt{Kirchschlager19, Kirchschlager23}, see \S~\ref{sec:dest}). At very early times, core-collapse SNe are the primary candidates for dust grain formation \citep[e.g.][]{Tielens98, Dwek98}. Observations of $z\approx7$ radio galaxies and quasars with $M_{\rm dust} > 10^7\; {\rm M_\odot}$ have been taken to suggest that SN on average produce $\approx 1 \; {\rm M_\odot}$ of dust \citep{Valiante09, Valiante11, Watson15}, while populations of galaxies exhibiting much lower dust-to-metal ratios have also been observed down to $z \approx 4$ \citep{Burgarella25}. 

It is less clear whether type Ia SNe (SNIa) net produce dust \citep[e.g.][]{Kozasa09, Nozawa11}. Some studies have proposed SNIa as a viable dust production mechanism \citep{Clayton97,Travaglio99}. However, observations of local SNIa remnants do not generally show evidence of significant net dust formation \citep[e.g.][]{Gomez12}, though evidence of dust formation in a SNIa remnant has been observed by \citet{Wang24}. \citet{Nozawa11} posits that this lack of evidence may be explained by a lower formation radius and subsequent rapid destruction in SNIa, preventing them from being an appreciable source of ISM dust grains. Given these findings, we omit dust production by SNIa, and consider only  core-collapse supernova (CCSN) production.

We adapt CCSN dust nucleation rates from \citet{Zhukovska08} to compute dust yields. These are computed using the overall element yields of \citet{Correa25}, where the mass in a particular grain species is

\begin{equation}
\label{eq:nucleation}
    M_{\rm G} = \eta_{\rm G} M_j\frac{A_{\rm G}}{A_j},
\end{equation}
where $M_{\rm G}$ is the mass returned for a given grain species, $M_j$ is the total mass returned of the ``bottleneck" element $j$, $\eta_{\rm G}$ is the species condensation efficiency for CCSN, $A_{\rm G}$ is the mass of the grain molecule in atomic units and $A_j$ is the atomic weight of element $j$. The bottleneck element is the element limiting the number of grain molecules that can be synthesized for a given SN, and for our assumed silicates varies by  mass and metallicity of the progenitor star. This is determined by minimising $N_j/i$ where $N_j$ is the number of elements produced and $i$ the number of atoms in one molecule of the silicate species \citep{Zhukovska08}.

While this approach dictates what fraction of each ejected material is deposited into dust grains, it makes no distinction between subspecies of grain that comprise like elements. Given no \textit{a priori} preference, we adopt a fixed ratio of 0.5 between Fe- and Mg-endmember grains. This yields an effective silicate molecule of FeMgSiO$_4$, treated as a single reservoir for grain seeding purposes. Note that while this approach does not allow the ratio of silicate grain types to free-float, as in accretion processes detailed below, the choice of ratio has little influence on our final dust composition, due to the importance of accretion in building the majority of dust mass in our  model (see \S \ref{sec:grow}).

\begin{figure*}
\centering
\includegraphics[width=\textwidth]{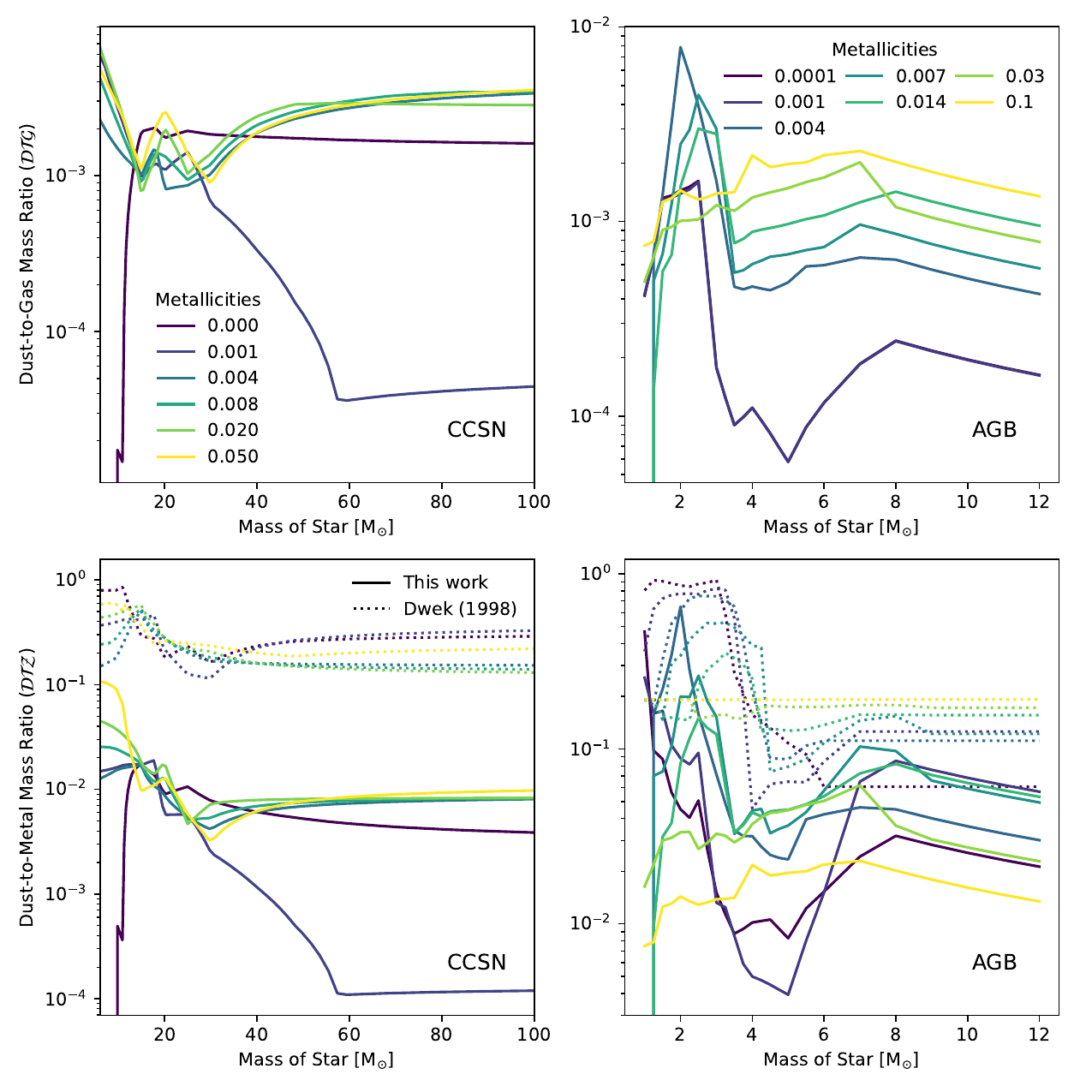}
\caption{Stellar dust yields used in this work, as a function of zero-age main sequence (ZAMS) stellar mass. The top row shows the dust-to-gas ratio in the ejecta computed for CCSN (left) and AGB (right) channels using our adopted yields (\textit{solid lines}, section~\ref{sec:yields}) for different absolute stellar metallicities (line colours). The bottom row shows the same, except that the y-axis is now the dust-to-metal ratio in the returned material, and the grain yields of \citealt{Dwek98} are provided for comparison (\textit{dashed lines}). We note that the two columns use different $y$-axis ranges. Yields are budgeted from the overall metal yields presented by \citet{Correa25}. For the contributions to the yields from elements passing through a star (i.e. present at birth), we assume the solar abundance pattern of 
\citetalias{Correa25}, scaled by the stellar metallicity relative to solar (0.0129). A striking feature of this plot is that the yields and dust-to-metal ratios of 
\citet{Dwek98} tend to be higher than ours \citep{Zhukovska08, DellAgli17} by about an order of magnitude. This difference is discussed further in the text.}
\label{fig:yieldcomp}
\end{figure*}

\subsubsection{AGN grain production}

Another possible avenue for grain nucleation is in the tori of AGN and their nearby environment. Partial motivation for this are studies identifying very dusty  high-redshift systems, requiring $\approx$1 M$_\odot$ of dust per CCSN \citep{Valiante09, Valiante11, Watson15}, while other studies identify very low levels of dust \citep[e.g.][]{Bouwens12}. This could point to a non-stellar channel, such as AGN, being important for nucleating dust. In addition, sight-lines to AGN are often better fit with dust extinction and emission models unlike those of the Galactic ISM \citep[e.g.][]{Sturm05, Srinivasan17}. While stellar sources are still needed to synthesise the metals that constitute grains, some models employ AGN tori as nucleation channels for dust \citep[e.g.][]{Sarangi19}. For simplicity, we neglect grain formation in AGN.

\subsubsection{Size distribution of seeded grains}
\label{sec:seedsize}
In addition to seeding the mass in each chemical species of grain, the initial size distribution of the seeded grains must also be set. We assume that seed grains are dominated by larger size ($a\approx0.1$~\textmu{}m). This reflects studies of AGB winds
\citep{Groenewegen97,Yasuda12,Asano13b}, and is consistent with the inference of $r_{\rm grain} \gtrsim 0.01$~\textmu{}m in core-collapse SNa 
\citep[e.g.][]{Nozawa07}. We initiate the size distribution with 90\% of the mass in large grains and the remaining 10\% in small grains.
This is an order-of magnitude estimate, chosen to be between the values seen in AGB winds \citep[e.g.][]{Asano13} and the observed Milky-Way (MW) mass distribution \citep[e.g.][]{Mathis77}. We note that the size distribution of seed grains, like their absolute masses, have little effect on the dust content of low-redshift galaxies that we present for out model, as this is set by the growth of grains and its equilibration with destructive processes in the ISM (see \S\ref{sec:isoevo}). However at higher redshift, or especially rapidly forming galaxies, the yield contribution can be more significant, with dust produced through yields before evolutionary processes can equilibrate\footnote{In section \ref{sec:cdd}, we show that in our model the total contribution from direct stellar yields is subdominiant to growth processes for $z \lesssim 11$ (\S~\ref{sec:cdd}, Fig.~\ref{fig:cdd}).}.

\subsubsection{Comparing dust yields from seeding channels}
\label{sec:seedcomp}

Given the considerable uncertainty and variation in the predictions for dust seeding by AGB and CCSNe, it is worthwhile to consider how our curated yields \citep{Zhukovska08, DellAgli17} may differ from other models. For comparison we consider the dust yields predicted by \citet{Dwek98}, which have been employed by various galaxy formation models \citep[e.g.][]{McKinnon15, McKinnon16, Dave19}.

In Fig.~\ref{fig:yieldcomp} we show the dust yields (top row) and returned dust-to-metal ratios (bottom row) for CCSN (left column) and AGB (right column) as functions of their ZAMS mass, $M_{\rm init}$. A striking feature is that the yields advocated by \citet{Dwek98} are higher than those we use by about an order of magnitude. In addition, the dust-to-metal ratio of returned material in the \citet{Dwek98} model is of the same order as that seen in the local ISM \citep[$\approx0.3$,][]{Inoue03}. This difference is profound as it implies different roles for evolutionary processes in reaching the dust content of galaxies like the Milky Way. Our model requires dust masses to grow by a factor $\sim 10$ in the ISM, while the \citet{Dwek98} yields require either no mass evolution, or an equilibrium of destruction and growth mechanisms. In particular, our lower level of seeding requires modelling of the cold, dense ISM conducive to efficient grain growth (see \S\ref{sec:grow}), as opposed to a level of seeding requiring only maintenance. We note that some studies use yields of order 10\% but still require growth to reproduce observables \citep[e.g. ][]{Lower24, Choban25}.

While the \citet{Dwek98} yields are empirically motivated, with some arbitrary choices with regards to grain nucleation properties, some more recent observations of post-shock region of supernovae support this high-yield paradigm \citep[e.g.][]{Shahbandeh24}. The \citet{Zhukovska08} yields are intended to be decoupled from subsequent moledcular cloud growth, and are consistent with the limited measurements of presolar grains in the solar system, as well as the lower end of supernovae observations.  Ultimately, these differences showcase the uncertainty in the direct yields of grains and differential role of growth in dust models, explored further in the following section.  

\subsection{Grain growth}
\label{sec:grow}

Once grains have been nucleated in the ejecta of stars, we can consider their evolution. One aspect of this is the growth of grains through the process of \textit{accretion}. A timescale argument comparing our dust injection into the ISM by stellar sources with dust destruction by sputtering suggests the need for significant growth of dust within the ISM to explain the observed dust content of galaxies \citep{Inoue03, Draine09, Zhukovska08, Pipino11, Valiante11, Asano13}. 

For accretion, we adapt the accretion timescale equations of \citet{Hirashita13}, using
\begin{equation}
\label{eq:acc}
\tau_{\rm acc} = \tau_{\rm G}
\left(\frac{r_g}{0.1 \text{\textmu m}}\right)
\left(\frac{\epsilon_{j,\odot}}{\epsilon_j}\right)
\left(\frac{10 \;{\rm cm}^{-3}}{n^{\prime}_{\rm H}}\right)
\left(\frac{10 \;{\rm K}}{T}\right)^{0.5}
\left(\frac{0.3}{S_{\rm acc}}\right),
\end{equation}


\noindent where ${n^\prime_{\rm H}}$ is the effective local number density of hydrogen, $T$ is the local gas temperature,  $S_{\rm acc}$ is the sticking probability for accretion, $\tau_{\rm G}$ is the accretion timescale normalisation for a given grain species, and $\epsilon_j$ and $\epsilon_{j,\odot}$ represent the local and solar abundance of the bottleneck abundance in the gas-phase relative to hydrogen, respectively. For our purposes, we assume a fixed  $S_{\rm acc}$ value of 0.3, with $\tau_{\rm G}$ for silicate and carbon grains taken from 
\citet{Hirashita13} and listed in Table~\ref{table:grains}. The fractional change in the mass of a given dust species due to accretion is computed over a timestep of $\Delta  t$ for each gas particle, as $\exp{(\Delta  t/\tau_{\rm acc})}$.

The main variation from \citet{Hirashita13} is the $\epsilon_{j,\odot}/\epsilon_j$ factor, replacing a $Z_{\odot}/Z$ factor. In the case of solar abundance patterns (assuming the \citealt{Wiersma09} abundances), these terms are equivalent, but considering individual element abundances allows us to go beyond this assumption in the ISM, exploiting the individual traced elements. Another advantage is that as the local abundance of a depleted element approaches zero, $\tau_{\rm acc}$ approaches infinity. This allows us to transfer mass in depleted elements from the gas  phase to the dust phase consistently, in the fixed proportions dictated by their chemical compounds (Table~\ref{table:grains}).

Equation~\ref{eq:acc} is applied in the context of our dust model, to the two overarching dust species (silicate and carbon). For carbon grains, this is straightforward; with only one, homonuclear, subspecies of grain the bottleneck abundance is always C, i.e. $\epsilon_{j} = \epsilon_{\rm C}$. For silicates, comprising O, Mg, Si and Fe, this is found by maximising the $\epsilon_{j,\odot}/\epsilon_j$ term,

\begin{equation}
    \frac{\epsilon_{j,\odot}}{\epsilon_j} = \max \left(\frac{\epsilon_{\rm O,\odot}}{\epsilon_{\rm O}}, \frac{\epsilon_{\rm Si,\odot}}{\epsilon_{\rm Si}}, \frac{\epsilon_{\rm Mg+Fe,\odot}}{\epsilon_{\rm Mg+Fe}}\right), 
\end{equation}

\noindent Such that $j$ represents the bottleneck abundance, where Mg+Fe is the composite abundance of the two silicate subspecies endmember elements Mg and Fe, i.e.

\begin{equation}
     \frac{\epsilon_{\rm Mg+Fe,\odot}}{\epsilon_{\rm Mg+Fe}}= 
     \frac{\epsilon_{\rm Mg,\odot} + \epsilon_{\rm Fe,\odot}}{\epsilon_{\rm Mg} + \epsilon_{\rm Fe}}.
\end{equation}

\noindent This is reflective of the fact that total accretion in the silicate species is freely divisible between  Mg and Fe endmembers, such that the total abundance of Mg and Fe can be considered a single reservoir of gas-phase metals available for accretion. With O and Si always present in silicates with a fixed abundance ratio, these elements can be treated as independent reservoirs.

This accretion for the overall silicate species is computed and the total accreted mass is divided between endmember species proportionally to the relative abundance ratio of gas-phase endmember elements in the ISM. 

Accretion proceeds most efficiently in cold, dense environments where dust constituent elements are available in the gas-phase. Accounting for how dust is distributed, the majority of accretion is posited to take place in the cold, neutral medium \citep[e.g.][]{Jenkins09, Hirashita12}.  This represents local densities of $n_{\rm H} \gtrsim 10\;{\rm cm}^{-3}$ \citep{Hirashita00}, where $n_{\rm H}$ is the hydrogen number density where $n_{\rm H} = m_{\rm p}^{-1} {\rm X_g} \rho$ for proton mass $m_{\rm p}$ and hydrogen mass fraction, ${\rm X_g}$. 
Efficient accretion in these contexts is supported by the enhanced depletion observed in dense clouds \citep{Savage96}. 
 For this reason, to properly account for accretion it is important that dense, molecular media are represented. \citet{Zhukovska08} attribute smaller dust nucleation rates compared to \citet{Dwek98} from stellar sources to a more explicit treatment of dust growth in molecular clouds. A conflation of stellar nucleation and accretion could be one explanation for the significantly higher dust yields displayed for \citet{Dwek98} in Fig.~\ref{fig:yieldcomp}, though higher dust yields now observed in individual CCSNe may support the higher level grain seeding paradigm \citep{Shahbandeh24} comparable to \citet{Dwek98}.  

 Dust yields that implicitly include some molecular cloud grain growth can be useful for simulations that do not attempt to model such dense gas structures \citep{McKinnon16, Dave19, Vijayan19}. However, our model has been developed for galaxy formation simulations that include multiphase gas and that can attempt (at sufficient resolution) to distinguish molecular gas clouds from the diffuse ISM, such that using yields that implicitly include an accreted component could effectively \textit{`double count'} accretion. Despite the relatively high densities reached in our simulations, we are still far from resolving the dense cores of molecular clouds where these grains are thought to accrete efficiently. As a result we apply a clumping factor, $C$, to boost the densities input into the accretion rate. Our fiducial scheme uses a variable boost factor that rises monotonically from $C=1$ to 100 between $-1 < \log_{10}{n_{\rm H}/{\rm cm^{-3}}} < 2$ and is clipped to the boundary values on either side. This is found to reproduce observed dust-to-metal ratios within gas at ISM densities ($n_{\rm H} \gtrsim 0.1\;{\rm cm^{-3}}$). This provides a means to calibrate across regimes in resolution. We investigate the clumping factor further in \S\ref{sec:clump}.
 For our fiducial clumping with representative properties of the cold, neutral medium ($T = 100$~K, $n_{\rm H} = 10\;{\rm cm}^{-3}$) at $0.1 \; Z_\odot$ a small carbon (silicate) grain in an environment where 5\% of the bottleneck element is available for depletion in the gas-phase gives $\tau_{\rm acc}=26.4$~Myr ($\tau_{\rm acc}=14.6$~Myr), of comparable order to the `enhanced' accretion model found necessary by \citet{Choban25}.  

\subsection{Grain destruction}
\label{sec:dest}

Just as grains may grow and accrete material in the ISM, grains may also be shrunken and destroyed through their interaction with local media. We employ a few distinct mechanisms\footnote{We opt for a minimal-necessary model. For our grain population, such as photo-evaporation of grains, are neglected as by theory minor contributors to grain destruction \citep[e.g.][]{Nanni20}.}, detailed below. For computed destruction timescales, the fractional mass change of a given grain is computed for each particle over a timestep of length $\Delta t$ as $\exp(-\Delta t/\tau)$. 

\subsubsection{Thermal Sputtering}
Thermal sputtering is the erosion of dust grains through high-velocity interactions with gas particles in hot gas. We use the thermal sputtering prescription of \citet{Tsai95}, which uses the grain shrinkage rate
\begin{equation}
    \frac{{\rm d} r_{\rm grain}}{{\rm d} t} = -3.2\times 10^{-18} \; {\rm cm \; s^{-1}} \left(\frac{n_{\rm H}}{1\;{\rm cm^{-3}}} \right)
    \left[1 + \left(\frac{T}{T_0}\right)^{-2.5}\right]^{-1}
\end{equation}
where the timescale for sputtering the dust mass for uniform spherical grains can then be computed as
\begin{equation}
    \tau_{\rm sp} = 0.85 \; {\rm Myr} \left(\frac{r_{\rm grain}}{0.1\;\text{\textmu m}} \right)\left(\frac{n_{\rm H}}{1\;{\rm cm^{-3}}} \right)^{-1}
    \left[1 + \left(\frac{T}{T_0}\right)^{-2.5}\right]
\end{equation}
where $T_0$ is the temperature normalisation, set to
$2\times 10^6$~K. We note that the sputtering timescale is highly sensitive to the temperature, such that dust in gas heated to temperatures $T > T_0$ will be destroyed very quickly at ISM densities. In hot gas at the lowst CGM densities the timescale can become long again. Note that because thermal sputtering occurs in the diffuse-gas phase, we use the unmodulated density, $n_{\rm H}$, for sputtering (as opposed to $n_{\rm H}^\prime$, modulated by a clumping factor, $C$).

\subsubsection{Astration}
\label{sec:astration}

Another process to consider is the destruction of grains upon their reincorporation into stellar atmospheres, also known as \textit{astration}. In our model, the dust mass that is present in a gas particle as it is being converted to a stellar population is transferred back to the \textit{gas-phase} element mass fraction array stored in each particle, in the proportions dictated by each grain compound. These gas-phase elemental mass fractions then come to represent the \textit{total} elemental mass fractions for the resultant star particle.    

\subsubsection{SN shocks}
\label{sec:sndest}

The effect of SN shocks on grains may be considered a complement to the thermal sputtering we implement to account for resolved heating processes. The shocks of individual SNe are not resolved in galaxy-scale simulations. Instead, in COLIBRE bulk feedback and star formation events are modelled stochastically (\S\ref{sec:sf}, \S\ref{sec:fb}).
For consistency, we implement a stochastic approach linked directly to the feedback module - dust is totally destroyed in the particles stochastically selected to receive SN feedback\footnote{For either thermal or kinetic implementations of SN feedback.}. This is handled in the same way as astration - on being hit, mass in dust is redistributed into the gas-phase of the constituent elements, and dust abundances are set to zero, all within the struck particle. The practical, stochastic implementation of this approach paired with the complex varying properties of the gas receiving feedback make it difficult to compute a timescale from first principles. To estimate the timescale for dust destruction by SNe in our simulations, we compare the average feedback-hit particles in the last 100~Myr (before $z=0$) with the $z=0$ dust properties in our fiducial model (see \S~\ref{sec:sims}), the ratio of which provides a cosmic timescale of around 420~Myr. 

\subsection{Grain size evolution}
Other processes have been identified that conserve the mass in dust, but modify the grain size distribution. Despite not affecting the grain mass directly, this has a knock-on effect on the evolution of grains, with the accretion and sputtering timescales being sensitive to grain size, with small grains exhibiting shorter timescales due to their higher surface to volume ratios. 

\subsubsection{Shattering}
Mass can transfer from the large to small grain component via \textit{shattering}, where energetic grain-grain collisions break their constituents into many smaller grains. For shattering, we follow the formula  of \citet{Aoyama17}, adapted by \citet{Granato21},

\begin{equation}
\tau_{\rm sh} = 54.
1~{\rm Myr} \; \left(\frac{\mathcal{DTG}}{0.01}\right)^{-1} \left(\frac{a_L}{0.1\;\text{\textmu m}} \right) \times \begin{cases}
\left(\frac{n_{\rm H}}{{\rm cm^{-3}}}\right)^{-1} & \text{for } \frac{n_{\rm H}}{{\rm cm^{-3}}} < 1, \\ 
\left(\frac{n_{\rm H}}{{\rm cm^{-3}}}\right)^{-\frac{1}{3}} & \text{for } \frac{n_{\rm H}}{{\rm cm^{-3}}} \geq 1,
\end{cases}
\end{equation}
where the timescale normalisation value assumes a material density of 3~${\rm g \; cm^{-3}}$ (intermediate between typical silicate and carbon grains), and $a_L$ is the radius of large grains (0.1~\micrometer{} in our model). The \citet{Granato21} prescription differs from previous applications in that the shattering does not shut off abruptly above a gas density of $n_{\rm H} = 1 \; {\rm cm^{-3}}$, but instead increases more gradually above this density. Note, as in sputtering, we are using the unmodulated density, $n_{\rm H}$ (as opposed to $n_{\rm H}^\prime$), as sputtering is a diffuse-gas process. 

\subsubsection{Coagulation}
The collisions of grains can conversely lead to mass transfer from small to large sizes, as the constituents of lower velocity grain-grain collisions can \textit{coagulate} to form a single larger grain. To represent these processes, we again follow the prescription of \citet{Aoyama17}, modified to,

\begin{equation}
\tau_{\rm co} = \tau_{\rm co,\, 0} \; \left(\frac{a_S}{0.1\;\text{\textmu m}}\right) \left(\frac{\mathcal{DTG}}{0.01}\right)^{-1} \left(\frac{v_{\rm co}}{0.1 \; {\rm km \; s^{-1}}}\right)^{-1} \left(\frac{n^\prime_{\rm H}}{10 \;{\rm cm}^{-3}}\right)^{-1},
\label{eq:coag}
\end{equation}
where the timescale normalisation, $\tau_{\rm co,\, 0}$ is taken to be $27.1~{\rm Myr}$ and incorporates an assumed material density of 3~${\rm g \; cm^{-3}}$ for grains, and $a_S$  is the radius of small grains (0.01~\micrometer). We also assume a fixed coagulation velocity of $v_{\rm co}=0.2 \; {\rm km \; s^{-1}}$. In the original prescription of \citet{Aoyama17} this process is assumed to take place in molecular clouds, with a fixed density of $10^3~{\rm cm^{-3}}$. We adapt the \citet{Aoyama17} equation to depend on an effective density, $n^\prime_{\rm H}$, i.e. the local simulated gas density boosted by a clumping factor to represent unresolved structure (see \S~\ref{sec:clump}), as with the other collisional processes assumed here. Equation~\ref{eq:coag} makes use of the higher densities afforded by the higher resolution and cold gas physics of our simulations, and does not depend on a sub-resolution criterion for molecular gas content. This allows us to maintain a self-consistent approach across the dust evolution processes. This is particularly important when dust is coupled to cooling and molecule formation processes (\S\ref{sec:couple}).

Despite our higher resolution and cold gas physics, we cannot resolve the interiors of molecular clouds, as addressed in section~\ref{sec:grow}. The coagulation rates are therefore also subject to a variable clumping factor, explained further in \S\ref{sec:clump}. In addition to the clumping factor, we note that the coagulation timescale $\tau_{\rm co,\, 0}$ is relatively uncertain, so calibration of $\tau_{\rm co,\, 0}$, to match second-order dust properties, such as the size distribution of grains, is a reasonable approach. In particular, a lower $\tau_{\rm co,\, 0}$ could be argued for, particularly due to the assumption of spherical grains; more prolate/oblate grains could increase grain-grain collision rates, and increase the growth of grains at fixed density. We will consider the option of boosting this coagulation rate in future work.

\subsection{Grain transport \& diffusion}
\label{sec:diff}

In addition to the processes that govern how dust forms and evolves in each gas particle, we also model how dust mixes through the gas. This mixing occurs across multiple scales; from large-scale gas flows to the fine, fractal structure of turbulent motions. At scales we can resolve, dust mixing is naturally handled by the hydrodynamics: gas motions carry dust out of enriched regions and redistribute it throughout galaxies. For smaller, unresolved scales, we use an explicit diffusion model, that follows the same implementation to the one used for chemical elements \citep{Correa25}. Our approach of dust mixing adapts the diffusion equation, ${\rm{d}}X_{i}/{{\rm{d}}t} =\rho^{-1}\nabla(D\nabla X_{i})$, into the SPH framework \citep{Monaghan05}. Here, $X_i$ is the dust fraction of gas particle $i$ and $D$ is the diffusion coefficient. The coefficient $D$ depends on the local gas velocity shear, following a Kolmogorov turbulence model \citep{Smagorinsky63}, to represent turbulent mixing. Its normalization is set by a free parameter, $C_{\rm d}$. The value of $C_{\rm d}$ varies significantly in prior studies from values $C_{\rm d} \approx 0.003$, representing a minimal subgrid diffusion level that limits extreme abundances in forming stars \citep{Escala18}, to $C_{\rm d}\sim0.1$ as suggested by Kolmogorov theory directly \citep[e.g.][]{Smagorinsky63, Shen10, Su18}.

Diffusion of individual elements and dust are treated independently. For simplicity, we assume no relative boost in dust diffusion rates with respect to the diffusion of metals.  We adopt the fiducial value of metal diffusion coefficient $C_{\rm d} = 0.01$ \citep{Correa25}, with the exception of variations explicitly varying $C_{\rm d}$, finding that a higher relative dust diffusion rate has only a small influence on dust evolution.

Turbulent diffusion helps to mitigate sampling noise in abundances of dust brought about by the discrete enrichment events. Particularly at low dust abundances, this allows gas that has not been directly enriched by star particles to accrue some dust and begin dust evolution. 

We do not account for other kinematic effects of dust on gas, for instance through radiation pressure or the effect of gas-dust drag. Dust drag can be an important process at high densities ($n_{\rm H} \gg 10^3 \; {\rm cm^{-3}}$), particularly those seen at circumstellar scales and protoplanetary disks \citep{Weidenschilling77, Birnstiel16}. Dust drag may also be important internally for dense, molecular clouds \citep[e.g.][]{Padoan06}. However, for the density scales that we resolve ($n_{\rm H} \lesssim 10^3 \; {\rm cm^{-3}}$), the drag timescale considered here is long enough ($\tau_{\rm drag} \gg 1 \; {\rm Myr}$) that the stopping times of grains are long with respect to the gas dynamical times, and we may assume dust motions trace the gas motions well \citep{Draine11, Hirashita13}.

\subsection{Coupling to COLIBRE hybrid cooling and heating}
\label{sec:couple}

\begin{figure*}
    \centering
    \includegraphics[width=0.98\textwidth]{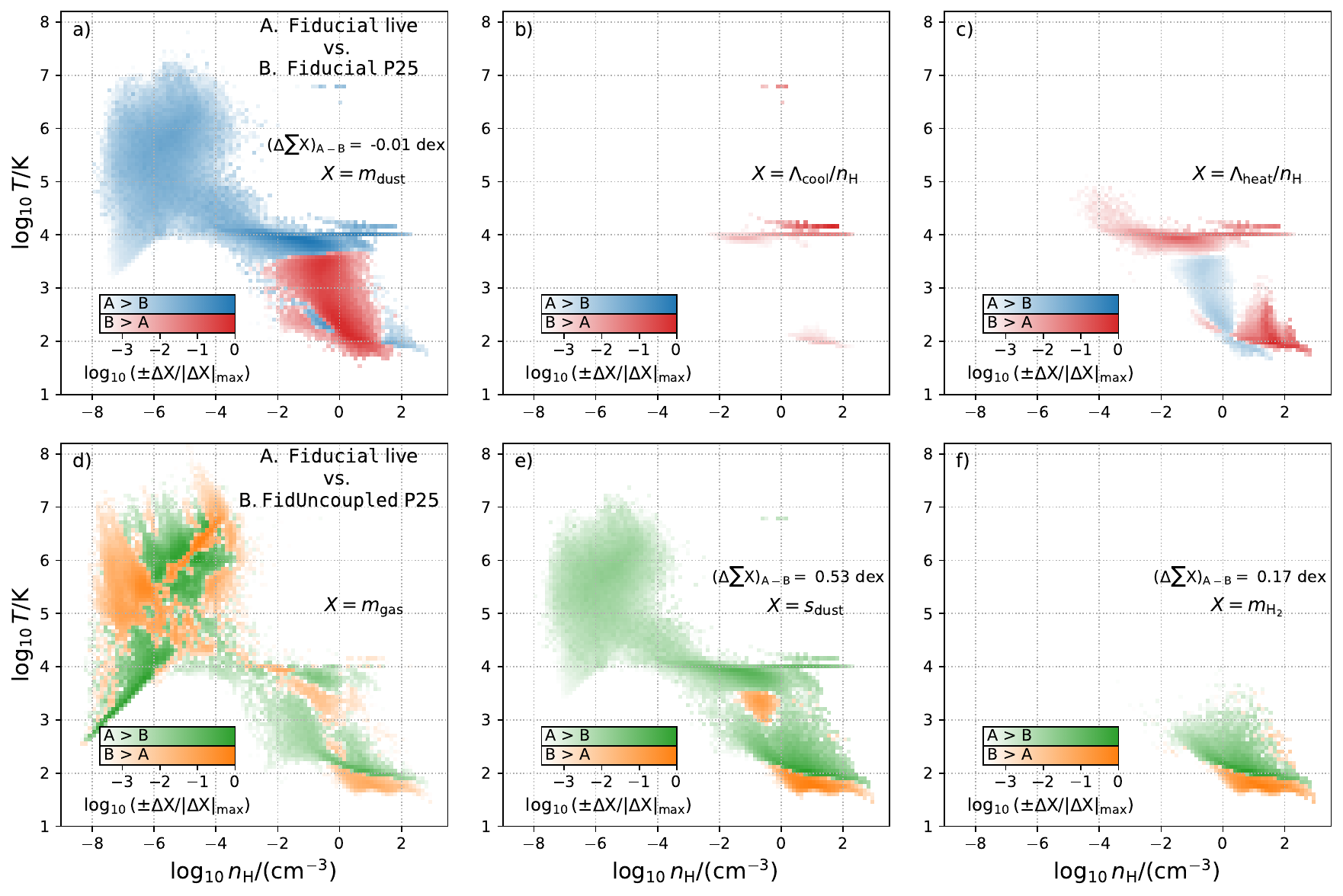}
    \caption{Differential density-temperature ($n_{\rm H}$-$T$ or \textit{`phase'}) diagrams for particles at $z=0$, comparing intra- and inter-simulation gas and dust properties. The \textit{top row} compares properties within the \texttt{Fiducial} run; comparing our live dust model, to values associated with the instantaneous, \citetalias{Ploeckinger25} dust (computed by interpolating the \citetalias{Ploeckinger25} tables). Comparing within the same simulation is intended to isolate differences, given an otherwise identical state of the gas. From left to right we compare total dust content, and elemental cooling and heating rates given these different depletions. The \textit{bottom row} compares between the \texttt{Fiducial} and \texttt{FidUncoupled} runs, considering live and \citetalias{Ploeckinger25} dust for each, respectively. From left to right, this compares the gas mass distribution, the local grain collisional cross-section, and the H$_2$ mass distribution. For the top row, the absolute difference cell-by-cell (for $100\times100$ cells) between the live (A) versus \citetalias{Ploeckinger25} dust (B) properties within \texttt{Fiducial} are taken, and shaded blue or red depending on whether there is an excess in the former or latter case, respectively, revealing where significant differences exist. For the bottom row comparing \texttt{Fiducial} live (A) vs \texttt{FidUncoupled} \citetalias{Ploeckinger25} (B) dust, we instead use orange and green. Where significant, the difference in the quantity summed over all particles is written, where positive values indicate and excess in the \texttt{Fiducial} live dust case. In particular we see higher grain cross sections (by 0.53~dex) and H$_2$ masses (by 0.17~dex) in the \texttt{Fiducial} run using live dust, when compared to the \texttt{FidUncoupled} using \citetalias{Ploeckinger25} dust.}  
    \label{fig:coupling}
\end{figure*}

The evolutionary life-cycle model of dust described above emphasises the influence of stars and gas in the ISM. Given this model, we can in turn account for the influence of local dust properties on various heating and cooling processes, such that dust may in turn influence gas physics generally and even star formation. In \S\ref{sec:cool} we detail the relevant processes influenced by live dust grains. The relevant quantity for cooling processes linked to grains is the local dust cross-section per unit volume, $n_{\rm d}\sigma_{\rm d}$. The hybrid cooling module assumes a constant dust mix internally, through an \textit{`effective size'} translating to an $A_V/N_{\rm H}$ value (see \S\ref{sec:cool}). Having fixed attenuation curves parametrised solely by their normalisation in the $V$-band, $A_V$, means it is less complex and computationally expensive to compute heating and cooling rates. With mass-weighted average grain sizes of the MW ($a\sim 0.1$~\textmu m, see e.g. \citealt{Draine03}) consistent with the effective size assumed for our \texttt{OneSize} model (Table~\ref{table:sims}, taken as our large-grain size in \texttt{Fiducial} and elsewhere) it is straightforward to simply scale extinctions and collisional cooling processes particle-by-particle using our modelled dust-to-gas ratio $\mathcal{DTG}_{\rm model}$. This is done via a rate scaling factor $s_{\rm dust}$ where $s_{\rm dust} = \mathcal{DTG}_{\rm model} / \mathcal{DTG}_{\rm MW}$, where we take $\mathcal{DTG}_{\rm MW} = 6.6\times 10^{-3}$ (following the ISM grain model in Cloudy, \citealt{Chatzikos23}).

By contrast, our \texttt{Fiducial} dust model uses two grain size bins to capture local variations in the size distribution. At a fixed dust mass and material density, dust-associated heating and cooling rates scale with the surface-to-volume ratio, which scales in inverse proportion with the grain radius ($\propto a^{-1}$). Accounting for the two sizes gives a new expression for $s_{\rm dust}$:

\begin{equation}
\label{eq:sdust}
s_{\rm dust} = \frac{a_{\rm MW}\left(\frac{1}{a_{\rm L}}D_{\rm L} + \frac{1}{a_{\rm S}}D_{\rm S}\right)}{\mathcal{DTG}_{\rm MW}},
\end{equation}

\noindent where $D_{\rm L}$ and $D_{\rm S}$ are the fraction of a particle's total mass in small and large grains, respectively, and $a_{\rm MW}$ is the effective size of dust grains in the Milky Way.

We note that for collisional processes (e.g. H$_2$ formation) the $s_{\rm dust}$ scaling simply reflects the increased surface area per unit mass contributed by small grains. For heating and cooling processes, however, this scaling may strictly depend upon an additional extinction efficiency, $\langle Q_{\rm ext} \rangle$, i.e. the ratio of the extincting cross section (the sum of scattering and absorption cross sections) to the physical cross section of grains \citep[e.g.][]{Weingartner01}. This can vary with wavelength and differs depending on the material and size properties of grains. While $V$-band attenuation $A_V$ is used as a standard normalisation for extinction curves, the processes relevant to the dust coupling described here (e.g. self shielding, photoelectric heating) actually depend on the FUV, ionising or Lyman-Werner photons. While extinction efficiencies of large grains, $\langle Q_{{\rm ext},L} \rangle$, dominate at optical wavelengths \citep[e.g.][]{Draine01}, in the UV the small-grain extinction efficiency, $\langle Q_{{\rm ext},S} \rangle$, becomes comparable. For simplicity, we neglect the influence of $\langle Q_{\rm ext} \rangle$, assuming $\langle Q_{{\rm ext},S} \rangle / \langle Q_{{\rm ext},L} \rangle \approx 1$ and marginal differences between material species. We look into detailed extinction and attenuation properties of grain in more detail in upcoming studies post-processing COLIBRE galaxies using radiative transfer, and justify our assumptions around extinction efficiency further in Appendix~\ref{sec:qext}.

The live dust model may also impact radiative cooling and heating processes more indirectly, through the different depletion of gas-phase elements, affecting gas-phase cooling channels (such as metal-line cooling and heating). The element-by-element cooling channels for gas-phase metals in the hybrid cooling scheme accounts for the assumed \citetalias{Ploeckinger25} dust depletion, via a MW-like relative depletion pattern \citep{Jenkins09} that scales with $Z_{\rm gas}$ (see \S\ref{sec:depletion} for a comparison of equilibrium and live dust depletions). The fractions of metal elements in the gas- and dust-phase of enriched gas are handled implicitly, scaling the tabulated rates and leading to different element-by-element cooling and heating between the coupled and uncoupled cases for gas of a fixed thermodynamic state and metallicity, owing to the differences between our live dust model and the \citetalias{Ploeckinger25} dust.

In Fig.~\ref{fig:coupling} we show differences between dust, cooling and gas properties, comparing our live dust with the instantaneous \citetalias{Ploeckinger25} dust, both within the \texttt{Fiducial} run, and when incorporating the coupled physics effects when comparing to  \texttt{FidUncoupled} runs. In the top row we see the influence of the dust
properties input into the heating and cooling modules alone, by comparing the \citetalias{Ploeckinger25} dust to the live dust model within \texttt{Fiducial}. This shows how, for like physical conditions, the influence of dust on cooling processes
differs across $n_{\rm H}$-$T$ space when live dust is included. The shading in this figure shows absolute pixel differences normalised by the maximum positive and negative difference pixel in each panel. This nuanced representation allows us to show where  bulk properties differ in phase space. The overall distribution of dust depletion fractions in phase space can be seen more directly in upcoming \S~\ref{sec:isoevo} with overall phase-spce distributions for COLIBRE presented in \citet{Schaye25}.

In Fig.~\ref{fig:coupling}, we see that for relatively cool ($\log_{10}T/{\rm K} \lesssim 3.5$) and moderately dense ($\log_{10}n_{\rm H}/{\rm cm^{-3}} > -2$) gas, the \citetalias{Ploeckinger25} model tends to yield lower \dtg{} ratios (panel a), though with small, exceptional regions (around $\log_{10}n_{\rm H}/{\rm cm^{-3}} \approx -0.5$, $\log_{10}T/{\rm K} \approx 2.5$ and $\log_{10}n_{\rm H}/{\rm cm^{-3}} \gtrapprox 2$). For hotter or more diffuse gas, the live dust model dominates, particularly due to the propagation of dust outside of the ISM and into the CGM and other phases that is assumed to be completely destroyed in the \citetalias{Ploeckinger25} model. This residual dust that avoids destruction by sputtering and direct annihilation in shocks is seen within H\textsc{ii} regions (horizontal sequence at higher density and $\log_{10} T/{\rm K} \approx 4$) and in the diffuse hot gas, where despite the high temperatures, dust destruction timescales can remain long enough due to the lower densities. We highlight the few cells in hot, denser gas showing evidence of live dust (around $\log_{10}T/{\rm K} \approx 7$, $\log_{10}n_{\rm H}/{\rm cm^{-3}} \approx 0$). These few particles are those caught in the simulation output \textit{between} the feedback and dust evolution steps (where this dust is rapidly destroyed), due to the order of operations in COLIBRE. The total dust content of the \texttt{Fiducial} volume when using \citetalias{Ploeckinger25} or live dust is very similar (within 0.01~dex or $\approx 2 \%$), though this level of consistency is coincidental, varying more in other volumes and environments. Recomputing the instantaneous \citetalias{Ploeckinger25} elemental cooling and heating rates of \texttt{Fiducial} using \citetalias{Ploeckinger25} dust (panels b and c respectively) yields complex differences, particularly higher metal-line cooling and heating rates in H\textsc{ii} regions (due to more gas-phase metals) and heating in the densest ISM. However, in overall cosmic terms, the difference in cooling and heating rates per unit density ($\Lambda_{\rm cool}/n_{\rm H}$, $\Lambda_{\rm heat}/n_{\rm H}$) are marginal ($\lesssim 0.2\%$).

The bottom row of Fig.~\ref{fig:coupling} compares the live dust of \texttt{Fiducial} to the \citetalias{Ploeckinger25} dust of the \texttt{FidUncoupled} run, including the physical differences and run-run variations. To illustrate this, we first show the different gas distributions (panel d). Along with the variations in hot, diffuse gas owing to differing and stochastic feedback histories of each run, there are also differences in the ISM; most notably the existence of an excess of colder gas in the \texttt{FidUncoupled} run at $\log_{10} T/{\rm K} \lessapprox 100$. Taking this into account, we see that the local grain cross sections are generally higher in the live dust model, and by 0.53 dex ($\approx 3\times$) overall, excepting this cold sequence and a region of the WNM, at $\log_{10}n_{\rm H}/{\rm cm^{-3}} \approx -1$, $\log_{10}T/{\rm K} \approx 3.25$. This is attributable to our variable size distributions and abundance of small grains, particularly outside the coldest, densest gas. Another significant difference to highlight is that H$_2$ extends to slightly warmer temperatures and lower densities in the coupled runs relative to the uncoupled ones (panel f). We attribute this to lower cooling rates in the dense ISM, paired with more small grains at intermediate density, boosting the local grain cross-section for H$_2$ formation. Ultimately, the distribution of H$_2$ in $\rho$-$T$ space leads to higher cosmic H$_2$ mass densities (by 0.17~dex), boosting galactic H$_2$ mass fractions. Altogether, this emphasises the complex differences that emerge from using a live dust model\footnote{Alongside any stochastic run-to-run variation \citep{Borrow23}.}, where dust evolution is influencing gas thermodynamics throughout cosmic time.

\subsection{Clumping factors}
\label{sec:clump}

As with gas physics in the ISM, dust evolution is a highly multi-scalar process, ranging from dense molecular clouds to the diffuse CGM. In particular, the dense cores of molecular clouds provide sites for the growth of grains through frequent collisions in relatively cold gas \citep{Ormel09}. The core densities of molecular clouds typically range from $n_{\rm H} = 10^4 \;{\rm cm}^{-3}$ up to proto-stellar densities. The spatial scales associated with the densest molecular gas is below what we resolve in our cosmological volume simulations. Inability to represent these denser structures skews the densities experienced by gas to lower values and inhibits the accelerated dust evolution inferred for observed molecular clouds. To represent the presence and structure of unresolved collapsed clouds, we experiment with a `clumping factor', $C$. This factor can boost input densities such that the effective gas densities (effective hydrogen number density), $\rho^\prime$ ($n^\prime_{\rm H}$), used by dense-gas processes (such as accretion and coagulation) are related to the physical densities in our simulations, as

\begin{equation}
    \rho^\prime = C\rho, 
\end{equation}
or
\begin{equation}
    n^\prime_{\rm H} = Cn_{\rm H},
\end{equation}

\noindent where $\rho$ (or $n_{\rm H}$) is the input physical gas density (or hydrogen number density) and $C>1$. With the $C$ factor tied to the resolution and small-scale physics limitations of the simulations, we treat $C$ as a free parameter of the model, which must be calibrated. We experiment with both a constant clumping factor as well as a density-dependent clumping factor, $C(\rho_{\rm phys})$. In the density dependent case 

\begin{equation}
  C(n_{\rm H})=
\begin{cases}
1, & \text{if} \; n_{\rm H} \leq n_{\rm H,\, min} \\
\left(\frac{n_{\rm H}}{n_{\rm H,\, min}} \right)^m , & \text{if} \; n_{\rm H,\, min} < n_{\rm H} \leq n_{\rm H,\, max} \\
C_{\rm max}, & n_{\rm H} > n_{\rm H,\, min}
\end{cases}
\end{equation}

\noindent where $n_{\rm H,\, min}$ and $n_{\rm H,\, max}$ are the minimum and maximum densities of the clumping transition, 
$C_{\rm max}$ is the maximum clumping value (see Table~\ref{table:sims}), and $m = {\log_{10}(C_{\rm max})/\log_{10}(n_{\rm H,\, max}/n_{\rm H,\, min})}$. For a gradual, 
transition we take $n_{\rm H,\, min} = 0.1 \; {\rm cm^{-3}}$ and $n_{\rm H,\, max} = 100  \; {\rm cm^{-3}}$, as well as a fiducial $C_{\rm max}=100$ (yielding $m = 2/3$). 

\begin{figure}
    \includegraphics[width=0.48\textwidth]{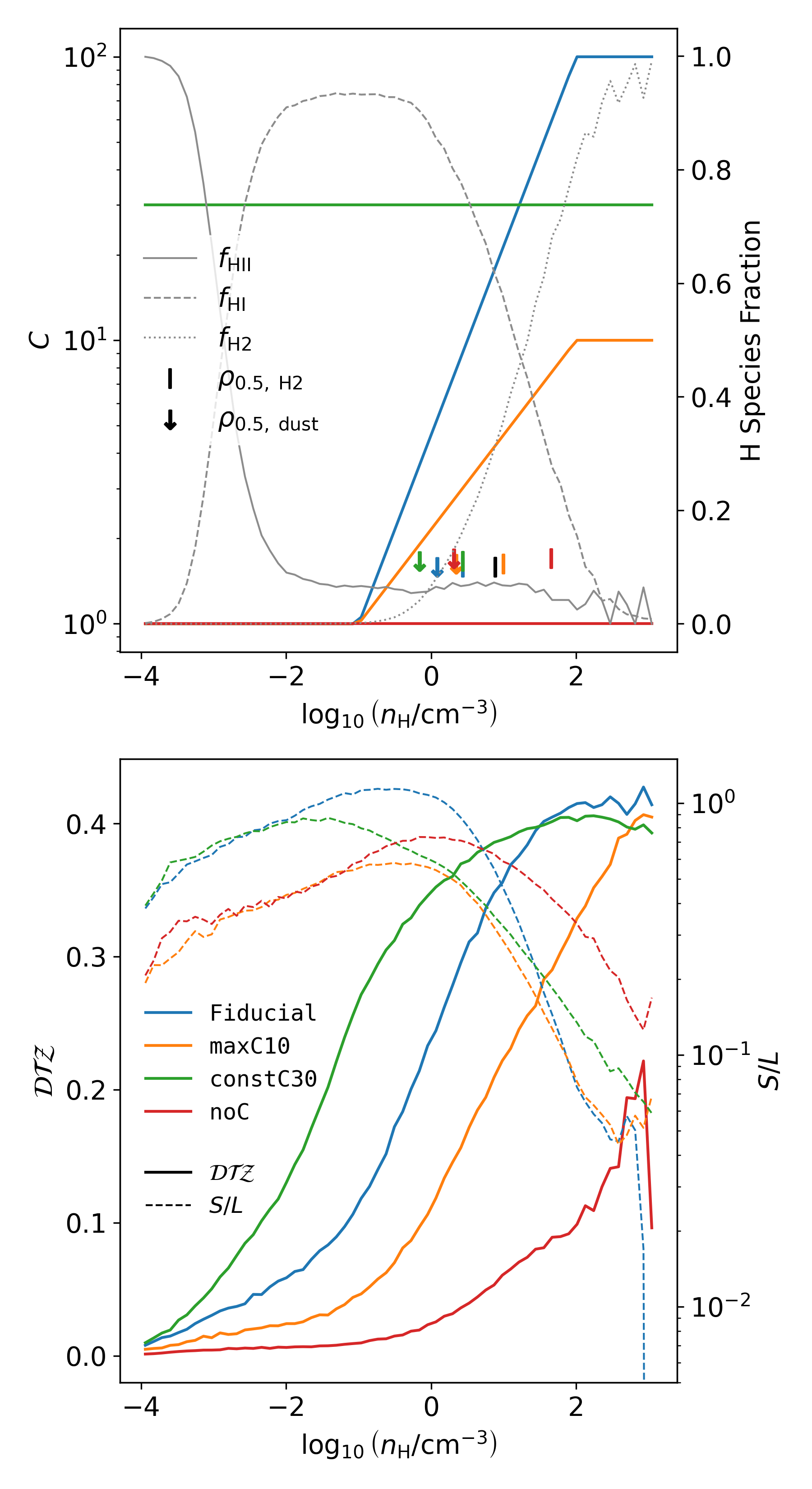}
    \caption{Density-dependent properties of dust models assuming differing  subgrid clumping factors. The top panel shows the clumping factor $C$ as a function of $n_{\rm H}$ for the ${\tt Fiducial}$ run, alongside a number of variations (thick coloured lines, left $y$-axis). For reference the hydrogen species fractions of the  \texttt{FidUncoupled} run are co-plotted. The dust (H$_2$) transitional densities (densities enclosing half the cosmic dust mass), are plotted as downward arrow marks (vertical line marks). Markers are given slight vertical offsets for distinguishability. The bottom panel shows properties of dust grains as a function of gas density, displaying the total \dtz{} (solid lines, left $y$-axis) and small-to-large grain mass ratio ($S/L$, dotted lines, right $y$-axis) in each bin. We see strong variations in the dust properties associated with different treatments of $C$, but also convergence at the highest densities for all but the \texttt{noC} run.} 
    \label{fig:clump}
\end{figure}

The runs varying the subgrid clumping factor, $C$ (see Table~\ref{table:sims}), are plotted together in Fig.~\ref{fig:clump}, showing properties as a function of $n_{\rm H}$. Here, the top panel indicates $C$ for each variation (coloured lines, left axis), with H species fractions for the \texttt{Fiducial} run (grey lines, right axis). The bottom panel then shows \dtz{} (solid lines, left axis) and the ratio of mass in small vs. large grains. To indicate how different runs transition from low \dtz{} at lower densities to the saturated regime in denser gas, we compute the density above which half of the dust in the entire cosmological volume is enclosed, $\rho_{0.5,\,{\rm dust}}$. These transition densities are indicated in the top panel of Fig.~\ref{fig:clump} using downward arrow markers. We can also see how varying $C$ affects gas properties; the density above which more than half of the gas is in the molecular phase, $\rho_{0.5,\,{\rm H2}}$, is model-dependent. For each run, $\rho_{0.5,\,{\rm H2}}$ is indicated using vertical line marks. 

For the \texttt{noC} and \texttt{maxC10} runs, the clumping factor is lower than or equal to that of the \texttt{Fiducial} run at all densities. We see that for \texttt{maxC10}, the transition to higher \dtz{} is pushed to higher densities by $\approx 1$~dex, while grain sizes are systematically larger for $\log_{10}(n_{\rm H}/{\rm cm^{-3}}) \lesssim 1.5$ (bottom panel). We note that the \texttt{maxC10} model does not reach the saturated regime, with an approximately constant \dtz{} ratio of $\approx 40\%$ at high ISM densities, as seen in \texttt{Fiducial}. For \texttt{noC}, we see a more extreme shift in the \dtz{} transition to higher densities. Interestingly, for $\log_{10}(n_{\rm H}/{\rm cm^{-3}}) \gtrsim 1.5$ the \texttt{noC} model yields smaller dust grains (dashed lines in the bottom panel, right axis). These trends in the grain size distribution are indicative of the lasting imprint of the grain sizes assumed for seed grains in low density gas where dust evolution proceeds slowly, i.e. reflective of seed grains having 90\% in large grains by mass (\S~\ref{sec:seedsize}), followed by efficient shattering at intermediate densities and coagulation in dense gas, balanced with the higher rates of gas-phase accretion per unit mass for smaller grains.

In terms of $\rho_{0.5,\,{\rm H2}}$ (top panel), we see that relative to \texttt{Fiducial}, \texttt{maxC10} is pushed to densities $\approx 0.6$~dex higher, while \texttt{noC} is pushed a further $\approx 0.6$~dex higher, indicative of the lower rate of molecule formation in less dusty ISM gas for a given density. The differences in  $\rho_{0.5,\,{\rm dust}}$ are less intuitive; the \texttt{maxC10} and \texttt{noC} values are both $\approx 0.2$~dex higher than fiducial. The lack of difference between the two variations could be a balance  of a more concentrated \dtz{} with a higher proportion of dust contributed by grain seeding and pushed outside the ISM.

The \texttt{constC30} model has a lower clumping factor than fiducial for $\log_{10}(n_{\rm H}/{\rm cm^{-3}}) \gtrsim 1$, but stronger clumping at lower densities. In the bottom panel we see how dust extends to significantly lower values in the constant clumping factor case, due to boosted growth at lower densities. The ratio of mass in the small grain bin to that in the large grain bin, $S/L$, is lower (i.e. larger grains) at intermediate densities, but falls more slowly towards high density as rates of coagulation rise less dramatically without an increasing $C$ value. We see that the \texttt{Fiducial} run yields approximately the same $\rho_{0.5,\,{\rm H2}}$ as the constant $C$ model, but a lower $\rho_{0.5,\,{\rm dust}}$. 

\subsection{Depletion}
\label{sec:depletion}

\begin{figure}
    \centering
    \includegraphics[width=\columnwidth]{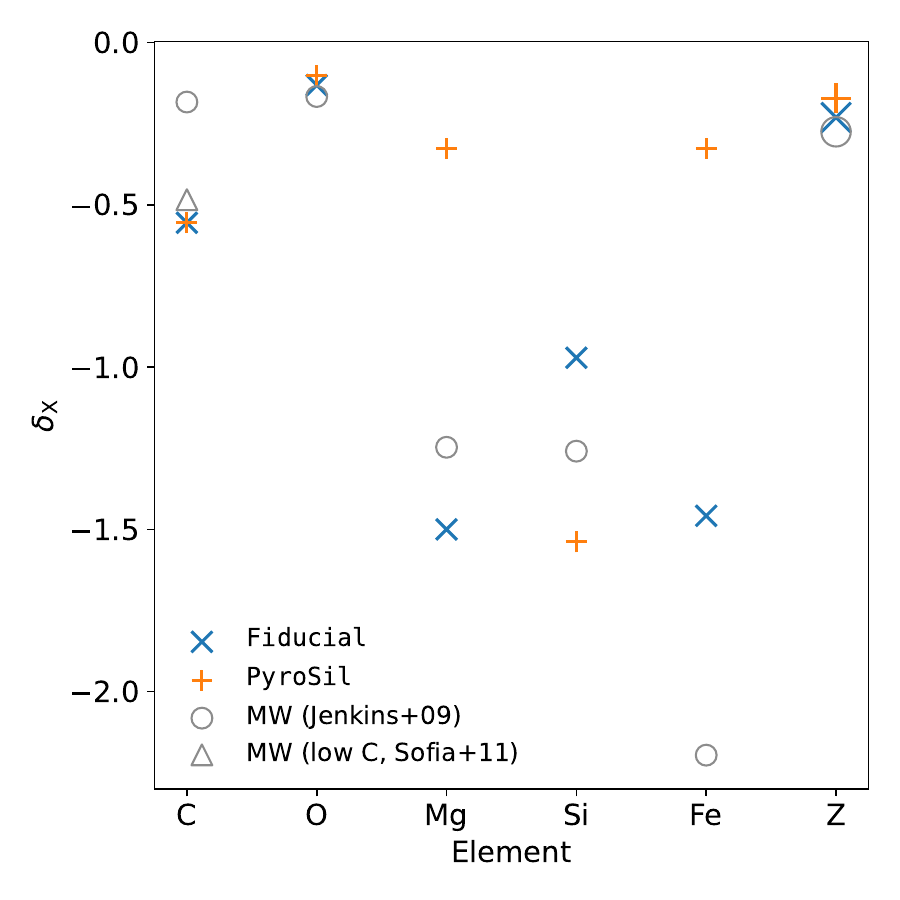}
    \caption{Elemental depletion and total metal depletion of dust constituents for our model. We compare the \texttt{Fiducial} run (\textit{blue $\times$ markers}), using graphite and olivine grain chemical species, to the grain chemistry variation \texttt{SilPyro} (\textit{orange + markers}), with graphite and pyroxene chemistry. For comparison, we plot the Milky Way ISM values as \textit{empty grey squares} (\citealt{Jenkins09}, corrected for our assumed $Z_\odot=0.0134$), as well as values for a factor 2 reduction in $\delta_{\rm C}$ \citep{Sofia11, Parvathi12}. C is depleted into the homonuclear graphite grains, while O, Mg, Si and Fe are depleted into the heteronuclear silicates. For visibility, the $\delta_{\rm Fe,\,MW}$ is plotted as an upper limit, given the value of -2.2. We also plot the total metal depletion $\delta_{Z}$. We see that the \texttt{Fiducial} run shows best MW-depletion agreement compared to \texttt{PyroSil}, with the exception of Si, which is comparable.} 
    \label{fig:depletion}
\end{figure}

In our development of the dust model, we prioritised total dust content (in terms of absolute masses, \dtg{} and \dtz{} ratios), but the depletion of individual elements provides more detailed information about the composition of the dust. For element X, depletion is the ratio between gas-phase and total (gas + dust) abundance $\delta_{\rm X}$, i.e.

\begin{equation}
    \delta_{\rm X} = [{\rm X/H}]_{\rm gas} - [{\rm X/H}]_{\rm total},
\end{equation}

\noindent such that more negative values are indicative of larger fractions of X being depleted onto dust grains. Strong depletion of elements is key to limiting grain growth in the ISM of galaxies, and can contribute to a saturation in the \dtz{} ratio and thus a near-constant \dtz{} ratio measured in massive galaxies \citep[e.g.][]{Zafar13, RemyRuyer14, DeVis19}. While a secondary consideration to the total dust content, elemental depletion  helps constrain our model.

We illustrate depletion patterns in  Fig.~\ref{fig:depletion} in order to compare how different elements bottleneck grain formation in different models, and how these compare to what we see in the dense ISM of the Milky Way (from \citealt{Jenkins09}, assuming the strong depletion case, $F_\star = 1$, corrected to our solar abundance as in \citealt{Ploeckinger20}). We aggregate gas of $n_{\rm H} > 50 \; {\rm cm^{-3}}$, to sample the depletion-limited regime as seen in the bottom panel of  Fig.~\ref{fig:clump}.

In this regime, the chosen grain chemistry and chemical enrichment set the dust content. We compare our \texttt{Fiducial} run to our silicate variant, \texttt{SilPyro}, which uses the \textit{pyroxene} chemical group for silicate grains, as opposed to our fiducial \textit{olivine}. We see that for \textit{total} metal depletion ($\delta_Z$, rightmost points) is slightly lower than the \citet{Jenkins09} observations for both models, but best fit by the grain chemistry choices of our \texttt{Fiducial} run.

\begin{figure*}
    \centering
    \includegraphics[width=0.98\textwidth]{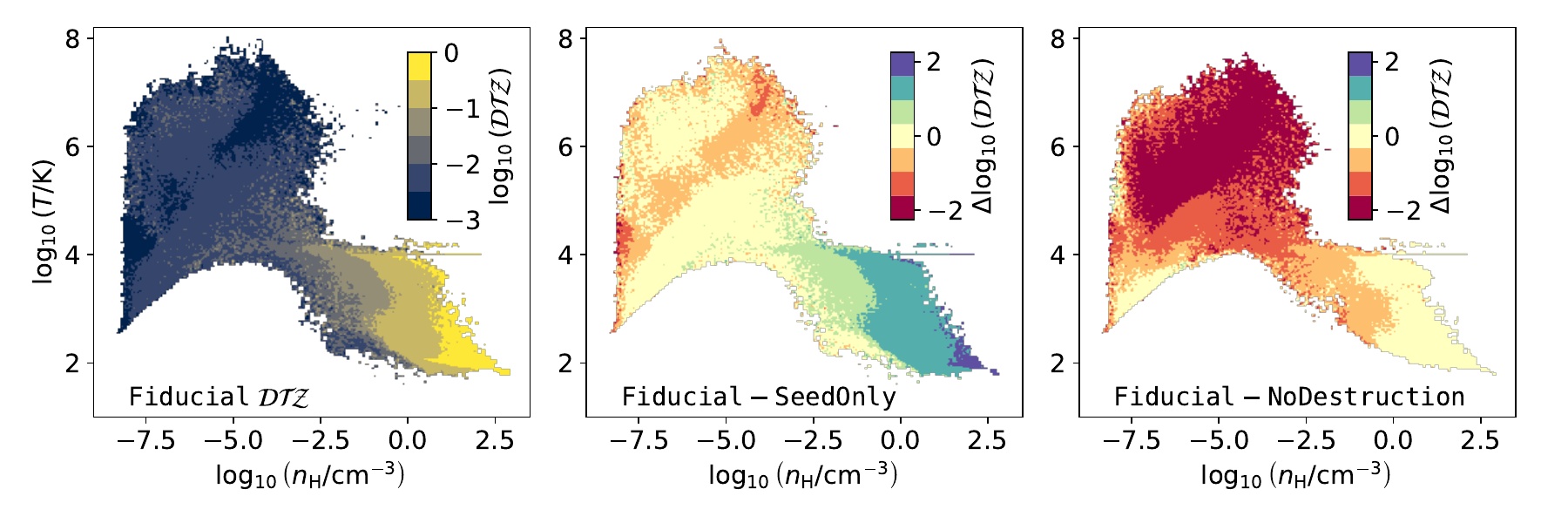}
    \caption{Density-temperature ($n_{\rm H}$-$T$ or \textit{'phase'}) diagrams for binned gas particles in our simulations, with stepped shading of cells to illustrate comparative dust properties. \textit{Left panel} shows the total $\log_{10}$ dust-to-metal ($\mathcal{DTZ}$) ratio for each $n_{\rm H}$-$T$ bin in the \texttt{Fiducial} run. The \textit{middle panel} shows the difference in $\log_{10} \mathcal{DTZ}$ of the \texttt{Fiducial} run relative to \texttt{SeedOnly}, with the \textit{right panel} comparing to \texttt{NoDestruction} in place of \texttt{SeedOnly}. We see that in our \texttt{Fiducial} run, $\mathcal{DTZ}$ is higher in relatively dense, cool ($T<10^{4}{\rm K}$) gas, boosted strongly above the $\mathcal{DTZ}$ in \texttt{SeedOnly}. Meanwhile, the \texttt{Fiducial} run shows a strong reduction in dust in relatively hot, diffuse gas relative to \texttt{NoDestruction}.}  
    \label{fig:rhoT}
\end{figure*}

\subsubsection{Silicate grain depletion}
In the case of heteronuclear silicate grains, the depletion pattern of constituent elements (O, Mg Si and Fe) depends on the grain chemistry, and its interplay with the yields and gas-phase abundance of those elements in the ISM. As such, depletion patterns can provide motivation for our choice of effective grain molecule. We consider both the \textit{pyroxene} and our fiducial \textit{olivine} groups as candidate silicate molecules, with the form (Mg,Fe)\textsubscript{2}Si\textsubscript{2}O\textsubscript{6} and (Mg,Fe)\textsubscript{2}SiO\textsubscript{4}, respectively. 

Comparing the $\delta_{\rm X}$ values of Fig.~\ref{fig:depletion} for silicate constituents shows that the \textit{olivine} chemistry of the \texttt{Fiducial} model is closer to the observed MW depletions
 \citep{Jenkins09} for all elements, excepting Si. We see that Si is over-depleted for \textit{pyroxene} (\texttt{SilPyro}) grains. It appears this has the effect of limiting the depletion of the pyroxene Fe and Mg endmembers\footnote{\textit{Ferrosilite} and \textit{enstatite}, respectively.}, leaving their depletion levels far below observed MW values (i.e. around 50$\%$ in the gas-phase, as opposed to just 5$\%$).

 While O is observed to be the least depleted silicate constituent, it can still contribute significantly to the silicate mass due to its higher abundance compared to the other constituent elements. We find that using higher \textit{olivine} and \textit{pyroxene} silicates both slightly under-depletes O, with the olivine molecule faring slightly better. This is intriguing; olivines represent the most oxygen-rich viable silicate candidates, so a pure olivine model still under-depleting O could be indicative of a potential tension between the metallicities and depletions, perhaps via the metallicity calibration examined further in Appendix~\ref{ap:zcal}. This difficulty in explaining O depletions has been noted previously \citep[e.g.][]{Whittet10}.

The two models differ qualitatively for Si; olivine silicate chemistry under-depletes Si, while it is over-depleted for pyroxenes. This is perhaps indicative of Si being the bottleneck element in the more silicon-rich pyroxene grains of the model, and explains the strong under-depletion of Mg and Fe in the pyroxene models.

It is notable that the MW Fe depletion is considerably stronger than we achieve in our models, with only around 0.6$\%$ left in the gas-phase compared with 3$\%$ for our \texttt{Fiducial} model. However, this reflects only a small change in the dust mass, which is the dominant cooling channel for colder gas (Fig.~9 of \citealt{Ploeckinger20}).

We note that laboratory studies comparing to observations in the X-ray can provide evidence for certain (olivine or pyroxene) silicates or some relative mix between the two \citep[see e.g.][]{Zeegers19, Rogantini20, Psaradaki23}. While a single silicate group is clearly a simplification, and tracking more grain chemistry species (e.g. including metal oxides, silicon carbide grains), may help to balance the depletion of elements, this would complicate our model\footnote{For example, having multiple grain types containing the same elements requires some decision as to how this depletion is split.} and contribute to its memory overhead. We therefore use olivine as a viable, effective molecule that we find to reasonably reproduce depletion through its fixed ratio of constituent elements is therefore our preference.

\subsubsection{Graphite grain depletion}
In the case of graphite, depletion is much simpler as it is a mononuclear grain. However, we found that when depletion of C is uninhibited, it depletes too readily, such that close to 100\% of carbon is locked in dust grains in the ISM. As gas-phase molecular C can be an important ISM coolant, this is both undesirable and unphysical.

In observed molecular clouds, the majority of gas-phase carbon is in CO \citep{Fuente19}. The triple bond in CO is the strongest molecular bond and CO only solidifies at temperatures $T\lessapprox 70~{\rm K}$. We take these properties together to mean that carbon in the CO reservoir, which our simulations do not resolve and which we do not track in our coupled simulations, is not readily available for depletion onto solid grains. We therefore limit carbon depletion by making some fraction of the gas-phase element unavailable for depletion. We choose a maximum dust depletion factor of 3 (66\% of carbon in dust), following the empirical study of local molecular clouds of \citet{Fuente19}. We note that this limit is only relevant for carbon, and oxygen depletion is instead bottle-necked by the availability of silicate constituent elements. 

A similar need to limit accretion of carbonaceous dust, and appeal to a CO reservoir, is employed by \citet{Choban22} for their simulations implemented in \textsc{fire-2}. \citet{Choban22} also highlight that C depletion is particularly poorly constrained by observations, owing to the limited environments probed and scarcity of sight-lines in \citet{Jenkins09}, with the suggestion that the observed C depletion may be too high \citep{Sofia11, Parvathi12}. We find that our $\delta_{\rm C}$ is indeed too low compared to \citet{Jenkins09}, but incorporating this correction factor brings $\delta_{\rm C}$ values into good agreement with our model. For simplicity, and given the limited resolution of our simulations, we use our empirical treatment which avoids tracking CO formation directly. However, we note that employing a CHIMES chemical network that includes C and O would allow us to track this self-consistently to limit the pure C accretion. 

\subsection{Resolution}
\label{sec:res}

While we focus on the m6 resolution in this work, with gas (DM) particles of $1.84\times10^6 \; {\rm M_\odot}$ ($2.42\times10^6 \; {\rm M_\odot}$) , the dust model presented here is intended to be used across a range of resolutions, and
as a single component of the multi-resolution COLIBRE simulation suite. In particular, we defer plots showing the convergence properties of dust (particularly the $M_\star$-$M_{\rm dust}$ relation and grain size ratio as a function of $M_\star$) to the presentation of the COLIBRE simulations as a whole by \citet{Schaye25} in context with the convergence of other aspects of galaxies. With COLIBRE we target \textit{`weak convergence’} \citep[e.g.][]{Schaye15}, aiming to achieve a consistent repoduction of target datasets through the choice and calibration of key parameters. This calibration process happens in concert with the other modules that constitute COLIBRE, with feedback calibration detailed in \citet{Chaikin25}.

Elements of the modelling were designed to support a range of resolutions, for example by the implementation of a clumping factor (\S~\ref{sec:clump}) intended to represent unresolved structures like molecular clouds and their dense cores at intermediate and low resolutions, as opposed to an explicit sub-grid model of molecular clouds in simulation. The dust properties in COLIBRE are generally converged for galaxies resolved by $\gtrsim$1000 star particles, or $\gtrsim 2\times10^9 \; {\rm M_\odot}$ at m6 resolution. At the highest masses, $\gtrapprox 10^{10.7} \; {\rm M_\odot}$, convergence deteriorates somewhat owing to the role of AGN feedback calibration. Dust convergence properties are discussed in more detail in \S~7.2.5 of \citet{Schaye25}.

\section{Results}
\label{sec:results}

Here, we present results from our simulations (Table~\ref{table:sims}), and compare to observations. In \S\ref{sec:intragal} we first present and discuss some resolved dust maps for some of our \texttt{Fiducial} run galaxies. In \S\ref{sec:isoevo}, we present phase diagrams shaded according to \dtz{}, isolating the role of different evolutionary processes in the proliferation of dust. We then consider the evolution of cosmic dust density with redshift in \S\ref{sec:cdd}, comparing our \texttt{Fiducial} run to observations, alongside other variation runs isolating different effects. In \S\ref{sec:gdmf}-\ref{sec:sizes}, we then  compare galaxy dust relations from our cosmological volume simulations at $z=0$ with low-redshift observational data, focussing on our \texttt{Fiducial} model.

\subsection{Resolved properties}
\label{sec:intragal}

\begin{figure*}
    \centering
    \includegraphics[width=\textwidth]{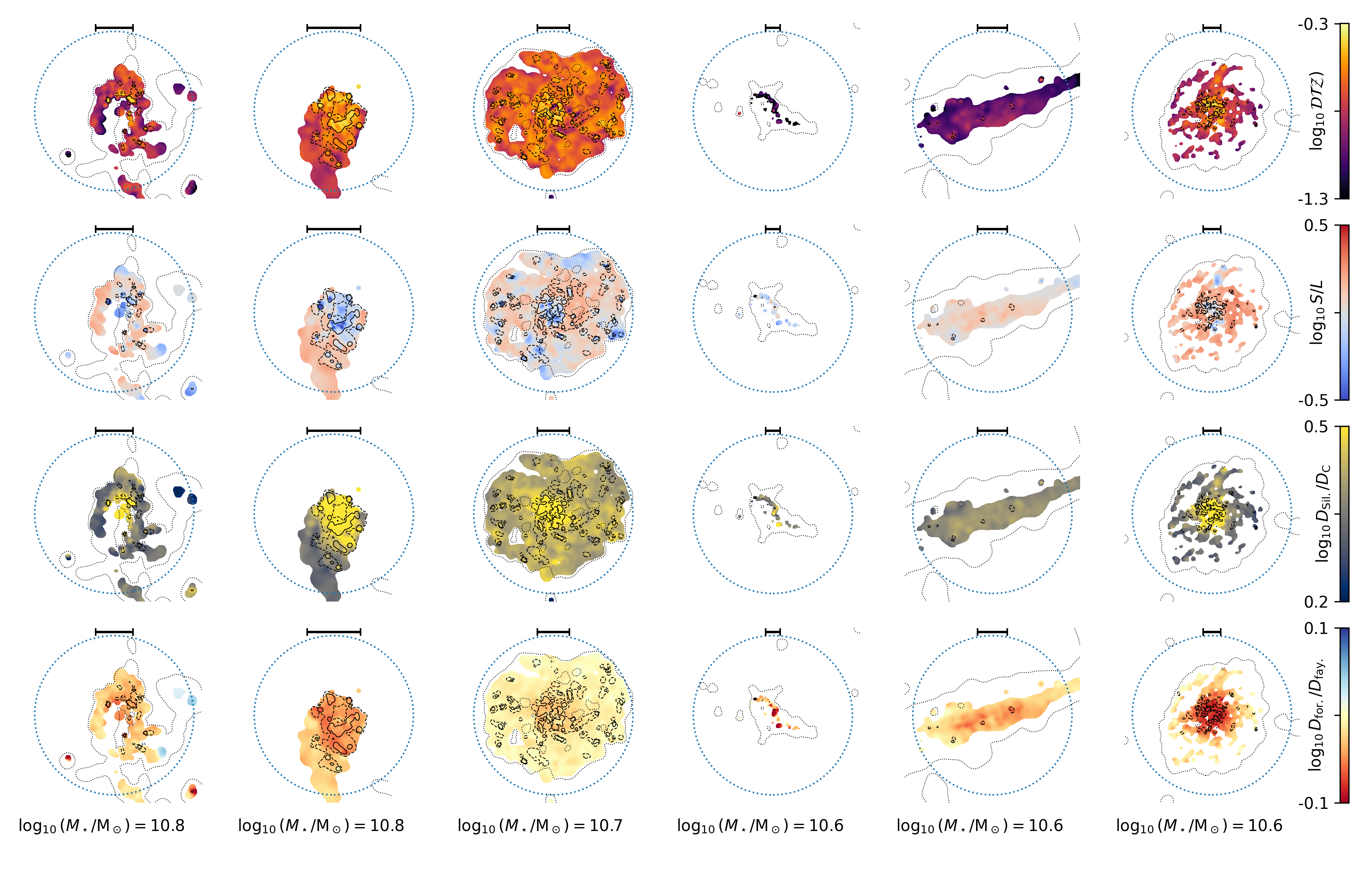}
    \caption{A gallery of the six most massive galaxies with more than 10$\%$ of their baryons in gas from our \texttt{Fiducial} run, projected in box coordinates (i.e. randomly-oriented), illustrating the heterogeneity of dust morphologies and the variation in the grain population across their diverse ISM. Each column shows a different galaxy, scaled to 1.5 times the 3D gas half-mass radius ($R_{\rm 0.5, \, gas}$), indicated by the dotted circles, where the constant physical scale of 5~ckpc is indicated above each projection using black whiskers. H$_2$ surface densities are also indicated in each image using contours to map gas species (\textit{dot} and \textit{dashed} enclose 50\% of hydrogen in H\textsc{i} and H$_2$ respectively, with \textit{solid} 90\% H$_2$). The dust distribution above the 60th percentile in dust density is shaded by different grain property ratios by mass in each row. From top to bottom, these show the \dtz{} ratio, the small-to-large grain ratio, the ratio of silicate to graphite grains, and the ratio of \textit{forsterite} (Mg-endmember) to \textit{fayalite} (Fe-endmember) silicates. Significant intra- and inter-galaxy variations can have important implications for the processing and re-emission of radiation.}
    \label{fig:gallery}
\end{figure*}

First, we briefly present a gallery of randomly oriented (i.e. box-projected) dust maps for a handful of galaxies taken from our \texttt{Fiducial} model in Fig.~\ref{fig:gallery}. These are the 6 highest stellar mass galaxies, defined within a 50~ckpc spherical aperture, that also have $>10\%$ of their baryonic mass constituted by gas. For each galaxy, we map the highest dust surface density ($\Sigma_{\rm d}$) pixels with colour shading, using a kernel density projection of the SPH gas particles, via the \texttt{swiftgalaxy}\footnote{\href{https://github.com/SWIFTSIM/swiftgalaxy}{\tt github.com/SWIFTSIM/swiftgalaxy}} tool. These are plotted alongside neutral and molecular hydrogen contours to show the relatively cold-dense gas regions. Colour shadings illustrate various properties of the dust population; the local \dtz{} ratio, the small-to-large grain ratio, the ratio of silicate to graphite/carbon grains and the ratio of forsterite (Mg-endmember) to fayalite (Fe-endmember) silicates.    

The main qualitative feature we aim to get across here is the heterogeneity of these populations and their spatial configuration; we see diverse dust morphologies between galaxies, with distinct discs, as well as an irregular and compact morphologies. We see some common features; for example, large-grain dominated dense, molecular regions, and small-grain dominated diffuse regions as well as higher depletion and more silicate-heavy central regions of galaxies. There are also distinct differences, for example overall \dtz{} variations, or dust chemistry differences. For example we see the clumpy and asymmetric property distributions of the first column galaxy, compared to the radial gradients and disc morphology of the last column galaxy.

\subsection{Isolating evolutionary processes}
\label{sec:isoevo}

In order to show the influence of different dust evolution processes, we compare runs where various processes are isolated. In particular, the \texttt{SeedOnly} run has no ISM dust evolution, only injecting grains from stellar channels. The \texttt{NoDestruction} run then turns on accretion and size transfer processes, leaving processes converting dust mass back into gas-phase metals off. These runs are compared to \texttt{Fiducial} in Fig.~\ref{fig:rhoT}, where we show gas phase diagrams ($T$ as a function of $n_{\rm H}$), the left panel shading by total $\mathcal{DTZ}$ ratio, and centre and right panels shading by the difference in $\mathcal{DTZ}$ with respect to the \texttt{Fiducial} run of \texttt{SeedOnly} and \texttt{NoDestruction}, respectively. 

We see an increasing gradient in the \texttt{Fiducial} run $\mathcal{DTZ}$ towards  denser, cooler gas, reaching values of $\mathcal{DTZ} \gtrsim 10 \%$ for $T \lesssim10^4~{\rm K}$, $n_{\rm H} \gtrsim 1 \; {\rm cm^{-3}}$. We see that this gradient is driven by accretion in the ISM; the $\mathcal{DTZ}$ is over an order of magnitude higher in the \texttt{Fiducial} run relative to \texttt{SeedOnly} (centre panel). We see that in the cool-dense gas, destruction processes have little influence over this, with the \texttt{NoDestruction} run showing similar $\mathcal{DTZ}$ to \texttt{Fiducial} (right panel).

By contrast, in the plume of hot, diffuse gas we see in Fig.~\ref{fig:rhoT} (i.e. $T > 10^4~{\rm K}$, $n_{\rm H} < 10^{-2.5} \; {\rm cm^{-3}}$) we see $\mathcal{DTZ} \lesssim 0.01$, and of order the level directly injected from stellar channels (centre panel Fig.~\ref{fig:rhoT}, comparing with Fig.~\ref{fig:yieldcomp}). However, we see that dust destruction plays an important role here; including destruction effects reduces $\mathcal{DTZ}$ by more than an order of magnitude relative to runs with only accretion and size-transfer processes (\texttt{Fiducial} vs \texttt{NoDestruction} comparison, right panel). This shows dust destruction processes mitigate dust growth through accretion in the ISM, leading to higher $\mathcal{DTZ}$ in hot, diffuse gas.

\subsection{Cosmic dust density evolution}
\label{sec:cdd}

\begin{figure*}
    \centering
    \includegraphics[width=0.98\textwidth]{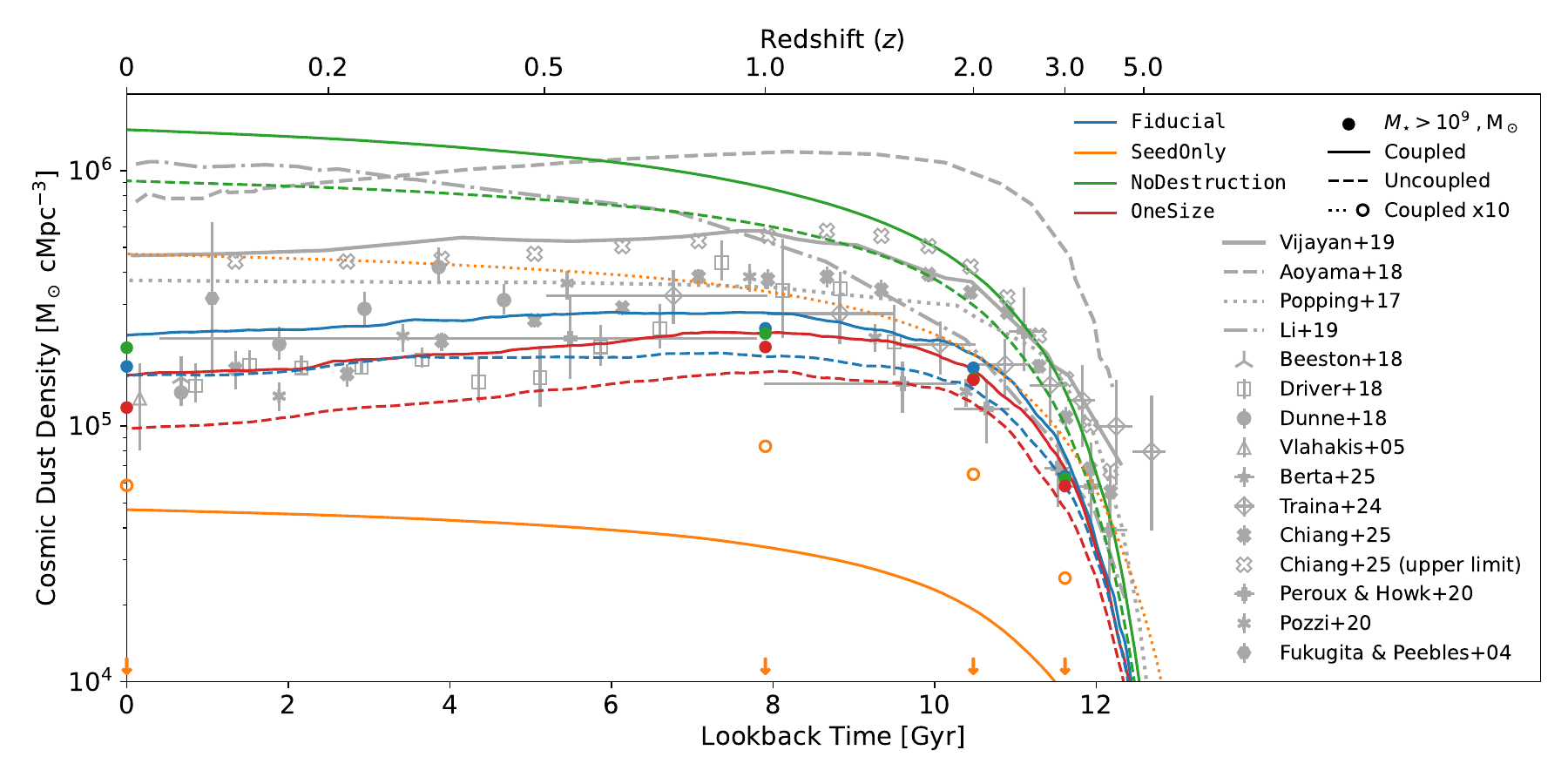}
    \caption{Co-moving cosmic dust mass density ($\rho_{\rm c, dust}$) as a function of lookback time, compared to observationally derived values. Comparing the \texttt{Fiducial} simulation (\textit{blue}) with the \texttt{SeedOnly} (\textit{orange}), \texttt{NoDestruction} (\textit{green}) and \texttt{OneSize} (\textit{red}) runs. Lines show both the cooling coupled (\textit{solid}) and uncoupled  (\textit{dashed}) versions of each run (excepting \texttt{SeedOnly} which is always uncoupled). \textit{Filled circles} include only dust in galaxies with stellar masses $\log_{10}(M_\star / {\rm M_{\odot}}) > 9$ in 50~pkpc apertures. Observations span a range of redshifts and approaches \citep[SED fitting, elemental depletions, attenuations;][]{Pozzi20, Peroux20, Beeston18, Driver18, Dunne18, Vlahakis05}. The total  $\rho_{\rm c, dust}$ of the \texttt{Fiducial} run lies between the \texttt{SeedOnly} and \texttt{NoDestruction} runs, while being comparable to the various data sets. However, when considering just dust in galaxies (\textit{coloured data points}) the \texttt{Fiducial} and \texttt{NoDestruction} runs are comparable, emphasising how the excess dust in \texttt{NoDestruction} is outside galaxies (see Fig.~\ref{fig:rhoT}), while the small impact on \texttt{Fiducial} implies that in the ISM, dust fraction are set by the balance between accretion and metal depletion.}
    \label{fig:cdd}
\end{figure*}

For insight into evolutionary processes further to \S\ref{sec:isoevo}, we investigate how the total dust content evolves with redshift in our simulations. Fig.~\ref{fig:cdd} shows the cosmic dust density, $\rho_{\rm dust}$, evolution (total dust mass density per co-moving volume, as a function of cosmic lookback time). Observational data \citep[grey markers,][]{Fukugita04, Vlahakis05, Beeston18, Driver18, Dunne18, Peroux20, Pozzi20, Traina24, Berta25, Chiang25} and alternative dust models \citep[grey lines,][]{Popping17, Aoyama17, Vijayan19, Li19}, are plotted here for comparison. Following Fig.~\ref{fig:rhoT}, \texttt{SeedOnly} yields $\rho_{\rm dust}$ values lower than all the data and models, by $\approx0.5\;{\rm dex}$ at its closest point ($z \approx 0$), showing the necessity of ISM accretion in our model to reproduce cosmic dust densities. By contrast, \texttt{NoDestruction} produces $\rho_{\rm dust}$ much higher than typically observed, and notably above the upper limit cosmic infrared background estimates of \citet{Chiang25}  $z\lesssim 2$ and the cosmic mass budgeting of \citet{Fukugita04}. The \texttt{Fiducial} model shows $\rho_{\rm dust}$ intermediate between the two, and is consistent with the dust given the variance between the different observations. The combination of accretive and destructive processes in \texttt{Fiducial} also induces a turnover in the evolution of $\rho_{\rm dust}$ that is not seen in the other two models. Finally, we see that neglecting small grains, size-related effects and transfer processes (\texttt{OneSize}), leads to a $\approx 50\%$ reduction in  $\rho_{\rm dust}$ relative to \texttt{Fiducial}.

The variety of observational datasets represent a number of different approaches to inferring $\rho_{\rm dust}$ through cosmic time. Galaxy survey approaches \citep[e.g.][]{Vlahakis05, Beeston18, Dunne18, Driver18, Pozzi20, Traina24, Berta25}, summing values from FIR detected galaxies or integrating the galaxy dust mass function, require some form of completeness corrections for faint galaxies \citep[though see][]{Duivenvoorden20}. Another approach is integrating the total cosmic infrared background \citep{Chiang25}, which avoids the completeness correction. These FIR approaches can be sensitive to temperature if the shape of the SED is not well constrained, but this can be mitigated by sampling across the Rayleigh-Jeans tail \citep[e.g.][]{Traina24, Chiang25}. Another important observational assumption is the wavelength-dependent dust opacity, $\kappa(\lambda)$. Commonly this is taken to be $\kappa(850 {\rm \upmu m} )= 0.77 \; {\rm cm^{2}\; g^{-1}}$, though estimates can vary by an order of magnitude \citep{Clark19}. Alternative probes, such as measuring depletion in damped Ly$\alpha$ absorbers \citep{Peroux20} or cosmic budgetary arguments \citep{Fukugita04} provide independent estimates. Some studies infer significant dust in the CGM of galaxies through indirect observation and budgetary arguments for metal production ranging from more or comparable dust content in the CGM relative to the galactic ISM \citep[e.g.][]{Menard10, Fukugita11, Peek15, Meinke23} to $\sim 10 \%$ in the ISM \citep[e.g.][]{McCormick18, Romano24}. To facilitate a comparison with such estimates, we additionally compute $\rho_{\rm dust}$ within galaxies only, by applying a 50~pkpc exclusive spherical aperture about the most-bound particle of each subhalo (the fiducial aperture scale for COLIBRE calibration) for galaxies of $M_{\star} > 10^9\;{\rm M_\odot}$, at $z \in [0,1,2,3]$ for the coupled runs (coloured circles). We see that with this approach, the \texttt{Fiducial} and \texttt{NoDestruction} runs are much closer together, both agreeing well with the observed data, with the \texttt{OneSize} lower (showing a similar offset below the total $\rho_{\rm dust}$ as with \texttt{Fiducial}). This shows the large increase in CGM dust in the \texttt{NoDestruction} run, with  $\approx 90\%$ of dust outside galaxies, by mass. \textcolor{black}{The lack of destruction in \texttt{SeedOnly} also leads to a similar fraction of dust existing outside galaxies, such that aperture values all lay below the plotting range (downward arrows).} In contrast, the \texttt{Fiducial} run has $\lesssim 30\%$ of the dust mass in the CGM. \texttt{NoDestruction} (\texttt{SeedOnly}) overproduces (under-produces) CGM dust compared to all observations, while the \texttt{Fiducial} run is consistent with observations finding lower levels of CGM dust \citep[e.g.][]{McCormick18, Romano24}.

For these runs we also show uncoupled equivalents of runs (i.e. where dust model has no influence on other gas physics, see \S\ref{sec:couple}), indicated by dashed lines, and excepting the \texttt{SeedOnly} run which is only run in an uncoupled context. In other cases we see that the uncoupled runs yield a factor $\sim 2$ lower $\rho_{\rm dust}$ values by $z=1$, as the self-consistent dust-treatment leads to greater grain growth by accretion. In addition, for the \texttt{SeedOnly} run we also show an orange dotted curve illustrating $\rho_{\rm dust}$ boosted by an order of magnitude, and empty orange circles showing the same boost for only dust within the $M_{\star} > 10^9\;{\rm M_\odot}$ sample. This is intended as a rough approximation of dust yield levels comparable to e.g. \citet{Dwek98}. We see that a high-yield paradigm appears better at reproducing the observations in lieu of evolutionary processes. However, we see that only counting dust within $M_{\star} > 10^9\;{\rm M_\odot}$ galaxies is still significantly low relative to the data so would require dust growth. This shows that adopting higher dust yields (discussed in \S~\ref{sec:yields}) could be a means to reproduce observations regulated with milder evolutionary processes (e.g. a lower clumping factor, \S~\ref{sec:clump}), as found by e.g. \citet{Choban25}. 

\subsection{Galaxy dust-mass function}
\label{sec:gdmf}

\begin{figure}
\centering
    \includegraphics[width=0.45\textwidth]{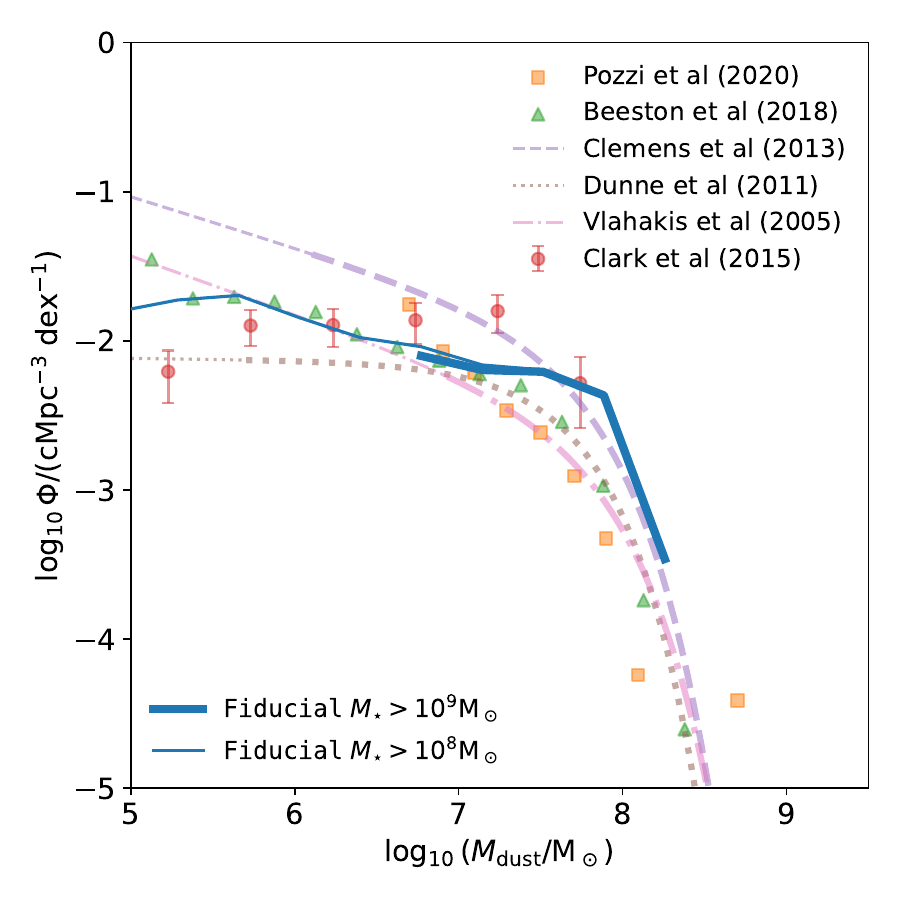}
    \caption{The galaxy dust mass function (GDMF). We plot values for our \texttt{Fiducial} simulation (blue, solid lines), indicating galaxies with stellar mass $\log_{10} \, M_\star/{\rm M_\odot} > 9$ (thick blue line) for bins where the selection is $\geq 85\%$ complete, as well as a broader selection with $\log_{10} \, M_\star/{\rm M_\odot} > 8$ (thin blue line). We plot observational data from a number of literature sources. Recent literature mass functions are plotted using individual data points for each $M_{\rm dust}$ bin \citep{Pozzi20, Beeston18, Clark15}, while those that are older are presented as single-Schechter fits \citep[patterned, translucent lines][]{Vlahakis05, Dunne11, Clemens13}, as collated by \citet{Clark15}. Where Schechter fits are extrapolated beyond the fitting range, a thinner line-style is used. We see a general consistency with observational data.}
    \label{fig:gdmf}
\end{figure}

In Fig.~\ref{fig:gdmf} we plot the \texttt{Fiducial} galaxy-dust mass function (GDMF), the cosmic number density of galaxies per unit logarithmic galaxy dust mass bin width (${\rm dex}^{-1}$), as a function of $M_{\rm dust}$, corrected to our assumed cosmology. To indicate the influence of resolution, we plot the mass function of galaxies with stellar masses $\log_{10} \, M_\star/{\rm M_\odot} > 9$ (i.e. resolved by $\gtrsim 500$ star particles) , as well as for a broader selection, $\log_{10} \, M_\star/{\rm M_\odot} > 8$, using thick and thin blue lines respectively. The narrower selection is plotted down to galaxy masses where it is $>85\%$ complete.

We plot a variety of observed GDMFs from the literature. For values published in the last 10 years, we indicate each bin value as individual data points. For older mass functions, we plot their best-fitting single Schechter functions \citep{Schechter76}, as collated by \citet{Clark15}. All Schechter functions are extended down to $\log_{10} \, M_{\rm dust}/{\rm M_\odot} = 5$, but we indicate values extrapolating the fitting range using a thinner line style.

We first note differences between the observational GDMF values. A variety of low-mass slopes and normalisations are exhibited, from close to flat (\citealt{Dunne11}, $\alpha=-1.01$) to negative (\citealt{Vlahakis05}, $\alpha=-1.67$) and even positive \citep{Clark15}. This can be attributed to the smaller observational volumes over which low-mass galaxies can be detected, inducing larger uncertainty in the low-mass GDMF. As a result, the Schechter functions are fit down to different limits and need to be extrapolated down to $\log_{10} M_{\rm dust} / {\rm M_{\odot}} = 5$, where they vary by $\approx 1$~dex. The number density of galaxies around the knee of the mass function (where the contribution to the cosmic mass density peaks) shows two distinct normalisations, with \citet{Pozzi20, Beeston18, Dunne11} and \citet{Vlahakis05} within $\approx 0.1$~dex of $\log_{10} \Phi / ({\rm cMpc^{-3} \; dex^{-1}}) = -2.5$ and \citet{Clemens13} and \citet{Clark15} at $\log_{10} \Phi / ({\rm cMpc^{-3} \; dex^{-1}}) \approx -2.1$. An important consideration for comparing the GDMF is the evolution observed in the late universe, with cosmic dust densities reducing by a factor $\approx 5$ over the past 3-5~Gyr for some observations \citep[][see Fig~\ref{fig:cdd}]{Beeston24, Pozzi20, Dunne11}. Here, we compare dust mass functions up to $z = 0.25$.

Comparing to these observations, we see that our \texttt{Fiducial} simulation GDMF exhibits general consistency with the range of observations. At lower dust masses ($5.5 \lesssim \log_{10} M_{\rm dust} / {\rm M_{\odot}} \lesssim 7.25$) the \texttt{Fiducial} GDMF traces the \citet{Beeston18} GDMF well, which extends well-constrained masses significantly below the limits of prior works ($\log_{10} M_{\rm dust} / {\rm M_{\odot}} \sim 4.5$) owing to the large number of galaxies in the optically-selected GAMA \citep{Driver11} galaxy sample.  At higher masses ($\log_{10} M_{\rm dust} / M_{\odot} \gtrsim 7.5$) we see that the knee of the GDMF extends to masses marginally higher than those bracketed by observations, and agrees better with the high-knee set of observations \citep{Clark15, Clemens13}. While the \texttt{Fiducial} model GDMF agrees well with observational sets in the high- and low-mass regime, it does not follow one observed GDMF consistently. The higher knee of the GDMF we predict appears consistent with the slightly higher $\rho_{\rm c,dust}$ we predict at low redshift.

\subsection{Galaxy dust-mass scaling relations}
\label{sec:scaling}

\begin{figure*}
  \centering
  \includegraphics[width=0.98\textwidth]{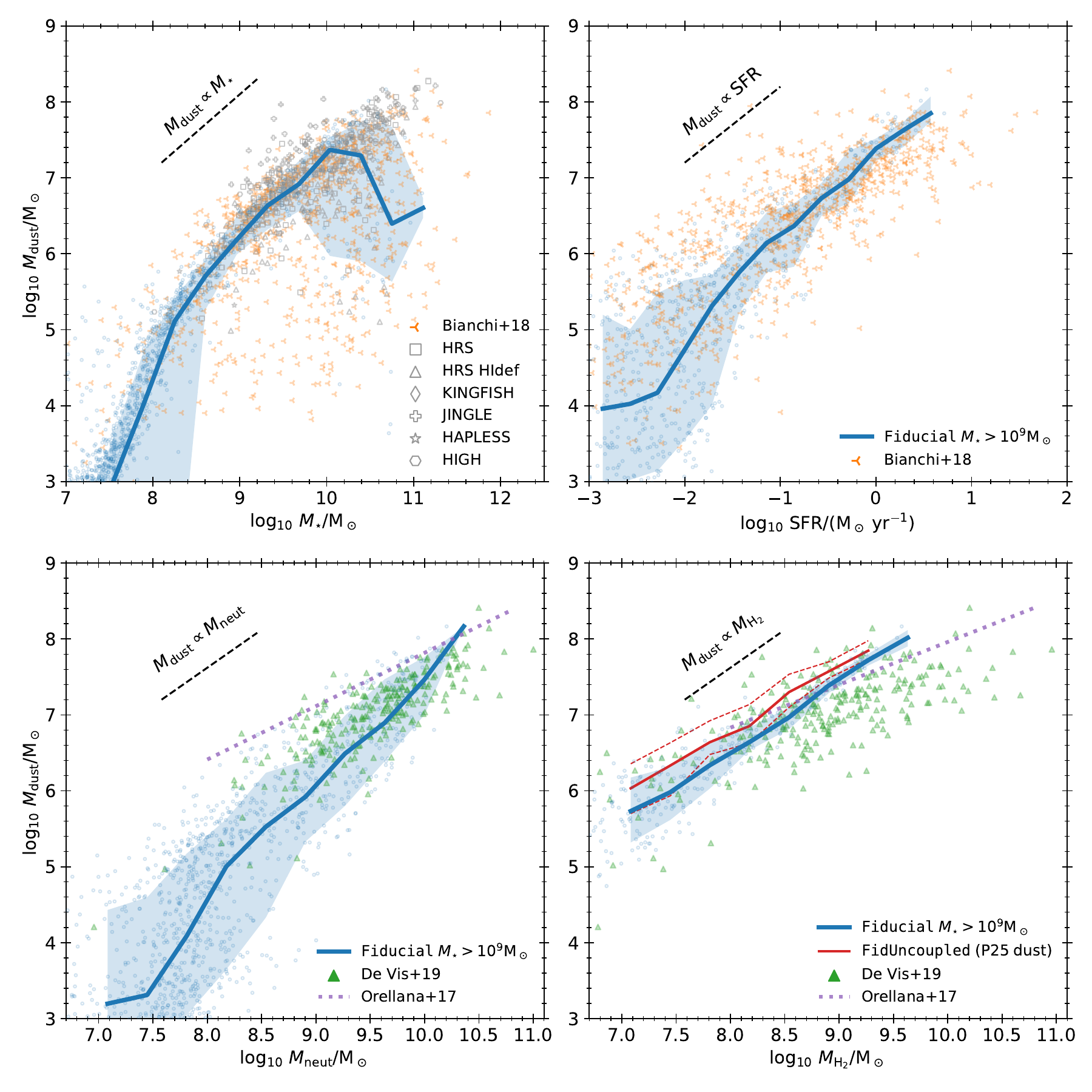}
    \caption{Galaxy dust scaling relations, showing the dust mass ($M_{\rm dust}$)  as a function of other baryonic (star and gas) properties of galaxies. In each panel, we show the median relation for galaxies taken from the \texttt{Fiducial} simulation (solid blue line), with shading indicating the 16th-84th percentile range (translucent blue). Individual galaxies are also plotted as low opacity blue data points. The \textit{top left} panel shows the  $M_{\rm \star}$-$M_{\rm dust}$ relation, with the \textit{top right} panel shows the ${\rm SFR}$-$M_{\rm dust}$ relation, comparing both to \textit{DustPedia} data presented \citet{Bianchi18}. We also compare $M_{\rm \star}$-$M_{\rm dust}$ specifically to the survey compilation of \citet{DeLooze20} (indicated for each contributing survey using grey markers). The \textit{bottom left} panel shows the $M_{\rm neut}$-$M_{\rm dust}$ relation, while the \textit{bottom right} panel shows $M_{\rm H_2}$-$M_{\rm dust}$, comparing to data using the masses of hydrogen phases in \textit{DustPedia} galaxies, as presented by \citet{DeVis19}, as well as a power-law fit to the Planck-2MRS data \citep{Orellana17}. To show if the tight $M_{\rm H_2}$-$M_{\rm dust}$ relation is driven by the dust coupling, we additionally plot medians in the bottom-right panel and scatter from the \texttt{FidUncoupled} runs, aggregating the \citetalias{Ploeckinger25} dust properties with SOAP.}
    \label{fig:scaling}
\end{figure*}

To further evaluate the relationship between dust masses and the other phases of the ISM in our modelling, we plot galactic dust masses ($M_{\rm dust}$) as a function of various salient properties in Fig.~\ref{fig:scaling}, comparing to the observed scaling relations from the \textit{DustPedia} catalogue, as presented by \citet{Bianchi18} and \citet{DeVis19}. On the observational side, these are derived by applying the \texttt{CIGALE} Bayesian SED-fitting code to galaxies in the Dustpedia catalogue. In particular, we compare to the galaxy stellar masses ($M_\star$), the ongoing galaxy star formation rates (SFRs), the neutral (i.e. non-ionised) gas masses ($M_{\rm neut} = M_{\rm HI}+M_{\rm H_2}$) and the H$_2$ masses ($M_{\rm H_2}$).

For the $M_\star$-$M_{\rm dust}$ relation, we see a generally concave increasing  relation, super-linear for $\log_{10} M_\star/{\rm M_\odot} < 8$, close to linear for $ 8 < \log_{10} M_\star/{\rm M_\odot} < 10$, and exhibiting a turnover for $\log_{10} M_\star/{\rm M_\odot} > 10$. This relation is relatively tight, with a scatter of $\lesssim 0.5$~dex at $\log_{10} M_\star/{\rm M_\odot} = 9$. At higher $M_\star$, the turnover is induced by an increasing scatter to low $M_{\rm dust}$ in $\log_{10} M_\star/{\rm M_\odot} = 10$ galaxies, increasing to $\sim 2$~dex. Physically, this scatter to lower $M_{\rm dust}$ owes to increasing passive fractions at high $M_\star$, as evidenced by the relatively tight relation displayed by the $M_{\rm dust}$-SFR relation (top right, discussed below), tightening with $M_\star$. Comparing to the observations of \citet{Bianchi18}, we see that the simulated galaxies are consistent with the data, tracing well the upper ridge line amongst the data points. These observations similarly exhibit a marked increase in scatter at higher masses ($M_\star/{\rm M_\odot} > 10$), though they appear to show a larger number of galaxies scattered to low $M_{\rm dust}$ extending to low-intermediate stellar masses ($\log_{10} M_\star/{\rm M_\odot} > 8.5$)\footnote{We note that the COLIBRE hybrid AGN feedback model, calibrated in larger volumes than presented here, enhances the high-mass scatter in this relation by 0.2-0.3~dex for galaxies of $\log_{10} M_\star/{\rm M_\odot} > 10$ in a 100$^3$~Mpc$^3$ volume at m6 resolution.}. We also compare to the multi-survey compilation of \citet{DeLooze20}, which exhibits less scatter at intermediate masses, agreeing very well with the \texttt{Fiducial} galaxies in the $8.5 \lesssim \log_{10} M_\star/{\rm M_\odot} \lesssim 10$ range. This compilation also exhibits increased scatter at high $M_\star$, though significantly less than seen in the \citet{Bianchi18} data points. This owes to the differing selection functions of the surveys, in particular the heavy sampling of Virgo and Fornax clusters given the median sample distance of 20~Mpc, and the stronger representation of early type galaxies in \textit{DustPedia} scattering to lower dust-to-stellar ratios \citep{Davies19}. The \texttt{Fiducial} simulation appears intermediate between these two samples.

The median $M_{\rm dust}$-SFR relation of Fig.~\ref{fig:scaling} (top right) is close to linear in the range $-2 < \log_{10} {\rm SFR}/({\rm M_\odot \; yr^{-1}}) < 0.5$, with increasing scatter towards lower SFR values, $\lesssim 0.5$~dex at $1 \; {\rm M_{\rm \odot} \; yr^{-1}}$, and reaching $\sim 2$~dex for $0.001 \; {\rm M_{\rm \odot} \; yr^{-1}}$. This relation follows the \citet{Bianchi18} data, albeit with a slightly steeper slope; on the lower side at $-3 < \log_{10} {\rm SFR}/({\rm M_\odot \; yr^{-1}}) < -2$, and on the higher side by $\log_{10} {\rm SFR}/({\rm M_\odot \; yr^{-1}}) \approx 0$. The observations show a similar trend for the scatter, with slightly larger amplitude. Additional scatter in the observations could be attributed to the larger uncertainties in measuring dim FIR sources corresponding to the lowest dust masses. 

Comparing to gas masses in the bottom row, the $M_{\rm neut}$-$M_{\rm dust}$ relation (bottom left) for median \texttt{Fiducial} galaxies exhibits a steadily increasing, slightly super-linear slope for $\log_{10}(M_{\rm neut}/{\rm M_\odot}) > 8.25$, with a steeper slope below this, owing to a transition to lower dust densities at $\log_{10}(M_{\rm neut}/{\rm M_\odot}) \approx 8$. This is likely due to a physical transition in galaxies at an ISM density threshold, where the \dtz{} relation transitions from being set by the seeding, to being set by a balance between accretion and the availability of gas-phase elements for depletion (see following section). 

Comparing to the data, we see a general consistency with the cloud of observed data points from \citet{DeVis19}. The median relation appears steeper than observed, with an excess of dust at the highest neutral masses ($\log_{10}(M_{\rm neut}/{\rm M_\odot}) \sim 10.25$) compared to \citet{DeVis19}. Comparing to the line representing the data from \citet{Orellana17}, a power-law fit to the values they derive for their 2MRS+Planck selection, we see a higher normalisation than \citet{DeVis19}, but actually agreeing well with \texttt{Fiducial} galaxies at $\log_{10}(M_{\rm neut}/{\rm M_\odot}) \approx 10.25$. However \citet{Orellana17} also shows a shallower, sub-linear relation than \texttt{Fiducial}, such that the simulated galaxies diverge increasingly towards lower masses. The higher $M_{\rm dust}$ values obtained for the \textit{Planck} observations is consistent with the picture  we see for the GDMF (Fig.~\ref{fig:gdmf}), where the \citet{Clemens13} GDMF shows a higher normalisation of the knee and a steeper low-mass slope compared to other observations, which agrees well with \texttt{Fiducial} at high $M_{\rm dust}$, but is above the simulation and other data at lower $M_{\rm dust}$. The \texttt{Fiducial} scatter appears comparable to that of \citet{DeVis19}, narrowing towards high $M_{\rm dust}$ and $M_{\rm neut}$, attributable to the limited sampling of massive galaxies in these 25$^3$~cMpc$^3$ volumes. 

The bottom right panel then plots dust mass against molecular-phase hydrogen mass ($M_{\rm dust}$-$M_{\rm H_2}$). Here the median \texttt{Fiducial} relation is distinctly sub-linear, where an increase in dust mass corresponds a greater proportional change in $M_{\rm H_2}$. This is perhaps unsurprising; alongside the regular mass trends, dust catalyses molecule formation, due to the crucial role of grain surfaces as formation sites for molecules and shielding gas from radiation. This relationship is notably tighter than that of $M_{\rm dust}$-$M_{\rm neut}$, with a scatter of 0.2-0.3~dex in $M_{\rm dust}$ for $\log_{10}(M_{\rm H_2}/{\rm M_\odot}) > 8.25$. This could imply that the physical relationship between $M_{\rm dust}$ and $M_{\rm H_2}$ is regulating the scatter. To test this, we also plot the fiducial run with our dust model de-coupled from the cooling (using the hybrid model \citetalias{Ploeckinger25} dust, see \S\ref{sec:cool}); \texttt{FidUncoupled}, where medians and 16-84th percentile range shown via red solid and dashed lines, respectively. Here we see that the relation is offset high by up to $\approx$0.3~dex for the \texttt{FidUncoupled} run when aggregating the implied \citetalias{Ploeckinger25} dust relative to \texttt{Fiducial} galaxies, with a comparable level of scatter. This suggests that the dust-model coupling boosts the H$_2$ mass at a fixed $M_{\rm dust}$. We again see steeper relationships than are observed, by \citet{DeVis19} and \citet{Orellana17}. This again leads to an excess of $M_{\rm dust}$ in \texttt{Fiducial} galaxies at high $M_{\rm H_2}$ relative to the \citet{DeVis19}, and one that is more pronounced than seen in the $M_{\rm neut}$ comparison (up to 0.5~dex above the data for $\log_{10}(M_{\rm dust}/{\rm M_\odot}) \approx 8$). The \citet{Orellana17} fit to Planck+2MRS also shows a familiar offset toward higher $M_{\rm dust}$, and again agrees better with \texttt{Fiducial} at the highest masses, and \texttt{FidUncoupled} at lower masses. 

\subsection{Dust-to-gas ratio}
\label{sec:dtg}

\begin{figure}
    \centering
    \includegraphics[width=0.49\textwidth]{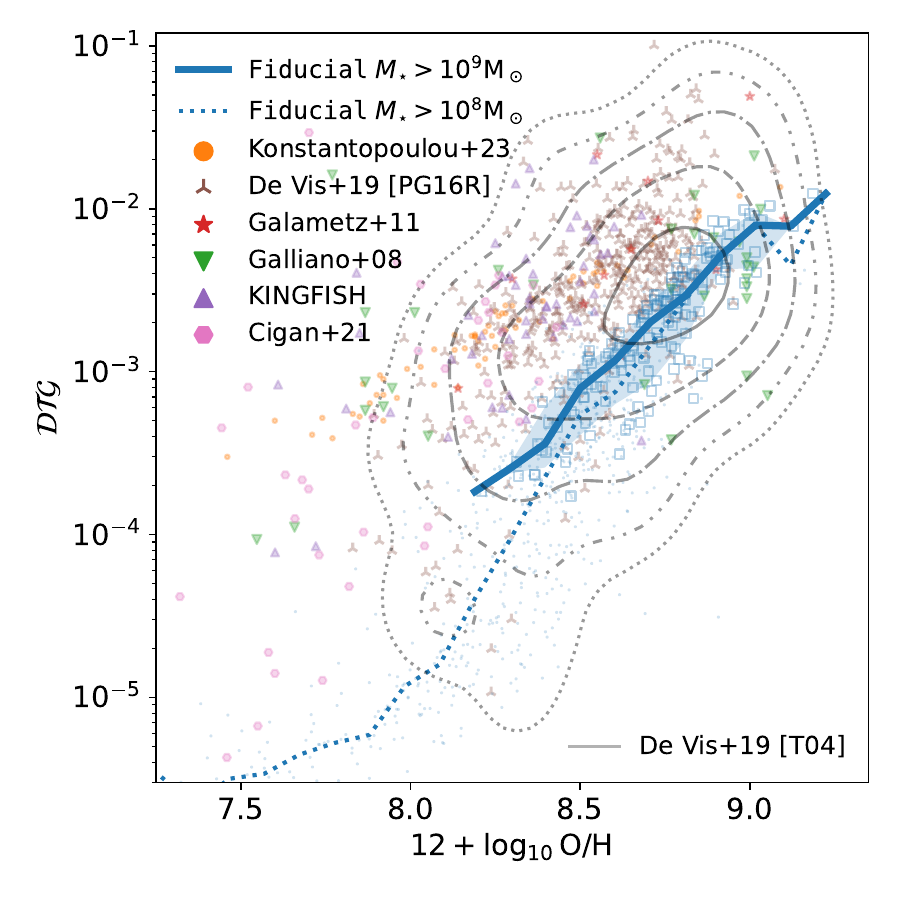}
    \caption{The dust-to-gas (\dtg{}) ratio of relatively \textit{active} (sSFR$= {\rm SFR} / M_\star > 0.01 \; {\rm Gyr}^{-1}$) galaxies as a function of their gas-phase oxygen abundances, measured in cool ($\log_{10}T/{\rm K} < 4.5$), dense ($\log_{10}n_{\rm H}/{\rm cm^{-3}} > -1$) gas, where ${\rm O/H}$ values are aggregated as the total linear ratio of oxygen to hydrogen in selected, in-aperture gas. These are compared to a compilation of observations \citep{Konstantopoulou23, Cigan21, DeVis19, Galametz11, Galliano08}. \textit{Blue lines} show median values while \textit{blue data points} show individual galaxies. The \textit{solid line} shows the median for galaxies with $M_{\star} > 10^9~{\rm M_\odot}$ and $\pm 1 \sigma$ ranges indicated by the shaded region and \textit{blue squares} showing individual galaxies. The \textit{dotted line} shows the median relation for all processed galaxies, with \textit{blue dots} indicating galaxies outside the high mass selection. We also use (KDE-smoothed) contours (with 0.46~dex spacing) to indicate the histogram of \citet{DeVis19} data using an alternative metallicity calibration (T04, \citealt{Tremonti04}), as opposed to their fiducial one (PG16R, \citealt{Pilyugin16}). Metallicity calibrations are investigated further in Appendix~\ref{ap:zcal}.} 
    \label{fig:dtg}
\end{figure}

To assess how the proportions of ISM material depleted into the solid dust phase in our simulated galaxies compare to observation, we present the galaxy dust-to-gas (\dtg{}) mass ratios in Fig.~\ref{fig:dtg}.

For \dtg{}, plotting values as a function of gas-phase metallicity can reveal how dust content varies with the ongoing enrichment of the ISM. Observationally, this metallicity is typically measured in the gas-phase via oxygen emission lines, taken as a proxy for total metallicity, $12+\log_{10}({\rm O/H})$. For our simulated galaxies, we compute this per subhalo for our simulated galaxies in 50~pkpc apertures about the most-bound particles, measuring the gas-phase oxygen fractions in relatively cold, dense gas ($\log_{10}T/{\rm K} < 4.5$, $\log_{10}n_{\rm H}/{\rm cm^{-3}} > -1$) gas, to approximate the regions probed observationally. Here, we limit the galaxy selection to actively star-forming galaxies (with specific star formation rates ${\rm SFR} / M_\star > 0.01 \; {\rm Gyr}^{-1}$), with $12+\log_{10}({\rm O/H})$ abundances aggregated linearly over the subhalo material (such that O/H is the ratio of total oxygen to hydrogen across all the cool-dense, in-aperture gas). Another important consideration for the comparison in this case is that with $12+\log_{10}({\rm O/H})$ as the independent variable, there can be considerable blending in particle resolution across the $x$-axis, particularly due to low-mass galaxies scattered to high metallicity bins. By default, we apply a stellar mass cut of $\log_{10}(M_\star/{\rm M_\odot}) > 9$ (corresponding to galaxies resolved by $\gtrapprox 500$ star particles) to our primary median relation (solid line), while also plotting a median relation using a lower $\log_{10}(M_\star/{\rm M_\odot}) > 8$ cut ($\gtrapprox 50$ star particles) to show the influence of less well resolved systems.

We see that the median $\mathcal{DTG}$ increases with metallicity as seen in the observations, following a slightly steeper relationship than the point cloud of observations. In this comparison, we also observe a noticeable offset between the data and the simulated $\mathcal{DTG}$ values at fixed $12+\log_{10}({\rm O/H})$, with the simulations $\approx 0.4-0.2$~dex lower $\mathcal{DTG}$ for a fixed metallicity. This is perhaps surprising, given that dust masses appear to be in good agreement or even offset above observed relations in previous plots (knee of the GDMF in Fig.~\ref{fig:gdmf}, $\log_{10}M_{\rm \star}/{\rm M_\odot} \approx 7.8$). Given the uncertainty in the absolute calibration of observationally-inferred metallicity values in the gas-phase \citep{DeVis19}, we also make use of the wider \textit{Dustpedia} catalogues provided by \citet{DeVis19}, tabulating numerous metallicity values per galaxy, via a variety of metallicity probes and calibrations from the literature. To show the extent of this variation, we also use contours to plot the \citet{DeVis19} O3N2 calibration of \citet{Tremonti04}, which is in good agreement with the simulated data at high metallicity. We expand on this calibration issue in Appendix~\ref{ap:zcal}, also showing similar levels of agreement with the \citet{Tremonti04} mass-metallicity relation. We note that the gas slection and aggregation of total O/H for galaxies can also help explain a portion of these differences. In lieu of a more observational, forward modelled measurement of O/H for our galaxies \citep[e.g. figure~2 of ][]{Nelson18}, we stick to a simple linear aggregation. 

Regarding the scatter in the relation between $\mathcal{DTG}$ and metallicity, we first see that the intrinsic scatter in our simulated galaxies appears significantly smaller than that of the cloud of observed data points, with a $\pm 1 \sigma$ range below 0.4~dex across the metallicity range for galaxies of $M_{\star} > 10^9~{\rm M_\odot}$. Considering these data sets individually, we see that a lot of this global scatter comes from systematic offsets between datasets, though the majority of observed sets individually exhibit scatter that is larger than predicted\footnote{The \citet{DeVis19} dataset can be seen isolated in Appendix~\ref{ap:zcal}.}. An exception is the relation of \citet{Konstantopoulou23}, which is significantly tighter. It is possible that a lack of scatter could be due to the limited resolution of the simulations, given we are unable to directly model ISM structure below the resolution scale, that could provide an additional source of variance between galaxies. However, variation in scatter between data sets, given intrinsic uncertainties associated with metallicity calibrations and dust measurements, could contribute significantly to the larger observed scatter.

Comparing the relation for better resolved ($\gtrapprox 500$ star particles) galaxies with a broader selection using a lower resolution threshold ($\gtrapprox 50$ star particles), we see that for ($12+\log_{10}({\rm O/H}) > 8.4$) the broader selection is marginally lower ($\lesssim 0.1$ dex) owing to the inclusion of less-resolved galaxies. We begin to see the broad selection $\mathcal{DTG}$s diverge to lower values at $12+\log_{10}({\rm O/H}) \sim 8.25$, falling super-linearly  to $12+\log_{10}({\rm O/H}) \lesssim 8.1$, and approximately linearly again below that. We note that this transition feature appears to be resolution-dependent, shifting to lower metallicities at higher resolution (Vijayan et al. \textit{in prep.}). For the narrower selection $12+\log_{10}({\rm O/H}) \lesssim 8.2$ we run out of galaxies, owing to a lack of low-metallicity massive objects.

\subsection{Dust grain sizes}
\label{sec:sizes}

\begin{figure}
    \centering
    \includegraphics[width=0.45\textwidth]{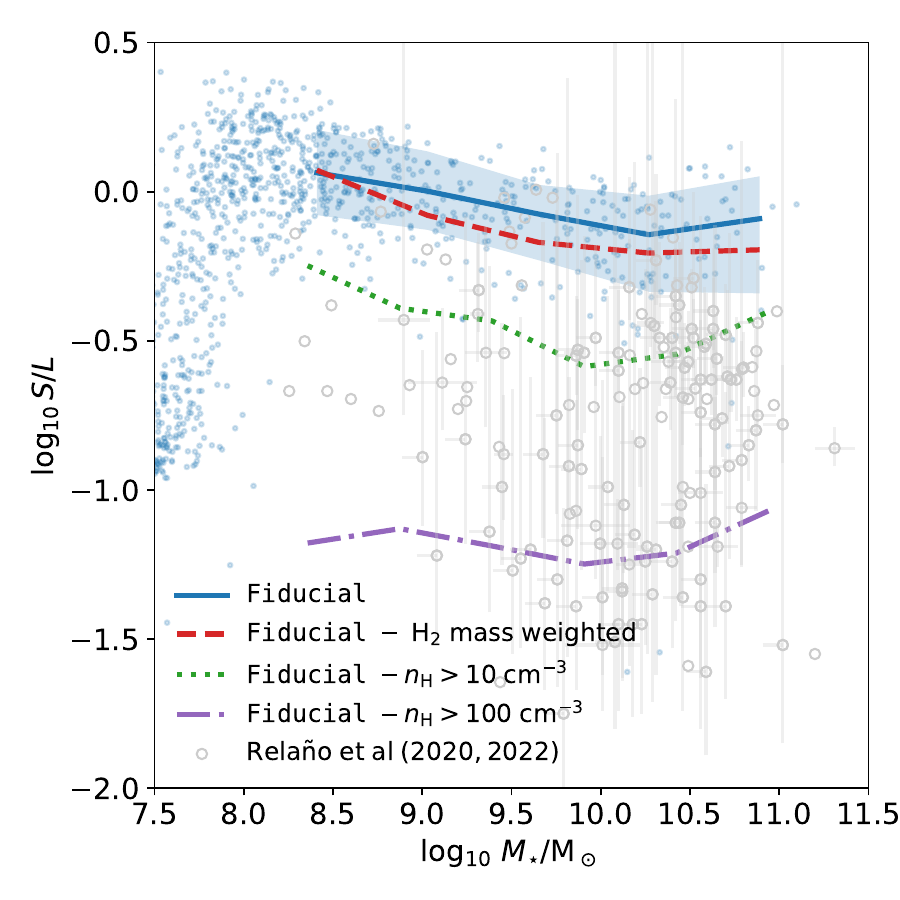}
    \caption{Ratio of depleted mass in small to large grains, as a function of galaxy stellar mass, measured within exclusive 50~pkpc apertures. We use the ratio of sizes in our 2 grain-size bins, $a_L = 0.1$~\micrometer{} and $a_S = 0.01$~\micrometer, to indicate the size distribution and compare it to the observationally inferred data of \citet{Relano20}, inferred through fitting large silicate grain (BG), very small grain and PAH components to the IR SEDs of local galaxies. The \textit{lines} show median $\log_{10} S/L$ values. For the \texttt{Fiducial} run we show mass weighted values among all bound aperture gas (\textit{blue solid}, with \textit{blue shaded} 16-84th percentiles), as well as medians computed within an imposed moderate-density ($\log_{10} n_{\rm H}/ {\rm cm^{-3}} > 1$, \textit{green dotted}) and high-density ($\log_{10} n_{\rm H}/ {\rm cm^{-3}} > 1$, \textit{purple dot-dashed}) gas selection. We also show H$_2$-mass weighted $S/L$ medians, across all aperture gas (\textit{red, dashed}).} 
    \label{fig:s2l}
\end{figure}

Another property we can probe with our dust evolution model is the grain size distribution. In Fig~\ref{fig:s2l}, we plot the logarithmic mass ratio of small grains relative to large grains for galaxies, $\log_{10} S/L$`. We first focus on ratios computed for all gas within the 50~pkpc exclusive aperture defining each galaxy (\textit{blue lines, shading} and \textit{points}). This provides a metric for the grain size distribution in galaxies, where positive values are small-grain mass dominated and negative values are large-grain mass dominated. We see that for our \texttt{Fiducial} model, median galaxy dust masses are dominated by large (0.1~\micrometer) grains for $\log_{10}(M_\star/{\rm M_\odot}) > 8.5$. In general, we see a weakly decreasing trend for our median values, such that the average galaxy goes from rough parity between the mass in small and large grains, to most massive galaxies having $\approx 60\%$ of their dust mass in large grains. This trend is significantly smaller than the scatter, however, which is of order 0.4-0.6~dex, and exhibiting diverse extremes; extending between close to an order of magnitude more mass in large grains, to around $50\%$ more mass in small grains. This diversity in the size distribution is profound, and will necessarily impact the extinction properties of grains (which are particularly relevant for post-processing observables from simulated galaxies), as well as the rates of dust-related cooling and heating processes at a fixed \dtg{} ratio.

Obtaining extragalactic dust sizes observationally is a challenging endeavour. These are very difficult to decode from the attenuation curves of external galaxies; the confounding influence of star-dust  geometries \citep{Fischera03, Narayanan18, Trayford20} alongside the actual extinction properties (shaped by the underlying population of grains). While certain grain populations may induce distinctive features in attenuation (e.g. the 2175\AA{} bump, \citealt{Noll09}), the overall mass-weighted grain distributions are more elusive. By also appealing to the shape of the FIR SED, and making assumptions about the heating of differently sized grains, we may be able to infer sizes indirectly. We make use of  values derived by \citet{Relano20, Relano22} for nearby galaxies. These are again presented as a single $\log_{10} S/L$ value per galaxy. \citet{Relano20} obtain their estimates through fitting three fixed-shape components to the FIR SEDs of galaxies; big (silicate) grains (BG), very small grains (VSG) and PAHs. These use the classical dust templates of \citet{Desert90}, and their defined size ranges of each component; with BGs of $a_{\rm BG} > 0.015$, VSGs of $0.0012 \; {\rm \upmu m} < a_{\rm VSG} < -0.015 \; {\rm \upmu m}$ and PAHs with $a_{\rm PAH} < 0.0012 \; {\rm \upmu m}$.

We see that while some observed galaxies are inferred to have a comparable mass in small grains relative to large, as exhibited by the all-gas median values (\textit{blue}) in our \texttt{Fiducial} simulation ($\log(S/L) \sim -0.1$), most observed galaxies are assigned significantly lower small-to-large grain mass fractions ($\log(S/L) \lesssim -0.5$). These inferred values do come with large error bars, typically $\approx 0.5$~dex
The data has a similarly weak trend with $M_\star$, exhibiting a large scatter in small-to-large grain ratio at fixed mass.

To investigate this apparent inconsistency further, we can look at the small-to-large grain ratio within a more stringent gas selection. With $S/L$ ratios showing strong trends in gas density and phase (see e.g. Fig.~\ref{fig:rhoT}), we over-plot medians for H$_2$-weighted average $S/L$ values (\textit{red}), as well a after imposing lower limits on the hydrogen number density, $n_{\rm H} > 10 \; {\rm cm}^{-3}$ and $n_{\rm H} > 100 \; {\rm cm}^{-3}$. We observe that considering H$_2$-weighted gas does generally push $S/L$ to lower values, but only by $\approx 0.1$~dex for $\log_{10}(M_\star / {\rm M_{\odot}}) > 9$. Considering high-density gas selections has a stronger impact, with $n_{\rm H} > 10 \; {\rm cm}^{-3}$ gas 0.3-0.5~dex lower than the all-gas ratios, and $n_{\rm H} > 100 \; {\rm cm}^{-3}$ with an offset of $>1$~dex below all-gas, consistent with the lower sequence of observed values. Along with the gas selection, we note that again resolution appears to play a role here, with the higher m5 (lower m7) resolution yields higher (lower) $S/L$ ratios respectively (Vijayan et al. \textit{in prep.}). 

This shows the influence of dense gas processes, particularly coagulation, in setting the grain sizes, and is suggestive that our fiducial coagulation timescales may not be short enough to reproduce observational size distributions. There are caveats with this assessment; our two discrete bin sizes do not correspond ideally to the dust templates used to fit observations, there are general challenges with obtaining these size distributions observationally and the selection effects associated with these observations. Dust in denser gas, particularly around young stellar populations, will tend to process significantly more radiation. However, black- or grey-body fits to IR SEDs are taken as a general measure of overall dust content. We await future results with the full COLIBRE model to explore size distributions further, particularly making use of radiative transfer post-processing, where we could reproduce the observational inference of sizes from FIR SEDs\footnote{Also considering how to use our two size bins to parametrise a continuous size distribution for radiative modelling.}, and explore the importance of dense vs. diffuse media. In the meantime, we consider an additional boost factor to the coagulation that can be applied to our lower resolution models in future work.

\section{Discussion \& Conclusions}
\label{sec:discussion}

We have presented a model for the life-cycle of dust and its interaction with the physics of the ISM, implemented using the SWIFT SPH code \citep{Schaller24}, as part of a network of physics modules constituting the COLIBRE suite of galaxy formation simulations \citep{Schaye25}. This model uses an established two-size (0.01~${\rm \upmu m}$ and 0.1~${\rm \upmu m}$) grain population paradigm \citep{Hirashita15}, as well as three chemical species of dust (carbonaceous grain, as well as Mg- and Fe-endmember silicate species), providing a model lightweight enough to be carried particle-by-particle in large SPH cosmological volume simulations. This dust is seeded from stellar channels, budgeting for consistency with overall metal enrichment, and can accrue mass from the gas-phase ISM (accretion), change its size distribution (shattering and coagulation) and ultimately be destroyed (sputtering, direct shock annihilation or astration). During this life-cycle, dust properties are coupled to the cooling and heating processes, accounting for the role of dust in e.g. absorbing and re-emitting ionising radiation, providing surfaces for molecule formation and depleting metals from the gas-phase that would otherwise contribute to cooling and heating processes. We presented results from (25~cMpc)$^3$ moderate resolution ($m_{\rm gas} \approx m_{\rm DM} \approx 10^6 \; {\rm M_\odot}$) simulations showing the effect of parameter and modelling choices on the distribution, depletion and evolution of dust, as well as results comparing galaxy dust distributions and scaling relations at redshift $z=0$ to observations. 

Generally, we find that our model produces agreement with a variety of observed properties. Our \texttt{Fiducial} run exhibits a cosmic dust density ($\rho_{\rm c, dust}$) evolution in good agreement with a wide range of observations across different redshift windows and observational probes (Fig.~\ref{fig:cdd}). We put particular stock in the recent \citet{Chiang25} results, which integrate the cosmic infrared background to probe all dust emission. We regard this as more representative of the $\rho_{\rm c, dust}$ value that we compute for the simulations, where we sum all the dust in the simulated volume. This shows a new level of agreement across the redshift range for cosmological models of dust evolution, compared to various literature models we consider. While it is arguable that the predicted $\rho_{\rm c, dust}$ evolution is flatter than observed over the range $0<z<2$, on the high side compared to low redshift observations and vice-versa, this is consistent with volume effects in a relatively small 25$^3$~cMpc$^3$ box that lacks the massive halos that assemble vigorously at high redshift but are largely evacuated of ISM at low redshift.

The $z=0$ galaxy-dust mass function (GDMF, Fig.~\ref{fig:gdmf}) also shows broad concordance with observations, favouring the set of observations that find more galaxies at the highest dust masses \citep[e.g.][]{Clark15, Clemens13}, consistent with being on the high side of low-$z$ $\rho_{\rm c, dust}$ observations. The simultaneous low-mass slope preference for lower-normalisation GDMFs \citep[e.g.][]{Beeston18}, however, may point to a different shape of the GDMF overall. Dust scaling relations (Fig.~\ref{fig:scaling}) convey a similar story, showing generally good agreement, agreeing better with higher normalised observations, particularly compared to atomic and molecular gas masses \citep[e.g.][]{Orellana17}. We note that the inclusion of small grains in our coupled dust model helps to improve the general agreement relative to the \citetalias{Ploeckinger25} dust model, increasing the cosmic H$_2$ densities (Fig.~\ref{fig:coupling}) and H$_2$ abundance at fixed dust mass (bottom right, Fig.~\ref{fig:scaling}).

A curious tension appears when considering the \dtg-Z relation (Fig.~\ref{fig:dtg}); despite the aforementioned dust comparisons suggesting similar or higher dust content in \texttt{Fiducial} compared to observations at $z=0$, the \dtg{}-Z relation is generally low compared to some of the observational data. Making use of the multiple metallicity calibrations of \citet{DeVis19}, however, we find that a calibration yielding higher metallicities \citep[e.g.][]{Tremonti04} confers excellent agreement with our simulations (explored further in Appendix~\ref{ap:zcal}). This combination of dust content (constrained well by FIR observations), depletion constraints (largely saturated in the ISM) and the \dtg{} ratio may suggest a preference for the higher gas-phase metallicity calibration.

A more elusive tension is seen for grain sizes (Fig.~\ref{fig:s2l}), where we see a systematic overabundance of small grains relative to the majority of \citet{Relano20, Relano22} observations. It is worth noting that inferring these observed grain sizes is difficult, requiring fitting multi-component fits to the FIR SED, with many potential systematic effects. However, we do find improved agreement using more exclusive selections of cold, dense gas where coagulation can be effective. It is unclear to what extent the observationally inferred values are more representative of dust in dense regions, and we look to investigate this connection by re-deriving size distributions using virtual FIR SEDs in future work. Other works in isolated galaxy simulations have found they can reproduce these more large-grain size dominated distributions, with MW-like net extinction curves\footnote{We defer analysis of extinction curves to future work, concerned with the post-processed extinction and attenuation properties of galaxies.} at the same time, with assumptions on how these simple two-size approximations constrain a broader size distribution \citep[e.g.][]{Hou19, Granato21, Dubois24}. Still, in our coupled cooling model these smaller grains also help increase the cosmic density of molecular gas, and its correlation with dust masses. We note that the grain size distributions are resolution-dependent, with increased (reduced) small-to-large grain size ratios for higher (lower) resolution. We will explore applying a coagulation boost to lower resolution runs in upcoming work as a means to reduce the small-to-large grain ratios.

Fig.~\ref{fig:gallery} shows the intra- and inter-galaxy variations in dust grain populations, that may have important implications for the observable properties, largely mediated by dust. Observationally, these effects may seem intractable from complex star-dust geometries, which can, by themselves, yield significant diversity in attenuation properties \citep[e.g.][]{Narayanan18, Trayford20}. However we can account for the grain population and geometric effects self-consistently using radiative transfer forward modelling, tracing the propagation of starlight through the dusty medium and accounting for absorption, scattering and re-emission. This was done for e.g. the EAGLE simulations using the SKIRT code \citep{Camps15,Camps20} assuming a fixed dust mix \citep{Camps16, Camps18, Trayford17}, but when coupled with a live dust model such as ours, it can offer broad insight into the properties shaping attenuation, and its potential effects on galaxy property inference. We will explore forward modelling with dust radiative transfer in upcoming work.

Detailed dust evolution within galaxy formation models that incorporate a cold phase, with molecular gas and dust physics, are emerging, but dust typically only has a \textit{passive} role  \citep[e.g. for {\sc fire-2} by][where radiation pressure, shielding, heating/cooling and prescriptions for molecule formation are unaware of the modelled dust]{Choban22}. Among dust evolution models in general, it is still uncommon to have dust \textit{coupling} to gas physics through the self-consistent influence on heating and cooling processes associated with grains \citep[though see e.g.][]{Vogelsberger19, Osman20}. \citet{Granato21} present such a model within cosmological zoom-in simulations, using the MUPPI framework, which implements semi-analytic models for multiphase ISM within GADGET SPH particles \citep{Murante10, Murante15, Valentini17, Valentini19, Valentini23}, and noting the important role of this coupling. This model shows the coevolution of dust and gas over cosmic time within simulations of MW-like galaxies, exhibiting universally lower small-to-large grain size ratios, in closer agreement with observational inference \citep{Relano20, Relano22} than we find here. Cosmological simulations adapting this dust model are presented in \citet{Parente22} and \citet{RagoneFigueroa24} at a baryonic mass resolution $m_g = 7 \times 10^{6} \; {\rm M_\odot}$. While the $z=0$ population of \citet{Parente22} presents remarkable concordance with the galaxy dust mass functions and consistency with \dtg{}-$Z$ observations, this is in the context where the mass-metallicity relation is too low and the cosmic star formation rate density and dust masses are under-predicted at high redshift, attributed to the direct application of the \citet{Granato21} model to cosmological simulations. \citet{RagoneFigueroa24} calibrate this model to improve the cosmic evolution for dust and gas properties in follow-up simulations, similar to those we present in our higher resolution runs with an explicit cold phase, with more results imminent from the larger and higher-fidelity volumes of the COLIBRE suite. 

Overall, we present a model generally consistent with a range of observations at low redshift, and able to reproduce the evolution of dust content well across cosmic time. While lightweight, the power of this model is in its self-consistent interaction with the panoply of sophisticated physics modules constituting the large-scale upcoming COLIBRE simulations, while setting the stage for novel insight into a profusion of contemporary dust-mediated observations through a forward modelling approach.

\section*{Acknowledgements}
We are thankful to our referee whose detailed suggestions have helped to improve a number of parts of the manuscript significantly.
We gratefully acknowledge everyone who has contributed to the COLIBRE project for their input.
This work used the DiRAC@Durham facility managed by the
Institute for Computational Cosmology on behalf of the STFC
DiRAC HPC Facility (www.dirac.ac.uk). The equipment was funded by BEIS capital funding via STFC capital grants ST/K00042X/1, ST/P002293/1, ST/R002371/1 and ST/S002502/1, Durham University and STFC operations grant ST/R000832/1. DiRAC is part of
the National e-Infrastructure. 
JT acknowledges support of his STFC \textit{Early Stage Research and Development grant} (ST/X004651/1). We are grateful to Maarten Baes, Andrea Gebek and Nick Andreadis for helpful discussions in the latter phase of this work. This work made generous use of the pipeline tools developed for the COLIBRE project, representing many years of work; including the \texttt{swiftsimio} \citep{Borrow20} and SOAP\footnote{\href{https://github.com/openjournals/joss-reviews/issues/7851}{\tt github.com/openjournals/joss-reviews/issues/7851}} tools. SP acknowledges support from the Austrian Science Fund (FWF) through project V 982-N. ABL acknowledges support by the Italian Ministry for Universities (MUR) program `Dipartimenti Di Eccellenza 2023-2027' within the Centro Bicocca dinCosmologia Quantitativa (BiCoQ), and support by UNIMIB's Fondo Di Ateneo Quota Competitiva (project 2024-ATEQC-0050). CSF acknowledges support from ERC Advanced Investigator Grant, DMIDAS (GA 786910). This project has received funding from the Netherlands Organization for Scientific Research (NWO) through research programme Athena 184.034.002. 

\section*{Data Availability}

The data presented in this article can be made available to individuals upon reasonable request. The \texttt{SWIFT} and CHIMES codes are already available at \href{http://www.swiftsim.com}{\tt www.swiftsim.com} and \href{https://richings.bitbucket.io/chimes/home.html}{\tt richings.bitbucket.io/chimes/home.html}, respectively. The COLIBRE code and data are scheduled for release some time after the initial COLIBRE papers.

\bibliographystyle{mnras}
\bibliography{example}

\appendix

\section{Additional Parameter Variations}
\label{sec:parvar}

Here, we expand upon the dust model parameter variations presented in the main text, particularly those that are not directly calibrated, using the additional runs of Table~\ref{table:sims}. Comparing these variations to our \texttt{Fiducial} run demonstrates how these parameters influence the model. 

\subsection{Smaller small grain sizes}
\label{sec:smallcomp}

\begin{figure}
    \includegraphics[width=0.49\textwidth]{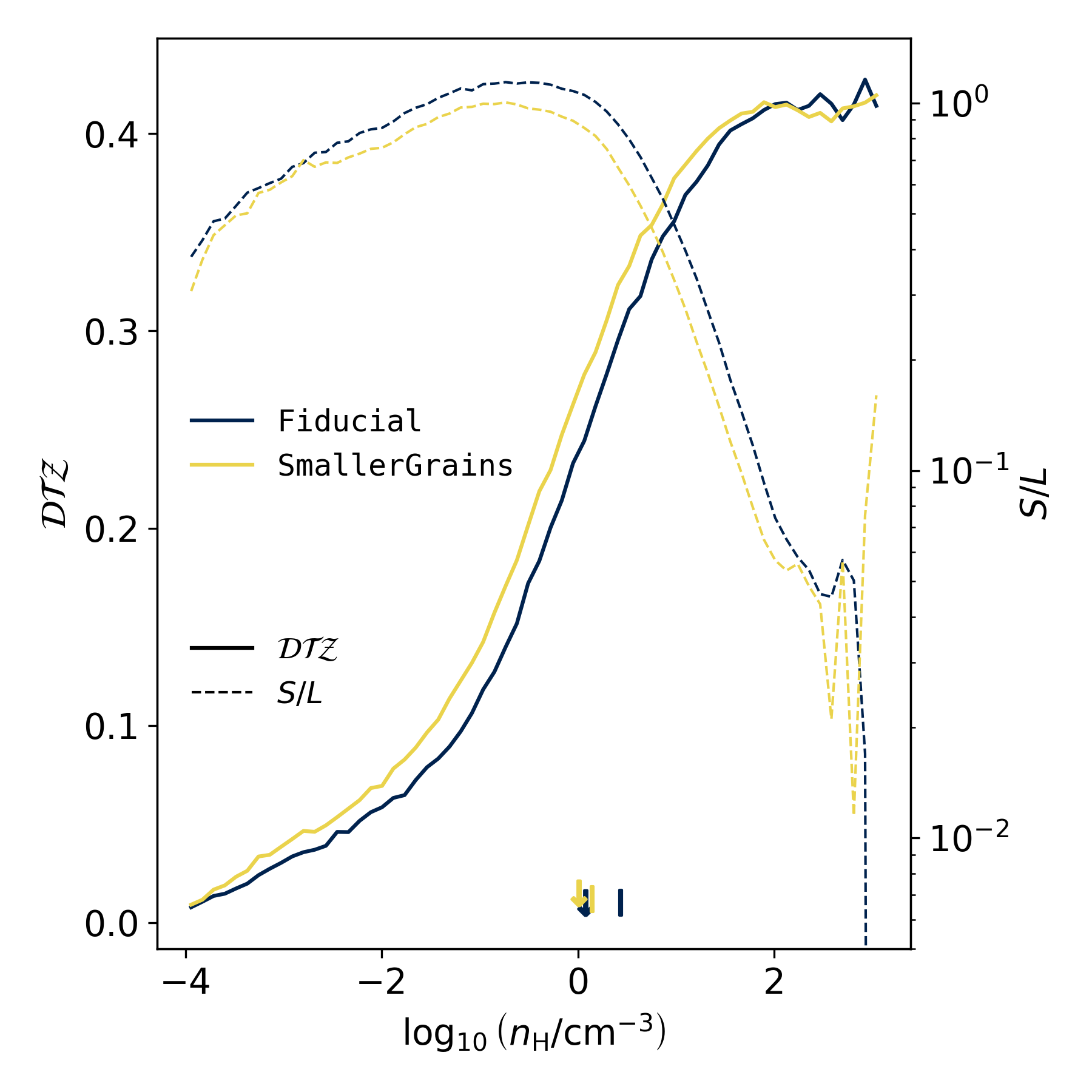}
    \caption{Properties of dust grains in logarithmic bins of gas density (as lower panel of Fig~\ref{fig:clump}), comparing the \texttt{Fiducial} run with the \texttt{SmallerGrains} run. Solid lines show the dust-to-metal (\dtz{}) ratios (left $y$-axis), with dashed-lines indicating the mass in small to large grains (right $y$-axis). Density transitions for molecular gas ($\rho_{\rm 0.5,\; H2}$) and  dust ($\rho_{\rm 0.5,\; dust}$). Using a small grain radius of 0.005 \micrometer{} (relative to the fiducial 0.01 \micrometer) slightly increases the \dtz{}, and increases $S/L$. The H\textsc{i}-H$_2$ transition density is reduced by $\approx0.3$~dex for the \texttt{SmallerGrains} run.
    }
    \label{fig:smallcomp}
\end{figure}

The choice of small-grain size influences all evolutionary process rates, as well as rates of dust-mediated cooling and surface nucleation. To show the effect of this choice, we compare to a run using a small-grain size of 0.005 \micrometer; a value used in a number of dust evolution models \citep[e.g.][]{Hirashita15, Hou19, Granato21}. We plot the \dtz{} and small-to-large grain mass ratios in logarithmic bins of density in Fig.~\ref{fig:smallcomp}.

We see that smaller grains shift the depletion-limited \dtz{} regime to lower densities, and boost the $S/L$ grain ratio by $\approx 0.1$~dex relative to \texttt{Fiducial} in $-2 \lesssim \log_{10} n_{\rm H}/{\rm cm^{-3}} \lesssim 2$. While the change in these \dtz{}-density relations is relatively small, we see that this acts in concert with the smaller grain size to produce a more marked shift in the H$_2$ transition (vertical line mark) to lower densities; a higher number of grains with higher surface-to-volume ratios at intermediate densities speeds up H$_2$ nucleation.

\subsection{Turbulent diffusion}
\label{sec:diffcomp}

\begin{figure}
    \includegraphics[width=0.49\textwidth]{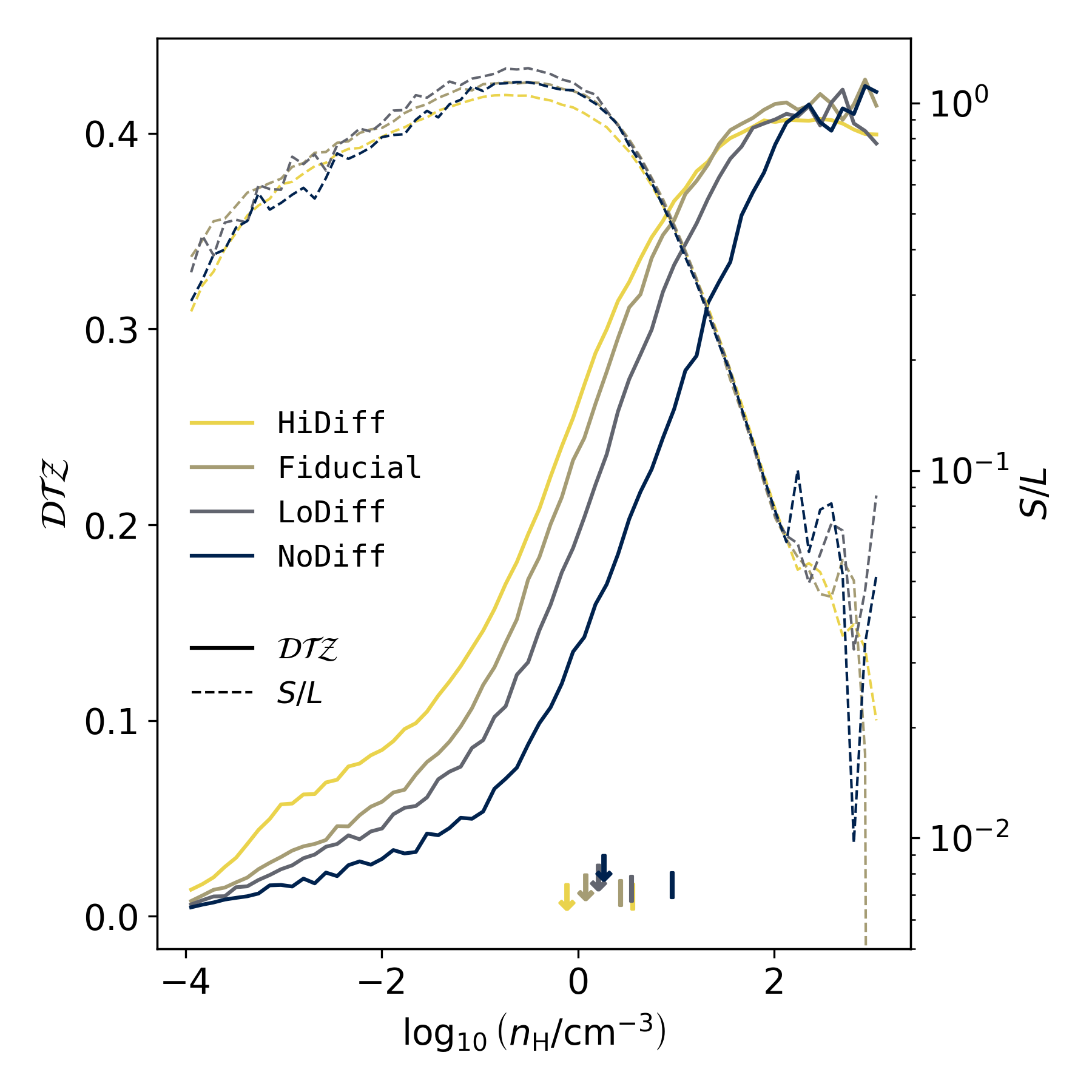}
    \caption{Properties of dust grains in logarithmic bins of gas density - following Fig.~\ref{fig:smallcomp}, now comparing runs with varying levels of turbulent diffusion. Lines from blue to yellow indicate increasing diffusion, from no diffusion, 0.1$\times$ fiducial, fiducial to 10$\times$ fiducial. We see that stronger diffusion leads to a lower transition density from low to high \dtz{}, with only a marginal change to the $S/L$ ratio. The transition density for dust (H$_2$) decreases by 0.3~dex (0.4~dex) between the no and high diffusion coefficient cases. 
    }
    \label{fig:diffcomp}
\end{figure}

Another process that can influence the evolution of the dust, and its associated influence on gas physics, is turbulent diffusion. As described in \ref{sec:diff}, the diffusion of dust follows that of gas-phase elements described in \citet{Correa25}. The amount of diffusion is modulated by a diffusion constant, $C_{\rm d}$, where elements or grains are diffused more readily for higher values. We compare a number of variations with different $C_{\rm d}$ values, shown in Fig.~\ref{fig:diffcomp}.

These runs are shaded from dark blue to light yellow,
in order of lowest to highest diffusion coefficient, where the effective $C_{\rm d}$ values are 0, 0.001, 0.01 and 0.1 respectively. We see that increased diffusion leads to a lower $\rho_{\rm 0.5 \; dust}$, pushing the saturated \dtz{} regime to lower densities. This is relatively intuitive; as growth by accretion is important in our model and is limited by the depletion in dense gas, diffusion of dust grains from dense gas into lower density gas with lower depletion allows this dust to grow further.

At ISM densities ($n_{\rm H} \gtrsim 1 \; {\rm cm^{-3}}$) we see the largest difference between the \texttt{NoDiff} and {\tt LoDiff} runs. This shows the important role even low-level of diffusion can have in simulating dust growth; diffusion allows transport of dust into gas particles that have not been directly enriched by stellar sources. As accretion relies on the dust-gas interaction, without these low seeding levels of grains, dust cannot grow to deplete elements at the levels we observe in galaxies. This demonstrates that while the dust mass in our simulated galaxies is much greater than directly seeded (see e.g. Fig.~\ref{fig:rhoT}), the production and distribution of seed grains is crucial to manifest the dust content of galaxies. We also note that the H\textsc{i}-H$_2$ transition density $\rho_{\, 0.5, \; {\rm H_2}}$ differs most for the \texttt{NoDiff} run; with dust only residing in gas particles that have been directly enriched, the important role of dust in nucleating H molecules is lacking in neighbouring gas.

We see that the differences between the \texttt{Fiducial} and \texttt{HiDiff} appears greater at low density. We attribute this to the direct effect of diffusion. While at the lowest densities gas cannot accrete efficiently, the increased flux of dust into these particles boosts their \dtz{} levels.

\section{Extinction Efficiencies and Dust Coupling}
\label{sec:qext}

In \S\ref{sec:couple} we detail how we ignore the influence of extinction efficiency, $\langle Q_{{\rm ext}} \rangle$, in deriving our dust size-dependent scaling factor $s_{\rm dust}$, for dust-dependent cooling rates. We justify this by assuming $\langle Q_{{\rm ext},S} \rangle / \langle Q_{{\rm ext},L} \rangle \approx 1$.

\begin{figure}
    \includegraphics[width=0.49\textwidth]{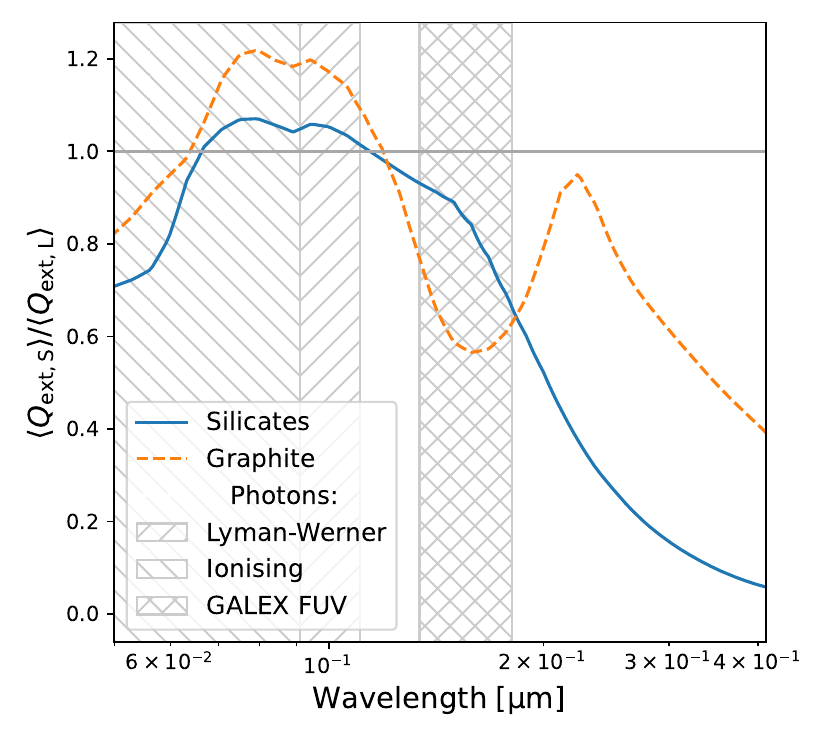}
    \caption{Ratio of extinction efficiencies for large and small grains as a function of the wavelength of incident light. The curves are derived for graphite and silicates taking the extinction ratio of large and small grains derived for a log-normal size distribution centred on the grain radius (0.1 and 0.01${\rm \upmu m}$, respectively) and with a standard deviation of 0.75~dex, using grain optical and calorimetric properties of \citet{Draine84}. Relevant photon energy ranges are hatched as indicated. We see that for Lyman-Werner and ionising photons, the extinction efficiency ratio is centred around $\langle Q_{{\rm ext},S} \rangle / \langle Q_{{\rm ext},L} \rangle \approx 1$ for both chemical species, suggesting we may ignore $Q_{\rm ext}$ for our purposes.}
    \label{fig:qext}
\end{figure}

In Fig.~\ref{fig:qext} we plot the $\langle Q_{{\rm ext},S} \rangle / \langle Q_{{\rm ext},L} \rangle$ ratio for silicate and carbonaceous grains from the \citet{Draine84} model as a function of the wavelength ($\lambda$) of incident light. We see that while this ratio is not constant, and can fall to values  $\langle Q_{{\rm ext},S} \rangle / \langle  Q_{{\rm ext},L} \rangle \ll 1$ for optical wavelengths ($\lambda \approx 0.4 \; {\rm \upmu m}$), it is closer to 1 for the Lyman-Werner and ionising radiation bands ($\lambda \lesssim 0.1 \; {\rm \upmu m}$) the extinction processes we consider important for dust physics, varying $\approx$20$\%$ about unity. These differences are small relative to the particle-particles differences in grain masses and species distributions, which justifies our simplifying assumption that extinction optical depths can be taken to scale with the surface-to-volume ratio of grains alone, and our use of the $s_{\rm dust}$ factor to scale dust rates (see equation~\ref{eq:sdust}).

\section{Dust and Metallicity Calibrations}
\label{ap:zcal}

A curious tension in the development of these models has been the diverging conclusions on the abundances of dust in our simulated galaxies that can be drawn through comparison to different data sets. While comparisons like the cosmic dust density, galaxy dust mass functions and dust scaling relations (see Figs.~\ref{fig:cdd}, \ref{fig:gdmf}, \ref{fig:scaling}) may point to a slight surfeit of dust, \dtg{} and \dtz{} values seem to imply a deficit of dust in the \texttt{Fiducial} run compared to observation (see e.g. Fig.~\ref{fig:dtg}). 

One way to reconcile these results could be via the gas-phase metallicity calibrations used for the observations; while absolute dust masses can be measured in a relatively reliable way through black- or grey-body fits to the infrared emission of galaxies, absolute gas-phase metallicity calibrations can be uncertain, with $\approx$~0.5~dex differences between average measured $Z_{\rm gas}$ for a given $M_\star$ \citep[see the distinct branches in the literature gas-phase mass-metallicity relations, e.g.][]{Kewley08}.

\begin{figure}
    \includegraphics[width=0.49\textwidth]{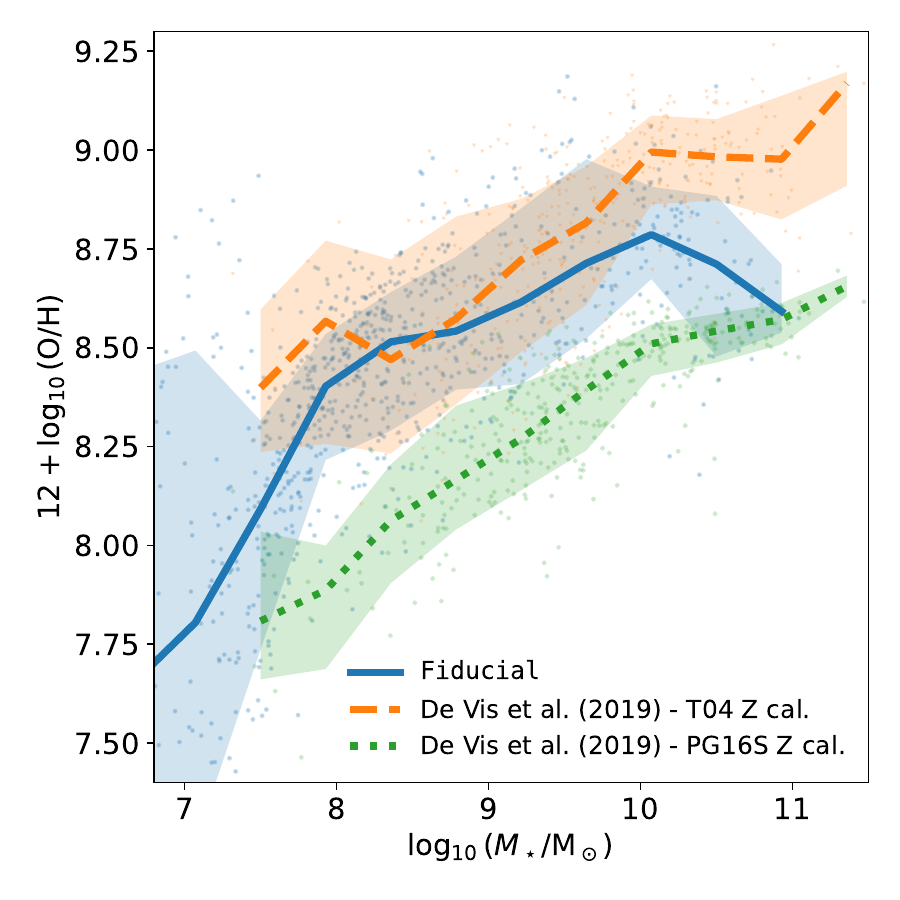}
    \caption{The gas-phase oxygen abundances as a function of stellar mass of galaxies. We plot the \texttt{Fiducial} run results, where oxygen abundances are measured in cold, dense gas ($\log_{10}T/{\rm K} < 4.5$, $\log_{10}n_{\rm H}/{\rm cm^{-3}} > -1$) and within a 50~ckpc aperture about the subhalo potential minima. \textit{Lines} show median values while \textit{data points} show individual galaxies. \textit{Shaded regions} indicate 16-84th percentile ranges. We compare the \texttt{Fiducial} run with the \textit{Dustpedia} results of \citet{DeVis19} for different observational calibrations; the [OII]-exclusive calibration of \citet[][PS16S]{Pilyugin16} and that of \citet[][T04]{Tremonti04}. We see the simulation data overlaps best with T04, particularly at intermediate metallicities.
    }
    \label{fig:mzgas}
\end{figure}

In Fig.~\ref{fig:mzgas} we make use of the \textit{Dustpedia} data presented in \citet{DeVis19}, which usefully provides a number of different inferred metallicity values (in the form of oxygen abundance) based on differing indices and calibrations. In particular, we plot the [OII]-exclusive calibration of \citet{Pilyugin16} (taken as the default in \citealt{DeVis19}) and the calibration of \citet{Tremonti04} \citepalias{Tremonti04}. We see a $\approx$0.5~dex offset between the two median relations, with the \citetalias{Tremonti04} calibration comparing best to our simulation for $\log_{10}(M_\star/{\rm M_\odot}) \lessapprox 10.25$.  

\begin{figure}
    \centering
        \includegraphics[width=0.49\textwidth]{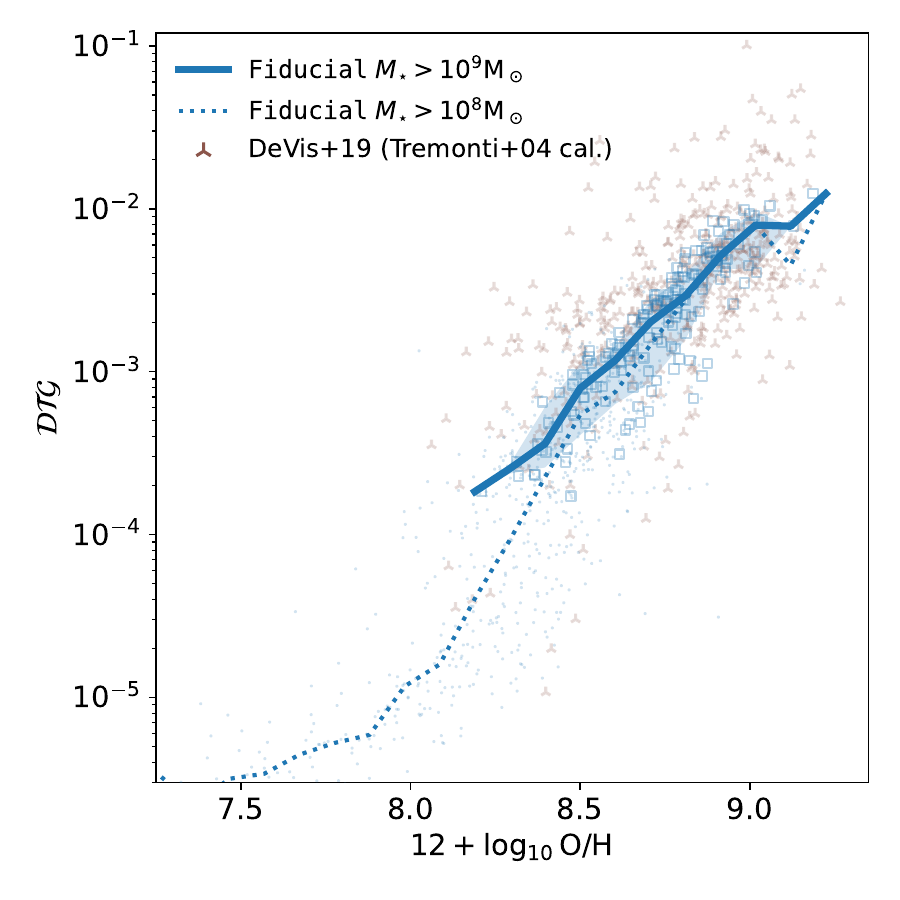}
        \includegraphics[width=0.49\textwidth]{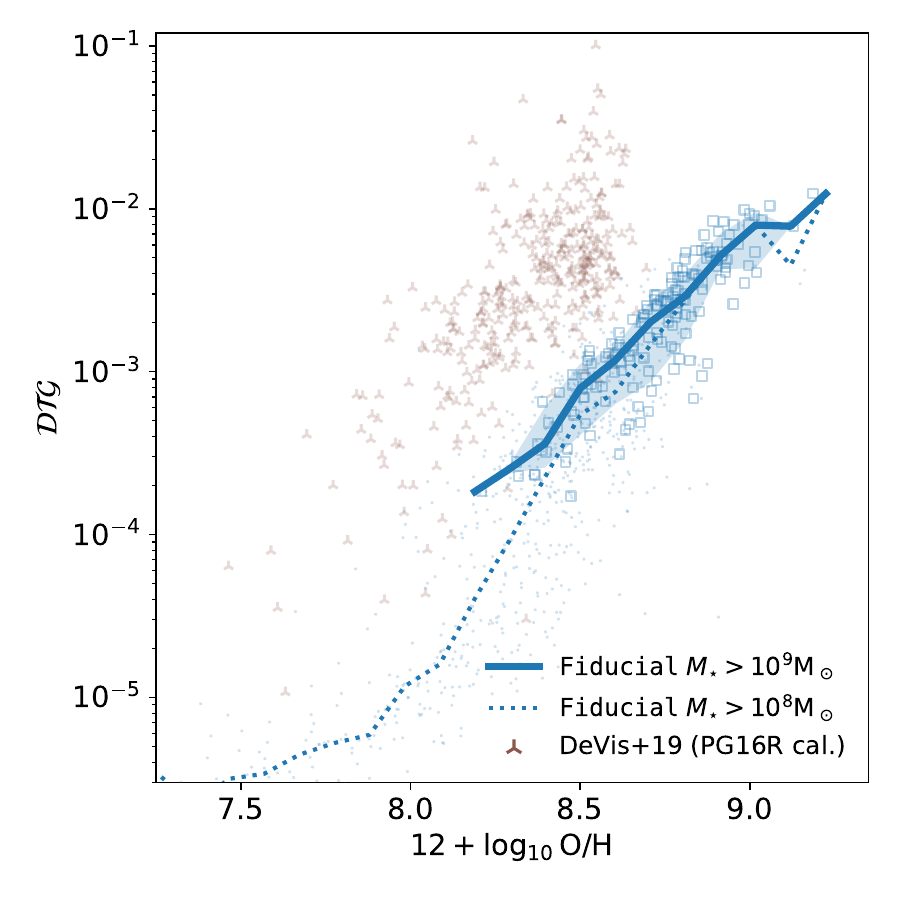}
    \caption{The dust-to-gas (\dtg{}) ratio of galaxies as a function of their gas-phase oxygen abundances, measured in cold, dense gas, with the same galaxy selection and ${\rm O/H}$ aggregation as described in Fig.~\ref{fig:dtg}. \textit{Blue lines} show median values while \textit{blue squares} show individual galaxies. The \textit{solid line} shows the median for galaxies with $M_{\star} > 10^9~{\rm M_\odot}$ and 16-84th percentile ranges indicated by the shaded region. The \textit{dotted line} shows the median relation for all processed galaxies. \textit{Top panel} shows comparison to the \citet{DeVis19} data calibrated using the [OII]-exclusive calibration of \citet[][PS16S]{Pilyugin16}. \textit{Bottom panel} compares instead of the \citet{DeVis19} data calibrated using the \citet[][T04]{Tremonti04} approach. We see that the \texttt{Fiducial} galaxies better reproduce the T04 calibration in this instance.
    } 
    \label{fig:dtg_cal}
\end{figure}

In Fig.~\ref{fig:dtg_cal}, we then plot the \dtg{} values of \texttt{Fiducial} simulation galaxies as a function of their gas-phase oxygen abundances, against the \citet{DeVis19} data for these two calibrations. We see that the simulated data agrees much better with the \dtg{} values when using the \citet{Tremonti04}-calibrated metallicities. This demonstrates how our modelling shows a preference for certain metallicity calibrations.

Given that dust in the ISM of our simulated galaxies is set largely by the balance between accretion and the (strong) depletion of metals (Fig.~\ref{fig:rhoT}), with little headroom to deplete more given viable dust grain chemistries (Fig.~\ref{fig:depletion} and associated discussion), our modelling suggests dust masses are a feasible means to help break the degeneracy in absolute metallicity calibrations. 

\bsp
\label{lastpage}
\end{document}